\documentclass[11pt]{article}

\usepackage[utf8]{inputenc} % Required for inputting international characters
\usepackage[T1]{fontenc} % Output font encoding for international characters
\usepackage{graphicx}
\usepackage{float} % locate figures
\usepackage{caption}
\usepackage{url}
\usepackage{mathtools} % load AMS maths
\usepackage{amsmath} % for math spacing
\usepackage{amssymb} % for math spacing
\usepackage{colortbl} % colour for table
\usepackage[table]{xcolor} % define colours
\usepackage{multirow} % multirow cell in table
\usepackage{array} % table cell, paragraph column aligned in the middle
\usepackage[margin=2.5cm]{geometry} % an easy way to change page layout
\usepackage{siunitx} % use this package to enter SI units
\usepackage{fancyhdr}
\usepackage[makeroom]{cancel}
\usepackage{amsthm}
\usepackage{subcaption}
\usepackage{tikz}
\usetikzlibrary{patterns,angles,quotes,arrows}
\usepackage[numbers]{natbib}
\usepackage{cite, enumitem}
\newtheoremstyle{mystyle}%    % Name
  {}%                         % Space above
  {}%                         % Space below
  {}%                         % Body font
  {}%                         % Indent amount
  {\bfseries}%                % Theorem head font
  {:}%                        % Punctuation after theorem head
  { }%                        % Space after theorem head, ' ', or \newline
  {}%                         % Theorem head spec (can be left empty, meaning `normal')
\newtheorem{remark}{Remark}
\newtheorem{theorem}{Theorem}
\newtheorem{prop}{Proposition}
\newtheorem{lemma}{Lemma}
\newtheorem*{pf}{Proof}
\newtheorem*{mydef}{Definition}
\newtheorem*{problem}{Problem}
\newtheorem{cor}{Corollary}

\let\emptyset\varnothing

\title{Quantitative Resilience of Linear Driftless Systems}
\author{Jean-Baptiste Bouvier, Kathleen Xu and Melkior Ornik\thanks{Jean-Baptiste Bouvier, Kathleen Xu and Melkior Ornik are with the Department of Aerospace Engineering and the Coordinated Science Laboratory, University of Illinois at Urbana-Champaign, Urbana, IL 61801, USA.\hfill \break
e-mail: bouvier3@illinois.edu \& ksxu2@illinois.edu \& mornik@illinois.edu \hfill \break
This work was supported by an Early Stage Innovations grant from NASA’s Space Technology Research Grants Program, grant no. 80NSSC19K0209. This material is partially based upon work supported by the United States Air Force AFRL/SBRK under contract no. FA864921P0123.}}

\date{}
\begin{document}
\maketitle

\begin{abstract}
    This paper introduces the notion of quantitative resilience of a control system. Following prior work, we study systems enduring a loss of control authority over some of their actuators. Such a malfunction results in actuators producing possibly undesirable inputs over which the controller has real-time readings but no control. By definition, a system is resilient if it can still reach a target after a partial loss of control authority. However, after a malfunction, a resilient system might be significantly slower to reach a target compared to its initial capabilities. We quantify this loss of performance through the new concept of quantitative resilience. We define such a metric as the maximal ratio of the minimal times required to reach any target for the initial and malfunctioning systems. Naive computation of quantitative resilience directly from the definition is a complex task as it requires solving four nested, possibly nonlinear, optimization problems. The main technical contribution of this work is to provide an efficient method to compute quantitative resilience. Relying on control theory and on two novel geometric results we reduce the computation of quantitative resilience to a single linear optimization problem. We illustrate our method on two numerical examples: an opinion dynamics scenario and a trajectory controller for low-thrust spacecrafts.
    % These examples show that quantitative resilience is a great tool to decide the redundancy of actuators in the design of a resilient system.
\end{abstract}

\section{Introduction}
When failure is not an option, critical systems are built with enough redundancy to endure actuator failure \citep{NASA_redundancy}.
The study of this type of malfunction typically considers either actuators locking in place \citep{actuator_lock} or actuators losing effectiveness but remaining controllable \citep{partial_LOC, partial_failure}.
However, when actuators can be subject to damage or hostile takeover, the malfunction may result in the actuators producing undesirable inputs over which the controller has real-time readings but no control.
This type of malfunction has been discussed in \citep{IFAC} under the name of \emph{loss of control authority} over actuators and encompasses scenarios where actuators and sensors are under attack \citep{actuator_attack}.

In the setting of loss of control authority, undesirable inputs are observable and can have a magnitude similar to the controlled inputs, while in classical robust control the undesirable inputs are not observable and have a small magnitude compared to the actuators' inputs \citep{Tubes1971, kurzhanski2002}.
The results of \citep{journal_paper} showed that a controller having access to the undesirable inputs is considerably more effective than a robust controller.  

After a partial loss of control authority over actuators, a target is said to be \emph{resiliently reachable} if for any undesirable inputs produced by the malfunctioning actuators there exists a control driving the state to the target \citep{IFAC}. However, after the loss of control the malfunctioning system might need considerably more time to reach its target compared to the initial system. 
In this work we thus introduce the concept of quantitative resilience for control systems in order to measure the delays caused by the loss of control authority over actuators. 
While concepts of quantitative resilience have been previously developed for water infrastructure systems \citep{water_qr} or for nuclear power plants \citep{nuclear_pp_qr}, such concepts only work for their specific application.

In this work we formulate quantitative resilience as the maximal ratio of the minimal times required to reach any target for the initial and malfunctioning systems.
This formulation leads to a nonlinear minimax optimization problem with an infinite number of equality constraints.
Because of the complexity of this problem, a straightforward attempt at a solution is not feasible.
While for linear minimax problems with a finite number of constraints the optimum is reached on the boundary of the constraint set \citep{max-min_programming}, such a general result does not hold in the setting of semi-infinite programming \citep{semi-infinite_programming} where our problem belongs.
However, the fruitful application of the theorems of \citep{liberzon, Neustadt} stating the existence of time-optimal controls combined with the specific geometry of our problem, allow us to derive two bang-bang results concerning some nonlinear optimization problems.
Then, the quantitative resilience of a driftless system is reduced to single linear optimization problem.

As a first step toward the study of quantitative resilience for linear systems we restrict this work to driftless systems. Indeed, we will see that even with these simple dynamics the theory is already sufficiently rich. Furthermore, one can find an abundance of driftless systems in robotics \citep{robotics}.

The contributions of this paper are fourfold.
First, we introduce the concept of quantitative resilience for systems enduring a loss of control authority over some of their actuators.
Secondly, in the course of solving our central problem, we determine a simple analytical solution to a related nonlinear optimization problem with applications not restricted only to control theory.
Thirdly, we provide an efficient method to compute the quantitative resilience of driftless systems by simplifying a nonlinear problem of four nested optimizations into a single linear optimization problem.
Finally, based on quantitative resilience and controllability we establish a necessary and sufficient condition to verify if a system is resilient.

The remainder of the paper is organized as follows. Section~\ref{section:preliminaries} introduces preliminary results concerning resilient systems and defines quantitative resilience. 
Section~\ref{section:optimization} establishes three optimization results that will prove crucial for the computation of quantitative resilience. 
To evaluate this metric we need the minimal time for the system to reach a target before and after the loss of control authority.
We calculate this minimal time for the initial system in Section~\ref{section:initial dynamics} and for the malfunctioning system in Section~\ref{section:malfunctioning dynamics}.
Section~\ref{section:r_q} is the pinnacle of this work as we design an efficient method to compute quantitative resilience and assess whether a system is resilient or not. 
In Section~\ref{section:examples} our theory is applied to an opinion dynamics scenario and on a linear trajectory controller for a low-thrust spacecraft.
Appendices~\ref{apx:proofs} and \ref{apx:proof with x_N and x_M} gather all the lemmas required to prove our central nonlinear optimization result.
The continuity of the minimal malfunctioning reach time is proved in Appendix~\ref{apx:continuity}.
Finally, we compute the dynamics of the low-thrust spacecraft in Appendix~\ref{apx:bar B}.

\vspace{2mm}

\emph{Notation:} 
We use $\partial X$ to denote the boundary of a set $X$ and its interior is denoted $X^\circ := X \backslash \partial X$.
Set $X$ is symmetric if for all $x \in X$, we have $-x \in X$.
The convex hull of a set $X$ is denoted with $co(X)$.
The set of integers from $1$ to $N$ is $[N] := \{1, \hdots, N\}$. 
We denote the set of nonnegative real numbers with $\mathbb{R}^+ := [0, \infty)$ and we use the subscript $_*$ to exclude zero, for instance $\mathbb{R}_*^+ := (0, \infty)$.
In the real $n$-dimensional space $\mathbb{R}^n$ we denote the Euclidean norm with $\| \cdot \|$ and the unit sphere with $\mathbb{S} := \{ x \in \mathbb{R}^n : \|x\| = 1\}$.
The ball of radius $\varepsilon$ centered on $x$ with $B_\varepsilon(x) := \big\{ y \in \mathbb{R}^n : \|y - x\| \leq \varepsilon \big\}$.
The scalar product of vectors is denoted by $\langle \cdot, \cdot \rangle$.
For $x \in \mathbb{R}_*^n$ and $y \in \mathbb{R}_*^n$ we denote as $\widehat{x, y}$ the signed angle from $x$ to $y$ in the 2D plane containing both of them. We take the convention that the angles are positive when going in the clockwise orientation.
We say that $x \in [x_1, x_2] \subset \mathbb{R}^n$ if there exists $\lambda \in [0,1]$ such that $x = \lambda x_1 + (1-\lambda)x_2$.
The infinity-norm of a vector $x \in \mathbb{R}^n$ is $\| x \|_\infty := \max \{ |x_i| : i \in [n] \}$.
The image of a matrix $A \in \mathbb{R}^{n \times m}$ is denoted $Im(A) \subset \mathbb{R}^n$, its rank is $rank(A) = \dim Im(A) \leq n$ and its norm is $\|A\| := \underset{x\, \neq\, 0}{\sup}\ \frac{\|Ax\|}{\|x\|}$.
Unless otherwise stated, the element at row $i$ and column $j$ of a matrix $A$ is denoted by $A_{i,j}$.
For square integrable functions $f: \mathbb{R} \rightarrow \mathbb{R}^n$, 
the $\mathcal{L}_2$-norm is defined as $\| f \|_{\mathcal{L}_2}^2 := \int_{t\, \in\, \mathbb{R}} \|f(t)\|^2\, dt$, and
the $\mathcal{L}_\infty$-norm is defined as $\| f \|_{\mathcal{L}_\infty} := \underset{t\, \in\, \mathbb{R}}{\sup}\ \|f(t)\|_\infty$. 
A set-valued function $\varphi$ from $X$ to $Y$ is denoted as $\varphi : X \twoheadrightarrow Y$ following \citep{inf_dim_analysis}.
The sequence $x_0, x_1, \hdots$ is denoted with $\{x_k\}$.

\section{Preliminaries and Problem Statement}\label{section:preliminaries}

As a first step toward linear systems, we begin with driftless systems governed by the differential equation
\begin{equation}\label{eq:original ODE}
    \dot{x}(t) = \bar{B} \bar{u}(t), \qquad \text{with} \quad x(0) = x_0 \in \mathbb{R}^n, \qquad \bar{u} \in \bar{U},
\end{equation}
where $\bar{B} \in \mathbb{R}^{n \times (m+p)}$ is a constant matrix. Let $u_{max} > 0$ be the bound on the input magnitude so that the set of allowable controls is 
\begin{equation}\label{eq:U bar}
    \bar{U} := \big\{  \bar{u} : \mathbb{R}^+ \rightarrow \mathbb{R}^{m+p} :  \|u\|_{\mathcal{L}_\infty} \leq u_{\max} \big\}.
\end{equation}
After a malfunction, the system loses control authority over $p$ of its $m+p$ initial actuators. Because of the malfunction the initial control input $\bar{u}$ is split into the remaining controlled inputs $u$ and the undesirable inputs $w$. Without loss of generality we always consider the columns $C$ representing the malfunctioning actuators to be at the end of $\bar{B}$. We split the control matrix accordingly: $\bar{B} = \big[ B\ C\big]$. Then, the dynamics become
\begin{align}\label{eq:splitted ODE}
    \dot{x}(t) &= Bu(t) + Cw(t), \qquad x(0) = x_0 \in \mathbb{R}^n, \quad u \in U, \quad w \in W,
\end{align}
with 
\begin{equation}\label{eq:function sets U and W}
    U := \big\{ u : \mathbb{R}^+ \rightarrow \mathbb{R}^m : \|u\|_{\mathcal{L}_\infty} \leq u_{max} \big\} \qquad \text{and} \qquad W := \big\{ w : \mathbb{R}^+ \rightarrow \mathbb{R}^p : \|w\|_{\mathcal{L}_\infty} \leq u_{max} \big\}.
\end{equation}
We will use the concept of \emph{controllability} of \citep{liberzon}.
\begin{mydef}
    A system following the dynamics \eqref{eq:original ODE} is controllable if for all target $x_{goal} \in \mathbb{R}^n$ there exists a control $\bar{u} \in \bar{U}$ and a time $T$ such that $x(T) = x_{goal}$.
\end{mydef}
We recall here the definition of the \emph{resilience} of a system introduced in \citep{journal_paper}.
\begin{mydef}
    A system following the dynamics \eqref{eq:original ODE} is resilient to the loss of $p$ of its actuators corresponding to the matrix $C$ as above, if for all undesirable inputs $w \in W$ and all target $x_{goal} \in \mathbb{R}^n$ there exists a control $u \in U$ and a time $T$ such that the state of the system \eqref{eq:splitted ODE} reaches the target at time $T$, i.e., $x(T) = x_{goal}$.
\end{mydef}

Notice that in previous work \citep{IFAC, journal_paper} the $\mathcal{L}_2$-norm of the inputs was constrained. In this work we consider instead $\mathcal{L}_\infty$ bounds because they are more widely used in applications. Therefore, most of the resiliency conditions of \citep{IFAC, journal_paper} do not directly apply here. 
We will establish a simple necessary condition for this new setting using only basic linear algebra.

\begin{prop}\label{prop:resilient full rank}
    If the system \eqref{eq:original ODE} is resilient to the loss of $p$ actuators, then the system $\dot x(t) = B u(t)$ is controllable.
\end{prop}
\begin{pf}
    Let $y \in \mathbb{R}^n$, $x_{goal} := y + x_0 \in \mathbb{R}^n$ and $w \in W$ such that $w(t) = 0$ for all $t \geq 0$. Since the system is resilient, there exist $u \in U$ and $T \geq 0$ such that
    \begin{equation*}
        x_{goal} = x(T) = \int_0^T \hspace{-3mm} \dot x(t)\, dt + x_0 = \int_0^T \hspace{-3mm} Bu(t)\, dt + x_0 = Bz + x_0 \qquad \text{with}  \qquad z := \int_0^T \hspace{-3mm} u(t)\, dt \in \mathbb{R}^m.
    \end{equation*}
    Then, $x_{goal} - x_0 = Bz = y \in Im(B)$, so $rank(B) = n$ and $\dot x(t) = B u(t)$ is controllable.  $\quad \blacksquare$
\end{pf}

By definition, a resilient system is still capable of reaching any target after losing control authority over $p$ of its actuators. 
However, the time for this malfunctioning system to reach a target might be considerably larger than the time needed for the initial system to reach the same target.
We introduce these two times for the target $x_{goal} \in \mathbb{R}^n$ and the target distance $d := x_{goal} - x_0 \in \mathbb{R}^n$.
\begin{mydef}
    The \emph{nominal reach time} $T_N^*$ is the shortest time required to reach the target for the initial system following \eqref{eq:original ODE}: 
    \begin{equation}\label{eq:nominal reach time}
        T_N^*(d) := \underset{\bar{u}\, \in\, \bar{U} }{\inf} \Big\{ T \geq 0 : \int_0^T \hspace{-3mm} \bar{B} \bar{u}(t)\, dt = d \Big\}.
    \end{equation}
\end{mydef}
\begin{mydef}
    The \emph{malfunctioning reach time} $T_M^*$ is the shortest time required to reach the target for the malfunctioning system following \eqref{eq:splitted ODE} when the undesirable input is chosen to make that time the longest:
    \begin{equation}\label{eq:malfunctioning reach time}
        T_M^*(d) := \underset{w\, \in\, W}{\sup} \Bigg\{ \underset{u\, \in\, U}{\inf} \Big\{ T \geq 0 : \int_0^T \hspace{-3mm} Bu(t) + Cw(t)\, dt = d \Big\} \Bigg\}.
    \end{equation}
\end{mydef}
By definition, if the system is controllable, then $T_N^*(d)$ is finite for all $d \in \mathbb{R}^n$, and if it is resilient, then $T_M^*(d)$ is finite. We only write the argument $d$ of $T_N^*$ and $T_M^*$ when their dependency on $d$ needs to be highlighted.

\begin{mydef}
    The \emph{ratio of reach times} in the direction $d \in \mathbb{R}^n$ is
    \begin{equation}\label{eq:t(d)}
        t(d) :=  \frac{T_M^*(d)}{T_N^*(d)}.
    \end{equation}
\end{mydef}
After the loss of control, the malfunctioning system can take up to $t(d)$ times longer than the initial system to reach the target $d + x_0$. Since the performance is degraded by the undesirable inputs, one can easily show that $t(d) \geq 1$.  We take the convention that $t(d) = +\infty$ whenever $T_M^*(d) = +\infty$, regardless of the value of $T_N^*(d)$.

\begin{remark}\label{rmk:d=0}
    The case $T_N^*(d) = T_M^*(d) = 0$ can only happen when $d = 0$, because $x(0) = x_0 = x_{goal}$. To make this case coherent with \eqref{eq:t(d)} and \eqref{eq:r_q} we choose $\frac{T_N^*(0)}{T_M^*(0)} = 1$.
\end{remark}

We now define the \emph{quantitative resilience} of a system.
\begin{mydef}
     The \emph{quantitative resilience} $r_q$ of a system following \eqref{eq:splitted ODE} is the inverse of the maximal ratio of reach times, i.e.,
     \begin{equation}\label{eq:r_q}
         r_q := \frac{1}{\underset{d\, \in\, \mathbb{R}^n}{\sup}\, t(d)} = \underset{d\, \in\, \mathbb{R}^n}{\inf}\ \frac{T_N^*(d)}{T_M^*(d)}.
     \end{equation}
\end{mydef}
Quantitative resilience can be defined in exactly the same way for general control systems, but we focus on linear driftless systems in this work.
For a resilient system, $r_q \in (0,1]$. The closer $r_q$ is to $1$, the smaller is the loss of performance caused by the malfunction.

Quantitative resilience $r_q$ depends on matrices $B$ and $C$, i.e., on the actuators that are producing undesirable inputs.
One could also define the quantitative resilience of a system to the loss of any $p$ actuators by taking the minimal $r_q$ over all configurations of malfunctions.

Computing $r_q$ requires solving four nested optimization problems over continuous constraint sets, with three of them being infinite-dimensional function spaces. A brute force approach to this problem is doomed to fail. Thus, we focus on the following problem.

\begin{problem}
    Establish an efficient method to compute $r_q$.
\end{problem}

\section{Optimization on Polytopes}\label{section:optimization}

In this section, we introduce three novel optimization results on polytopes that will be needed to compute quantitative resilience. The proofs rely heavily on geometric arguments. 

\begin{mydef}
    A \emph{polytope} in $\mathbb{R}^n$ is a compact intersection of finitely many half-spaces.
\end{mydef}
With this definition polytopes are considered to be convex. They are an $n$-dimensional generalization of planar polygons.
\begin{mydef}
    A \emph{vertex} of a set $X \subset \mathbb{R}^n$ is a point $x \in X$ such that if there are $x_1 \in X$ and $x_2 \in X$ with $x \in [x_1, x_2]$, then $x = x_1 = x_2$. 
\end{mydef}
With this definition, a vertex of a polytope corresponds to the usual understanding of a vertex of a polytope. We can now state our first optimization result on polytopes.

\begin{theorem}\label{thm:minimum on the vertices}
    Let $d \in \mathbb{S}$, $X$ and $Y$ two polytopes of $\mathbb{R}^n$ with $X \subset Y$. Then, there exists a vertex $v$ of $X$ such that $\|y^*(v) - v\| = \underset{x\, \in\, X}{\min}\ \|y^*(x) - x\|$, with
    \begin{equation}\label{eq:optimization over X and Y}
        y^*(x) := \arg \underset{y\, \in\, Y}{\max} \big\{ \|y-x\| : y-x \in \mathbb{R}^+ d\big\}, \quad \text{for}\ x \in X.
    \end{equation}
\end{theorem}
\begin{pf}
    First, we will show that the maximum in \eqref{eq:optimization over X and Y} exists and has a unique argument.
    For $x \in X$, the set $S(x) := \big\{y \in Y : y-x \in \mathbb{R}^+ d\big\}$ is compact since it is a closed subset of the compact set $Y$. Since $X \subset Y$, we have $x \in S(x)$ and so $S(x) \neq \emptyset$. The map $: y \mapsto \|y-x\|$ is continuous, so it reaches a maximum over $S(x)$. This maximum is reached at the point of $Y$ the furthest of $x$ in direction $d$, i.e., at a unique $y^*(x) \in \partial Y$.
    Then the map $y^* : X \rightarrow \partial Y$ introduced in \eqref{eq:optimization over X and Y} is well-defined.
    
    Note that $y^*(x)$ is the linear projection of $x$ along $+d$ onto $\partial Y$, so $y^*$ is continuous. Thus, the function $: x \mapsto \| y^*(x) - x \|$ is continuous and reaches a minimum over the compact and nonempty set $X$. This minimum is not necessarily achieved uniquely over $X$.
    
    Let $x^* \in X$ such that $\|y^*(x^*) - x^*\| = \underset{x\, \in\, X}{\min}\ \|y^*(x) - x\|$.
    Since $x^*$ must minimize the distance between itself and $y^*(x^*) \in \partial Y$, with $X \subset Y$ obviously $x^* \in \partial X$.
    For contradiction purposes assume now that $x^*$ is not on a vertex of $\partial X$. Let $S_x$ be the surface of lowest dimension in $\partial X$ such that $x^* \in S_x$ and $\dim S_x \geq 1$.
    
    Let $v$ be a vertex of $S_x$, $a := v - x^*$ and $x(\alpha) := x^* + \alpha a$ for $\alpha \in \mathbb{R}$. Notice that $x(0) = x^*$ and $x(1) = v$. Due to the choice of $v$, the convexity of $S_x$ and $x^*$ not being a vertex, there exists $\varepsilon > 0$ such that $x(\alpha) \in S_x$ for all $\alpha \in [-\varepsilon, 1]$.
    We also define the lengths $L(\alpha) := \|y^*\big(x(\alpha)\big) - x(\alpha)\|$ and $L^* := L(0)$.
    All these definitions are illustrated on Figure~\ref{fig:min x reached on a vertex}.
    
    \begin{figure}[htbp!]
        \centering
        \begin{tikzpicture}[scale = 1]
            \draw[] (-5, 0.5) -- (-4.5, 1) -- (0, 1) -- (1, 0.5);
            \node at (1.3, 0.5) {$\partial X$};
            \node at (-4.2, 1.2) {$S_x$};
            \node at (-4.5, 0.5) {$X$};
            \draw[] (-5, 3.5) -- (-2, 3.5) -- (0, 4) -- (1, 3);
            \node at (1.3, 3) {$\partial Y$};
            \node at (-4.5, 3) {$Y$};
            
            \draw[thick, |->] (-5.5, 1.5) -- (-5.5, 2.5);
            \node at (-5.3, 2) {$d$};
            
            \node at (-2, 0.7) {$x^*$};
            \draw[fill] (-2, 1) circle (0.05);
            \draw[thick] (-1, 0.9) -- (-1, 1.1);
            \node at (-1, 0.7) {$x(\alpha_0)$};
            \draw[thick] (-3.5, 0.9) -- (-3.5, 1.1);
            \node at (-3.5, 0.7) {$x(-\varepsilon)$};
            \draw[fill] (0, 1) circle (0.05);
            \node at (0, 0.7) {$v$};
            
            \draw[dashed] (-1, 1) -- (-1, 3.75);
            \node at (-0.5, 2.5) {$L(\alpha_0)$};
            \draw[dashed] (-2, 1) -- (-2, 3.5);
            \node at (-1.8, 2.5) {$L^*$};
            \draw[dashed] (-3.5, 1) -- (-3.5, 3.5);
            \node at (-3, 2.5) {$L(-\varepsilon)$};
            
            \draw[dotted] (-3.5, 3.5) -- (-1, 3.75);
            \draw[fill] (-2, 3.65) circle (0.05);
            \node at (-2, 3.9) {$z$};
        \end{tikzpicture}
        \caption{The convexity of $Y$ compels $x^*$ to be on a vertex.}
        \label{fig:min x reached on a vertex}
    \end{figure}
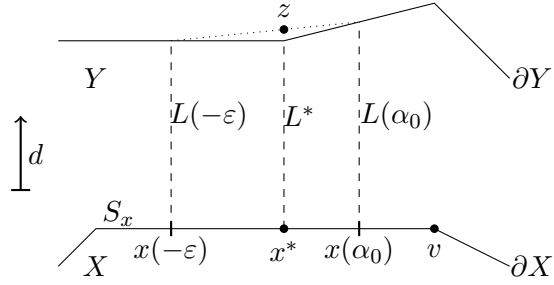
    
  Since $\|d\| = 1$ and $y^*\big(x(\alpha)\big) - x(\alpha) \in \mathbb{R}^+ d$, we have $L(\alpha) = \langle y^*\big(x(\alpha)\big) - x(\alpha), d \rangle$.
  By definition of $x^*$, we know that $L^* \leq L(\alpha)$ for all $\alpha \in [-\varepsilon, 1]$. Assume that there exists $\alpha_0 \in (0, 1]$ such that $L^* < L(\alpha_0)$. We introduce the convexity coefficient $\beta := \frac{\alpha_0}{\alpha_0 + \varepsilon}$ and then
  \begin{align*}
      L^* &= \beta L^* + (1 - \beta)L^* < \beta L(-\varepsilon) + (1-\beta)L(\alpha_0) \\
      &< \beta \langle y^*\big(x(-\varepsilon)\big) - x(-\varepsilon), d \rangle + (1-\beta) \langle y^*\big(x(\alpha_0)\big) - x(\alpha_0), d \rangle  = \langle z - x^*, d\rangle,
  \end{align*}
  with $z := \beta y^*\big(x(-\varepsilon)\big) + (1-\beta) y^*\big(x(\alpha_0)\big)$. Indeed, note that $\beta x(-\varepsilon) + (1-\beta) x(\alpha_0) = x^*$, and $z - x^* \in \mathbb{R}^+ d$. By convexity of $Y$, $z \in Y$, which contradicts the optimality of $x^*$. Thus, there is no $\alpha_0 \in (0,1]$ such that $L^* < L(\alpha_0)$. Therefore, for all $\alpha \in [0,1]$, $L(\alpha) = L^*$. By taking $\alpha = 1$, we have $x(\alpha) = v$, so the minimum $L^*$ is also reached on a vertex $v$ of $X$.     $\quad \blacksquare$
\end{pf}

Theorem~\ref{thm:minimum on the vertices} will help us calculate the malfunctioning reach time $T_M^*$ of resilient systems.
The following optimization result concerns a ratio of two optimization problems and will simplify the calculation of $r_q$.

\begin{prop}\label{prop:r(d) well-defined}
    For $d \in \mathbb{S}$, a compact set $Y \subset \mathbb{R}^n$ of dimension $n$ with $x \in Y^\circ$ and $-x \in Y^\circ$, the ratio 
    \begin{equation}\label{eq:r(d)}
        r_Y(d, x) := \frac{\underset{y\, \in\, Y}{\max} \big\{ \|y+x\| : y+x \in \mathbb{R}^+ d\big\}}{\underset{y\, \in\, Y}{\max} \big\{ \|y-x\| : y-x \in \mathbb{R}^+ d\big\} }
    \end{equation}
    exists and is finite.
\end{prop}
\begin{pf}
   The sets $S^{\pm}(x) := \big\{ y \in Y : y \pm x \in \mathbb{R}^+ d\big\}$ are both closed subsets of $Y$, so they are compact. They are nonempty because $x \in S^-(x)$ and $-x \in S^+(x)$. Functions $f^{\pm} : S^\pm \rightarrow \mathbb{R}$ defined as $f^\pm(y) := \|y \pm x\|$ are both continuous, so they each reach a maximum over respectively $S^{\pm}$. 
   Let $y^-$ be the argument of the maximum at the denominator of $r_Y(d,x)$. Because of its optimality $y^- \in \partial Y$. Since $x \in Y^\circ$, we have $\|y^- - x\| > 0$ for any $d \in \mathbb{S}$. Then, $r_Y(d,x)$ exists and is finite.
   $\quad \blacksquare$
\end{pf}

\vspace{2mm}

\begin{theorem}\label{thm:minimum of r(d) collinear with x}
    If $\, Y$ is a convex polytope in $\mathbb{R}^n$ with $\dim Y = n$, $x \in Y^\circ$ and $-x \in Y^\circ$, then
    $$\underset{d\, \in\, \mathbb{S}}{\max}\ r_Y(d, x) = \max \{r_Y(x, x),\ r_Y(-x,x)\}.$$
\end{theorem}
\begin{pf}
    Set $Y$ is compact because it is a polytope. Then, all the assumptions of Proposition~\ref{prop:r(d) well-defined} are satisfied and thus the ratio $r_Y(d,x)$ exists and is finite.
    Vector $x$ is fixed, so we write $r(d) := r_Y(d,x)$ to alleviate the notation.
    The proof of this theorem relies on numerous geometric arguments and is quite long. To help the reader, we divided the proof into several lemmas all gathered in Appendix~\ref{apx:proofs}.

    Let $d_0 \in \mathbb{S}$. Since $d_0$ and $x$ are two vectors of $\mathbb{R}^n$, there exists a two-dimensional plane $\mathcal{P}$ passing through the origin that contains both of $d_0$ and $x$.
    Let $d \in \mathbb{S} \cap \mathcal{P}$ and $y^+$, $y^-$ be the arguments of the two maxima in \eqref{eq:r(d)}, so that $r(d) = \frac{\|y^+ + x\|}{\|y^- - x\|}$.
    Because of their optimality, $y^+ \in \partial Y$ and $y^- \in \partial Y$. Additionally, $\pm x \in \mathcal{P}$ and $d \in \mathcal{P}$, so $y^\pm \in \mathcal{P}$. 
    
    We will study how $r(d)$ varies when $d$ takes values in $\mathbb{S} \cap \mathcal{P}$.
    To introduce all the necessary definitions we first consider the case where the rays directed by $d$, $y^-$ and $y^+$ all intersect the same face of $\partial Y$ as illustrated on Figure~\ref{fig:angles illustration}.

    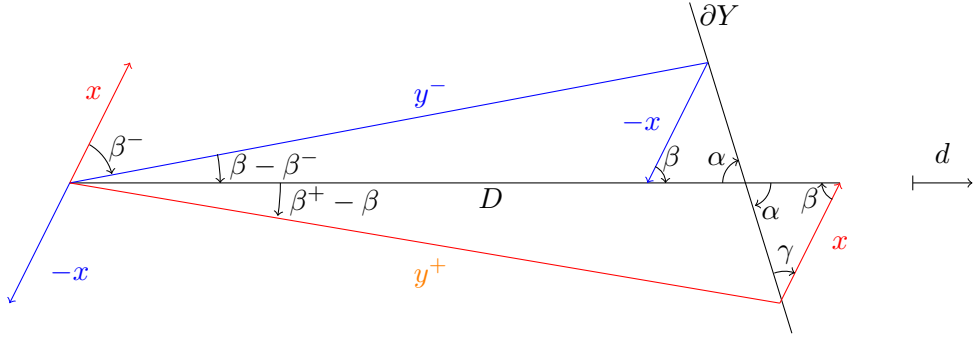
\begin{figure}[htbp!]
        \centering
        \begin{tikzpicture}[scale = 0.8]
            
            \draw[|->] (12,0) -- (13,0);
            \node at (12.5, 0.5) {$d$};
            \draw (-2, 0) -- (10.8, 0);
            \node at (5, -0.25) {$D$};
            
            \draw (8.3, 3) -- (10, -2.5);
            \node at (8.8, 2.8) {$\partial Y$};
            
            \draw[<-, blue] (-3, -2) -- (-2, 0);
            \draw[->, red] (-2, 0) -- (-1, 2);
            \node at (-1.6, 1.5) {\textcolor{red}{$x$}};
            \node at (-2, -1.5) {\textcolor{blue}{$-x$}};
            
            \draw[blue] (-2,0) -- (8.6, 2);
            \node at (4, 1.5) {\textcolor{blue}{$y^-$}};
            \draw[->, blue] (8.6, 2) -- (7.6, 0);
            \node at (7.5,1) {\textcolor{blue}{$-x$}};
            
            \draw[red] (-2,0) -- (9.8,-2);
            \node at (4,-1.5) {\textcolor{orange}{$y^+$}};
            \draw[->, red] (9.8, -2) -- (10.8, 0);
            \node at (10.8, -1) {\textcolor{red}{$x$}};
            
            \draw[<-] (-1.3, 0.15) arc (15:60:0.8);
            \node at (-1.05, 0.6) {$\beta^-$};
            \draw[->] (8.85, 0) arc (180:110:0.4);
            \node at (8.75, 0.4) {$\alpha$};
            \draw[<-] (10.05, -1.5) arc (70:110:0.5);
            \node at (9.9, -1.2) {$\gamma$};
            \draw[<-] (7.9,0) arc (0:65:0.3);
            \node at (8, 0.4) {$\beta$};
            \draw[<-] (10.5, 0) arc (180:245:0.3);
            \node at (10.3, -0.3) {$\beta$};
            \draw[->] (9.65, 0) arc (0:-70:0.4);
            \node at (9.65, -0.5) {$\alpha$};
            
            \draw[<-] (0.5, 0) arc (0:11:2.5);
            \node at (1.4, 0.3) {$\beta - \beta^-$};
            
            \draw[->] (1.5, 0) arc (0:-9:3.5);
            \node at (2.4, -0.3) {$\beta^+ - \beta$};
            
        \end{tikzpicture}
        \caption{Illustration of $y^+$ leading and outside with $y^-$ trailing and inside.}
        \label{fig:angles illustration}
    \end{figure}

    We introduce the signed angles $\alpha := \widehat{d, \partial Y}$, $\beta := \widehat{x, d}$, $\beta^+ := \widehat{x, y^+}$ and $\beta^- := \widehat{x, y^-}$. These angles are represented on Figure~\ref{fig:angles illustration} and they all take value in $[0, 2\pi)$. 
    Let $\alpha_0$ be the value of $\alpha$ when $\beta = 0$, i.e., when $d$ is positively collinear with $x$.
    
    \begin{mydef}
        We say that $y^+$ is \emph{leading} and $y^-$ is \emph{trailing} when $\beta^- < \beta < \beta^+$, and conversely when $\beta^+ < \beta < \beta^-$, we say that $y^-$ is \emph{leading} and $y^+$ is \emph{trailing}. 
    \end{mydef}
    
    If $\beta^+ = \beta$, then $y^+$ is collinear with $d$. So $x$ is also collinear with $d$ because $y^+ + x \in \mathbb{R}^+ d$. Then, $d$ and $y^-$ are collinear with $x$, so $\beta^- = \beta^+ = \beta \in \{0, \pi\}$. The same conclusion is reached when $\beta^- = \beta$.
    Thus, only when $\beta \in \{0, \pi\}$, neither $y^+$ nor $y^-$ are leading or trailing.
    For each $d \in \mathbb{S} \cap \mathcal{P}$ we define $D := \underset{y\, \in\, Y}{\max} \big\{ \|y\| : y \in \mathbb{R}^+ d \big\}$, whose existence is justified by the compactness of $Y$.
    \begin{mydef}
        We say that $y^\pm$ is \emph{outside} when $\|y^\pm \pm x\| > D$. Otherwise $y^\pm$ is \emph{inside}.
    \end{mydef}
    
    We parametrize all directions $d \in \mathbb{S} \cap \mathcal{P}$ by the angle $\beta$. Then, we will study how $r(d)$ varies when $\beta \in [0, 2\pi)$.
    We first establish in Lemma~\ref{lemma:r(d) cst on faces} of Appendix~\ref{apx:proofs} that the ratio $r(d)$ is constant on the faces of $\partial Y$. Then, $r(d)$ can only change when $d$ crosses a vertex.
    Prior to studying vertex crossings we need to find the range of $\beta$ for which $y^\pm$ is leading or trailing and outside or inside. 
    Lemma~\ref{lemma:leading and outside} establishes the following statements.
     \begin{itemize}[itemsep=2.5pt, topsep=2.5pt]
        \item If $\beta \in (0, \pi)$, then $y^+$ is leading. If $\beta \in (\pi, 2\pi)$, then $y^-$ is leading.
        \item If $\alpha + \beta \in (0, \pi)$, then $y^+$ is outside. If $\alpha + \beta \in (\pi, 2\pi)$, then $y^-$ is outside.
    \end{itemize}
    Based on Lemma~\ref{lemma:alpha + beta reparametrization} we can rewrite the above bullet list using only the angle $\alpha + \beta$.
     \begin{itemize}[itemsep=2.5pt, topsep=2.5pt]
        \item If $\alpha + \beta \in (\alpha_0, \pi)$, then $y^+$ is leading and outside.
        \item If $\alpha + \beta \in (\pi, \alpha_0 + \pi)$, then $y^+$ is leading and inside.
        \item If $\alpha + \beta \in (\alpha_0 + \pi, 2\pi)$, then $y^-$ is leading and outside.
        \item If $\alpha + \beta \in (2\pi, \alpha_0 + 2\pi)$, then $y^-$ is leading and inside.
    \end{itemize}
    The polygon $Y \cap \mathcal{P}$ can then be divided into four regions as illustrated by Figure~\ref{fig:global view}.
    
    \begin{figure}[htbp!]
        \centering
        \begin{tikzpicture}[scale = 0.7]
            
            \draw[dotted] (-8, 0) -- (8, 0);
            \draw[dotted] (1.25, 5) -- (-1.1, -4.5);
            
            \node at (-6, 4) {$\alpha + \beta \in (\alpha_0, \pi)$};
            \node at (6, 4) {$\alpha + \beta \in (\pi, \alpha_0 + \pi)$};
            \node at (6, -4) {$\alpha + \beta \in (\alpha_0 + \pi, 2\pi)$};
            \node at (-6, -4) {$\alpha + \beta \in (2\pi, \alpha_0 + 2\pi)$};
            
            \draw[<-, red] (-1, 0) -- (0, 0);
            \draw[->, blue] (0, 0) -- (1, 0);
            \node at (-1.4, 0.1) {\textcolor{red}{$x$}};
            \node at (1.6, -0.1) {\textcolor{blue}{$-x$}};
            
            \draw (-6, -2) -- (-2.5, 2.5) -- (1.23, 4.7) -- (3.8, 3) -- (6, -1) -- (3, -3) -- (-1, -4) -- (-6, -2);
            \node at (-3.6, 1.8) {$\partial Y$};
            \node at (0.8, 4.8) {$v_\pi$};
            \node at (-0.5, -4.2) {$v_{2\pi}$};
            \draw[<-] (-4, 0) arc (0:45:0.5);
            \node at (-3.7, 0.2) {$\alpha_0$};

            \draw (0,0) -- (-2, 4);
            \draw[blue] (0, 0) -- (-2.3, 2.6);
            \draw[->, blue] (-2.3, 2.6) -- (-1.3, 2.6);
            \draw[red] (0, 0) -- (-0.75, 3.57);
            \node at (-1.9, 1.5) {\textcolor{blue}{$y^-$}};
            \draw[->, red] (-0.75, 3.57) -- (-1.75, 3.57);
            \draw[->] (-0.2, 0) arc (180:117:0.2);
            \node at (0, 2.5) {\textcolor{red}{$y^+$}};

            \draw (0, 0) -- (6, 1.6);
            \draw[red] (0, 0) -- (4.9, 1);
            \draw[blue] (0, 0) -- (4.6, 1.5);
            \draw[->, blue] (4.6, 1.5) -- (5.6, 1.5);
            \node at (3.5, 1.5) {\textcolor{blue}{$y^-$}};
            \draw[->, red] (4.9, 1) -- (3.9, 1);
            \node at (4, 0.5) {\textcolor{red}{$y^+$}};
            \draw[->] (-0.3, 0) arc (180:20:0.3);

            \draw (0, 0) -- (6, -1.95);
            \draw[blue] (0, 0) -- (4.75, -1.85);
            \draw[red] (0, 0) -- (5.4, -1.4);
            \draw[->, blue] (4.75, -1.85) -- (5.75, -1.85);
             \node at (2.8, -1.5) {\textcolor{blue}{$y^-$}};
            \draw[->, red] (5.4, -1.4) -- (4.4, -1.4);
            \node at (3.9, -0.6) {\textcolor{red}{$y^+$}};
            \draw[->] (-0.4, 0) arc (180:-10:0.4);
            
            \draw (0, 0) -- (-6, -3.5);
            \draw[red] (0, 0) -- (-3.85, -2.85);
            \node at (-2, -1.9) {\textcolor{red}{$y^+$}};
            \draw[blue] (0, 0) -- (-5, -2.35);
            \node at (-3.8, -1.5) {\textcolor{blue}{$y^-$}};
            \draw[->, red] (-3.85, -2.85) -- (-4.85, -2.85);
            \draw[->, blue] (-5, -2.35) -- (-4, -2.35);
            \draw[->] (-0.5, 0) arc (180:-150:0.5);
            
            \node at (-0.7, 0.3) {$\beta$};

        \end{tikzpicture}
        \caption{Vector $y^\pm$ is leading or trailing and inside or outside depending solely on $\alpha + \beta$.}
        \label{fig:global view}
    \end{figure}
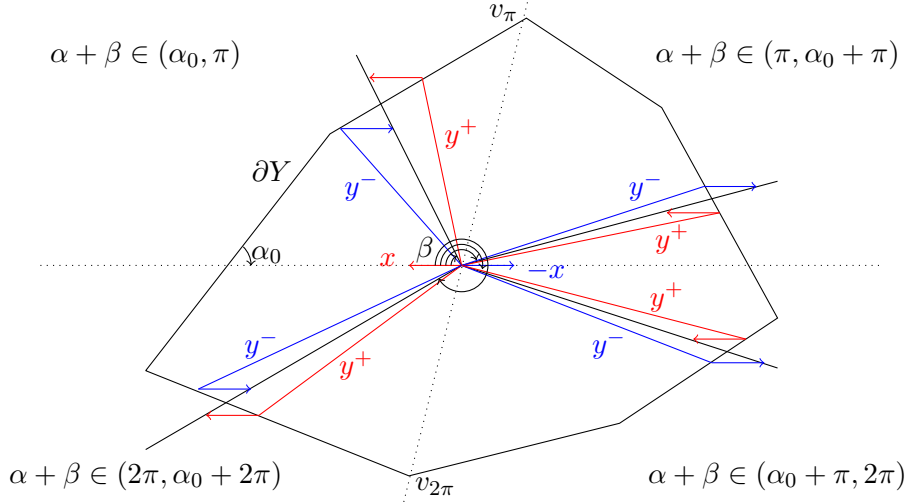
    
    According to Lemma~\ref{lemma: y^+ crossing} and Lemma~\ref{lemma: y^- crossing} in Appendix~\ref{apx:proofs}, $r(d)$ decreases during the crossing of a vertex when the leading vector $y^\pm$ is outside. This situation occurs for $\alpha + \beta \in (\alpha_0, \pi) \cup (\alpha_0 + \pi, 2\pi)$. Following Lemma~\ref{lemma:crossing with leading inside}, $r(d)$ increases during the crossing of a vertex when the leading vector $y^\pm$ is inside. This situation occurs for $\alpha + \beta \in (\pi, \alpha_0 + \pi) \cup (2\pi, \alpha_0 + 2\pi)$. The specific case of the vertices $v_\pi$ and $v_{2\pi}$ is tackled by Lemma~\ref{lemma:v pi crossing}.
    To summarize we have proved the following:
    \begin{itemize}[itemsep=2.5pt, topsep=2.5pt]
        \item if $\alpha + \beta \in (\alpha_0, \pi)$, then $y^+$ is leading and outside, so $r(d)$ is decreasing,
        \item if $\alpha + \beta \in (\pi, \alpha_0 + \pi)$, then $y^+$ is leading and inside, so $r(d)$ is increasing,
        \item if $\alpha + \beta \in (\alpha_0 + \pi, 2\pi)$, then $y^-$ is leading and outside, so $r(d)$ is decreasing,
        \item if $\alpha + \beta \in (2\pi, \alpha_0 + 2\pi)$, then $y^-$ is leading and inside, so $r(d)$ is increasing.
    \end{itemize}
    
    Then, the maximum of $r(d)$ over $\beta \in [0, 2\pi)$ happens when $\alpha + \beta \in \big\{\alpha_0, \alpha_0 + \pi\big\}$. This situation corresponds to $\beta \in \big\{0, \pi\big\}$, i.e., $d$ collinear with $x$.
    Then $\underset{d\, \in\, \mathcal{P} \cap \mathbb{S}}{\max}\ r(d) = \max \big\{ r(x), r(-x)\big\}$.
    Recall that we have worked with $d$ in the plane $\mathcal{P}$ generated by the vectors $x$ and $d_0 \in \mathbb{S}$.
    Therefore, $\underset{d\, \in\, \mathbb{S}}{\max}\ r(d) = \underset{d_0\, \in\, \mathbb{S}}{\max} \big\{ \underset{d\, \in\,  \mathbb{S} \cap \mathcal{P}(d_0)}{\max}\ r(d) \big\} = \max \big\{ r(x), r(-x) \big\}$.    $\quad \blacksquare$
\end{pf}

The ratio of optimization problems describing $r_q$ is actually more complex than the one solved in Theorem~\ref{thm:minimum of r(d) collinear with x} where the vector $x \in \mathbb{R}^n$ is fixed. Building on Theorem~\ref{thm:minimum of r(d) collinear with x} we will now introduce our optimization problems of interest.

\begin{prop}\label{prop: r_X,Y well-defined}
    Let $X$, $Y$ be two nonempty symmetric polytopes in $\mathbb{R}^n$ with $X \subset Y^\circ$ and $d \in \mathbb{S}$. Then, (i) $\underset{x\, \in\, X,\, y\, \in\, Y}{\max} \big\{ \|x + y\| : x + y \in \mathbb{R}^+d \big\}$ exists,
    (ii) $\lambda^*(x,d) := \underset{y\, \in\, Y}{\max} \big\{ \|x + y\| : x + y \in \mathbb{R}^+d \big\}$ exists for all $x \in X$,
    (iii) $\underset{x\, \in\, X}{\min} \big\{ \lambda^*(x,d) \big\}$ exists,
     and (iv) if $\dim Y = n$, then $\underset{x\, \in\, X}{\min} \big\{\lambda^*(x,d) \big\} > 0$.
\end{prop}
\begin{pf}
    \begin{enumerate}[label=(\roman*)]
        \item Let $S := \big\{ (x, y) \in X \times Y : x + y \in \mathbb{R}^+ d\big\}$. Set $S$ is a closed subset of the compact set $X \times Y$, so $S$ is compact. Since $X$ and $Y$ are nonempty, symmetric and convex, $0 \in X \cap Y$. Then, $(0,0) \in S$, so $S$ is nonempty. Function $f : S \rightarrow \mathbb{R}$ defined as $f(x,y) := \|x + y\|$ is continuous, so it reaches a maximum over $S$.
        
        \item For $x \in X$ define $S(x) := \big\{ y \in Y : x + y \in \mathbb{R}^+ d \big\}$. Since $S(x)$ is a closed subset of the compact set $Y$, $S(x)$ is compact. Since $-X \subset Y$, we have $-x \in S(x)$ and so $S(x) \neq \emptyset$. Function $f_x : S(x) \rightarrow \mathbb{R}$ defined as $f_x(y) = \|x + y\|$ is continuous, so it reaches a maximum over $S(x)$. 
        
        \item Let $y^*(x, d) := \arg \underset{y\, \in\, Y}{\max} \big\{ \|x + y\| : x + y \in \mathbb{R}^+ d \big\}$, uniquely defined as $y^*(x,d) = \lambda^*(x,d)d - x$ since $\|d\| = 1$. Lemma~\ref{lemma: lambda continuous} in Appendix~\ref{apx:continuity} shows that $\lambda^*$ is continuous in $x$ and $d$, so $y^*$ is also continuous in $x$ and $d$.
        Then, function $f : X \rightarrow \mathbb{R}$ defined as $f(x) = \|x + y^*(x,d)\|$ is continuous, so it reaches a minimum over the compact and nonempty set $X$. 
        
        \item Note that $y^*(x,d) \in \partial Y$ for all $x \in X$. Indeed, assume for contradiction purposes that there exists $\varepsilon > 0$ such that $B_\varepsilon\big(y^*(x,d) \big) \in Y$. We required $\dim Y = n$ to make this ball of full dimension, so that $z := y^*(x,d) + \varepsilon d \in Y$. Then, $x + z = \big(\lambda^*(x,d) + \varepsilon) d \in \mathbb{R}^+ d$ and $\|x + z\| = \lambda^*(x,d) + \varepsilon > \lambda^*(x,d)$ contradicting the optimality of $\lambda^*$.
        Thus, $y^*(x,d) \in \partial Y$. Since $-X \subset Y^\circ$, we have $\|x + y^*(x,d)\| > 0$ for all $x \in X$.        $\quad \blacksquare$
    \end{enumerate}
\end{pf}

Let $X$ and $Y$ be two nonempty symmetric polytopes in $\mathbb{R}^n$ with $X \subset Y^\circ$, and let $d \in \mathbb{S}$. We define
 \begin{equation}\label{eq:r_(X,Y)}
    r_{X,Y}(d) := \frac{\underset{y\, \in\, Y,\ x\, \in\, X}{\max} \big\{ \|x + y\| : x + y \in \mathbb{R}^+d \big\} }{ \underset{x\, \in\, X}{\min} \big\{ \underset{y\, \in\, Y}{\max} \big\{ \|x + y\| : x + y \in \mathbb{R}^+d \big\} \big\} }.
\end{equation}

\begin{theorem}\label{thm:varying x_M and x_N}
    If $X$ and $Y$ are two symmetric polytopes in $\mathbb{R}^n$ with $X \subset Y^\circ$, $\dim X = 1$, $\partial X = \{x, -x\}$ and $\dim Y = n$, then $\underset{d\, \in\, \mathbb{S}}{\max}\ r_{X,Y}(d) = r_{X,Y}(x)$.
\end{theorem}
\begin{pf}    
    Following Proposition~\ref{prop: r_X,Y well-defined}, $r_{X,Y}$ is well-defined. Reusing $y^*$ from the above proof, we introduce $x_M^*(d) := \arg\underset{x\, \in\, X}{\min} \big\{ \|x + y^*(x,d)\| \big\}$ and $x_N^*(d) := \arg \underset{x\, \in\, X}{\max} \big\{ \|x + y^*(x,d)\| : x + y^*(x,d) \in \mathbb{R}^+ d \big\}$. For some $d \in \mathbb{S}$ the $\arg\min$ and $\arg\max$ in the above definitions might not be unique; if so we take $x_M^*$ and $x_N^*$ to be any such argument.
    According to Theorem~\ref{thm:minimum on the vertices}, $x_M^*(d) \in \partial X$.
    We also define $y_N^*(d) := y^*\big( x_N^*(d), d\big)$ and $y_M^*(d) := y^*\big( x_M^*(d), d\big)$. Then,
    \begin{equation*}
        r_{X,Y}(d) = \frac{\underset{y\, \in\, Y}{\max} \big\{ \|y + x_N^*(d)\| : y + x_N^*(d) \in \mathbb{R}^+d \big\} }{ \underset{y\, \in\, Y}{\max} \big\{ \|y + x_M^*(d)\| : y + x_M^*(d) \in \mathbb{R}^+d \big\} } = \frac{\|x_N^*(d) + y_N^*(d)\|}{\|x_M^*(d) + y_M^*(d)\|}.
    \end{equation*}
    
    Since sets $X$ and $Y$ are symmetric, functions $y_N^*$, $y_M^*$, $x_N^*$ and $x_M^*$ are odd. Then, $r_{X,Y}$ is an even function, i.e., $r_{X,Y}(-d) = r_{X,Y}(d)$ for all $d \in \mathbb{S}$.
    
    Since $\dim X = 1$, we can take $\mathcal{P}$ to be a two-dimensional plane containing $X$. Then, we work with $d \in \mathcal{P} \cap \mathbb{S}$.
    In Lemmas~\ref{lemma: x_N^*(d) = - x_M^*(d) on faces}, \ref{lemma: x_N^*(d) = - x_M^*(d) before v_pi} and \ref{lemma: x_N^*(d) = - x_M^*(d) on regular vertices} of Appendix~\ref{apx:proof with x_N and x_M} we prove that $x_N^*(d)$ and $x_M^*(d)$ are constant and $x_N^*(d) = -x_M^*(d) \in \partial X$ for the directions $d$ not involved in the crossing of vertices $v_\pi$ and $v_{2\pi}$. These vertices were introduced in Lemma~\ref{lemma:alpha + beta reparametrization} and vertex crossing is defined in Lemma~\ref{lemma: x_N^*(d) = - x_M^*(d) before v_pi}.
    With $x_N^*(d) = -x_M^*(d)$ we have $r_{X,Y}(d) = r_Y\big( d, x_N^*(d) \big)$. Then, we can apply the proof of Theorem~\ref{thm:minimum of r(d) collinear with x} showing that the maximum of $r_{X,Y}(d)$ for $d$ not involved in the crossing of $v_\pi$ or $v_{2\pi}$ is achieved at either $x$ or $-x$.
    Lemma~\ref{lemma: x_N^*(d) crossing v_pi} states that $r_{X,Y}$ reaches a local minimum during the crossing of vertices $v_\pi$ and $v_{2\pi}$. Thus, the maximum of $r_{X,Y}$ over $d \in \mathcal{P} \cap \mathbb{S}$ is achieved at either $x$ or $-x$. 
    
    Then, $\underset{d\, \in\, \mathbb{S}}{\max}\ r_{X,Y}(d) = \underset{\mathcal{P}}{\max} \big\{ \underset{d\, \in\, \mathcal{P}\, \cap\, \mathbb{S} }{\max} r_{X,Y}(d) \big\} = \max\big\{ r_{X,Y}(x), r_{X,Y}(-x) \big\}$. Since $r_{X,Y}$ is even these two values are equal, leading to $\underset{d\, \in\, \mathbb{S}}{\max}\ r_{X,Y}(d) = r_{X,Y}(x)$.    $\quad \blacksquare$
\end{pf}

We will keep these optimization results under our belt for now and go back to the discussion of resilient systems.

\section{Dynamics of the Initial System}\label{section:initial dynamics}

We start with the initial system of dynamics \eqref{eq:original ODE} and aim to calculate the nominal reach time $T_N^*$. We introduce the set of constant inputs $\bar{U}_c := \big\{ \bar{u} \in \mathbb{R}^{m+p} : \|\bar{u}\|_\infty \leq u_{max} \big\}$.

\begin{prop}\label{prop:unperturbed time}
    For a controllable system \eqref{eq:original ODE} and $d = x_{goal} - x_0 \in \mathbb{R}^n$, the infimum $T_N^*(d)$ of \eqref{eq:nominal reach time} is achieved with a constant control input $\bar{u}^* \in \bar{U}_c$.
\end{prop}

\begin{pf}
    Dynamics \eqref{eq:original ODE} are linear in $x$ and $\bar{u}$. Set $\bar{U}$ defined in \eqref{eq:U bar} is convex and compact. 
    The system is controllable, so $x_{goal}$ is reachable.
    The assumptions of Theorem 4.3 of \citep{liberzon} are satisfied, leading to the existence of a time optimal control $\hat{u} \in \bar{U}$.
    Thus, the infimum in \eqref{eq:nominal reach time} is a minimum and $\int_0^{T_N^*} \bar{B} \hat{u}(t)\, dt = d$. If $d = 0$, then according to Remark~\ref{rmk:d=0}, $T_N^* = 0$ and we take $\bar{u}^* = 0$ so that $\bar{B}\bar{u}^* T_N^* = d$.
    Otherwise, $T_N^* > 0$, so we can define the constant vector $\bar{u}^* := \frac{1}{T_N^*} \int_0^{T_N^*} \hat{u}(t)\, dt \in \mathbb{R}^{m+p}$.
    Note that
    \begin{equation*}
        \| \bar{u}^* \|_\infty \leq \frac{1}{T_N^*} \int_0^{T_N^*} \|\hat{u}(t)\|_\infty \, dt \leq \frac{1}{T_N^*} \|\hat{u}(t)\|_{\mathcal{L}_\infty} T_N^* \leq u_{max},
    \end{equation*}
    since $\hat{u} \in \bar{U}$. Additionally, $\int_0^{T_N^*} \bar{B} \bar{u}^*\, dt = \bar{B} \bar{u}^* T_N^* = d$. $\quad \blacksquare$
\end{pf}

Following Proposition~\ref{prop:unperturbed time}, \eqref{eq:nominal reach time} simplifies to 
\begin{equation}\label{eq:T_N^* simplified}
    T_N^*(d) = \underset{\bar{u}_c\, \in\, \bar{U}_c}{\min} \big\{ T \geq 0 : \bar{B}\bar{u}_c\, T = d \big\}.
\end{equation}
The multiplication of the variables $\bar{u}_c$ and $T$ prevents the use of linear solvers.
Instead, we will consider
\begin{equation}\label{eq:linear program}
    T_N^*(d) = \left( \underset{\|\bar{u}\|_\infty\, =\, u_{max}}{\max} \big\{ \lambda : \bar{B} \bar{u} = \lambda d \big\} \right)^{-1},
\end{equation}
after using the transformation $\lambda = \frac{1}{T}$ in \eqref{eq:T_N^* simplified}.
Problem \eqref{eq:linear program} is linear in $\bar{u}$ so the optimal control input $\bar{u}^*$ belongs to the boundary of the constraint set \citep{max-min_programming} for $d \neq 0$.
The uninteresting case $d = 0$ has been treated in Remark~\ref{rmk:d=0}, so we consider $d \neq 0$. Then $\bar{u}^* \neq 0$, leading to $\|\bar{u}^*\|_\infty = u_{max}$.
The optimization variable is then $(\bar{u}, \lambda)$ and the constraints are the following
\begin{align*}
    \begin{bmatrix} -u_{max} \\ \vdots \\ -u_{max} \\ 0 \end{bmatrix} \leq \begin{bmatrix} \bar{u}_1 \\ \vdots \\ \bar{u}_{m+p} \\ \lambda \end{bmatrix} &\leq \begin{bmatrix} u_{max} \\ \vdots \\ u_{max} \\ -\end{bmatrix}\\
    \big[ \bar{B}\  -d \big] \begin{bmatrix} \bar{u} \\ \lambda \end{bmatrix} &= 0.
\end{align*}

We now introduce an interesting property of $T_N^*$ that will be needed later.

\begin{prop}\label{prop:proportional nominal reach time}
    The nominal reach time $T_N^*$ is an absolutely homogeneous function of $d$, i.e., $T_N^*(\lambda d) = | \lambda |\ T_N^*(d)$ for $d \in \mathbb{R}^n$, $\lambda \in \mathbb{R}$.
\end{prop}

\begin{pf}
    Let $d \in \mathbb{R}^n$, $\lambda \in \mathbb{R}$. The case $\lambda = 0$ is trivial since $T_N^*(0) = 0$, so consider $\lambda \neq 0$.
    The nominal reach time for $d$ is $T_N^*(d)$, so there exists $\bar{u}_d \in \bar{U}_c$ such that $\bar{B} \bar{u}_d T_N^*(d) = d$. Then, $\bar{B}\, (sign(\lambda) \bar{u}_d)\, |\lambda| T_N^*(d) = \lambda d$. The optimality of $T_N^*(\lambda d)$ to reach $\lambda d$ leads to $T_N^*(\lambda d) \leq |\lambda| T_N^*(d)$.
    
    There exists $\bar{u}_{\lambda d} \in \bar{U}_c$ such that $\bar{B} \bar{u}_{\lambda d} T_N^*(\lambda d) = \lambda d$. Then $\bar{B}\, (sign(\lambda)\bar{u}_{\lambda d})\, \frac{T_N^*(\lambda d)}{|\lambda|} = d$. The optimality of $T_N^*(d)$ to reach $d$ leads to $T_N^*(d) \leq \frac{T_N^*(\lambda d)}{|\lambda|}$.
    Thus, $|\lambda| T_N^*(d) \leq T_N^*(\lambda d)$.    $\quad \blacksquare$
\end{pf}

We have established that the nominal reach time is absolutely homogenous and can be achieved with a constant control input.
We can now tackle the dynamics of the malfunctioning system after a loss of control authority over some of its actuators.

\section{Dynamics of the Malfunctioning System}\label{section:malfunctioning dynamics}

We study the system of dynamics \eqref{eq:splitted ODE}, with the aim of computing the malfunctioning reach time $T_M^*$. 
We define the constant input sets
\begin{equation}\label{eq:constant sets U and W}
    U_c := \big\{ u \in \mathbb{R}^m : \|u\|_\infty \leq u_{max} \big\} , \qquad W_c := \big\{ w \in \mathbb{R}^p : \|w\|_\infty \leq u_{max} \big\},
\end{equation}
and $V_c$ as the set of vertices of $W_c$.

\begin{prop}\label{prop:u cst}
    For a resilient system, $d \in \mathbb{R}_*^n$ and $w \in W$, the infimum $T_M(w,d)$ of \eqref{eq:malfunctioning reach time} defined as
    \begin{equation}\label{eq:T_M}
        T_M(w,d) := \underset{u\, \in\, U}{\inf} \Big\{ T \geq 0 : \int_0^T \hspace{-2mm} Bu(t) + Cw(t)\, dt = d \Big\},
    \end{equation}
    is achieved with a constant control input $u_d^*(w) \in U_c$. 
\end{prop}
\begin{pf}
    First, we show that the infimum of \eqref{eq:malfunctioning reach time} is a minimum.
    Let $d \in \mathbb{R}^n$, $d \neq 0$ and $w \in W$. Then, 
    \begin{equation}\label{eq:inf U}
        T_M(w,d) = \underset{u\, \in\, U}{\inf} \big\{ T \geq  0 : \int_0^T \hspace{-3mm} Bu(t)\, dt = z \big\}, 
    \end{equation}
    with $z := d - \int_0^T Cw(t)\, dt \in \mathbb{R}^n$ a constant vector once $w$ is fixed. Since the system is resilient, any $z \in \mathbb{R}^n$ is reachable. Additionaly, $U$ is convex and compact, and \eqref{eq:inf U} is linear in $u$. Then, according to Theorem 4.3 of \citep{liberzon} a time-optimal control exists. Following the proof of Proposition~\ref{prop:unperturbed time}, we conclude that the infimum of \eqref{eq:inf U} is a minimum, the optimum $u_d^*(w)$ is independent of time and belongs to $U_c$.   $\quad \blacksquare$
\end{pf}

We can now work on the supremum of \eqref{eq:malfunctioning reach time}.

\begin{prop}\label{prop:w cst}
     For a resilient system and $d \in \mathbb{R}_*^n$, the supremum $T_M^*(d)$ of \eqref{eq:malfunctioning reach time} is achieved with a constant undesirable input $w^* \in W_c$.
\end{prop}
\begin{pf}
    We  will show first that we can restrict the constraint space to $W_c$ and then that the supremum of \eqref{eq:malfunctioning reach time} is a maximum. For $d \in \mathbb{R}_*^n$, following Proposition~\ref{prop:u cst}, \eqref{eq:malfunctioning reach time} simplifies to
    \begin{equation}\label{eq:T_M^* simplified}
        T_M^*(d) = \underset{w\, \in\, W}{\sup} \left\{ T :  B u_d^*(w) T + \int_0^{T} \hspace{-3mm} Cw(t)\, dt = d \right\},
    \end{equation}
    with $Bu_d^*$ from Proposition~\ref{prop:u cst}. Let $w \in W$ and consider 
    \begin{equation*}
        w_c := \int_0^{T_M(w,d)} \hspace{-2mm} \frac{w(t)}{T_M(w,d)}\, dt. \qquad \text{Then,} \quad \|w_c\|_\infty \leq \frac{1}{T_M(w,d)}\int_0^{T_M(w,d)} \hspace{-2mm} \|w\|_{\mathcal{L}_\infty}\, dt \leq u_{max}.
    \end{equation*}
    So, $w_c \in W_c$. Then, $B u_d^*(w) T_M(w,d) + \int_0^{T_M(w,d)} Cw(t)\, dt = d = \big(B u_d^*(w) + C w_c\big) T_M(w,d)$. Conversely, note that for all $w_c \in W_c$ and $T > 0$, we can define $w(t) := \frac{1}{T}w_c$ for $t \in [0,T]$ such that $\int_0^T Cw(t)\, dt = C w_c$ and $w \in W$.
    Therefore, the constraint space of \eqref{eq:T_M^* simplified} can be restricted to $W_c$.
    
    \vspace{2mm}
    
    We define the function $\varphi : W_c \rightarrow \mathbb{R}^n$ as 
    \begin{equation}\label{eq:phi}
        \varphi(w_c) := B u_d^*(w_c) + C w_c \qquad \text{for} \quad w_c \in W_c.
    \end{equation}
    When applying $w_c$ and $u_d^*(w_c)$ the dynamics become $\dot x = \varphi(w_c)$. 
    We now use the work from \citet{Neustadt} concerning the existence of optimal control inputs. 
    Neustadt defines in \citep{Neustadt} the attainable set from $x_0$ and using inputs in $W_c$ as 
    \begin{equation*}
        \mathcal{A}_{W_c} := \Big\{ (x_1, T) : \ \text{for}\ w_c \in W_c, \int_0^T \hspace{-2mm} \varphi(w_c)\, dt = x_1 - x_0 \Big\}.
    \end{equation*}
    Following Lemma~\ref{lemma: T continuous} in Appendix \ref{apx:continuity}, $\big(Bu_d^*(w_c) + Cw_c \big) T_M(w_c, d) = d$, then $\varphi(w_c) = \frac{1}{T_M(w_c, d)}d$, which is continuous in $w_c$.
    Set $W_c$ is compact, $t_0 = 0$ and $x_0 \in \mathbb{R}^n$ are fixed. Then, Theorem~1 of \citep{Neustadt} states that $\mathcal{A}_{W_c}$ is compact.
    
    Note that $T_M^*(d) = \sup \big\{ T : (x_{goal}, T) \in \mathcal{A}_{W_c} \big\}$, then $T_M^*(d)$ is the supremum of a continuous function over the compact set $\mathcal{A}_{W_c}$, so the supremum of \eqref{eq:T_M^* simplified} is a maximum achieved on $W_c$. $\quad \blacksquare$
\end{pf}

Following Propositions \ref{prop:u cst} and \ref{prop:w cst}, the malfunctioning reach time can now be calculated with
\begin{equation}\label{eq:T_M^* with W_c and U_c}
    T_M^*(d) = \max_{w_c\, \in\, W_c} \left\{ \underset{u_c\, \in\, U_c}{\min} \big\{ T \geq 0 : \big(Bu_c + Cw_c\big) T = d \big\} \right\}.
\end{equation}
The simplifications achieved so far were based on existence theorems from \citep{liberzon, Neustadt} upon which the bang-bang principle relies. The logical next step is to show that the maximum of \eqref{eq:T_M^* with W_c and U_c} is achieved by the extreme undesirable inputs, i.e., at the set of vertices of $W_c$, which we denote by $V_c$. However, most of the work on the bang-bang principle considers systems with a linear dependency on the input \citep{LaSalle, liberzon, Sussmann}, while $\varphi$ introduced in \eqref{eq:phi} is not linear in the input $w_c$. 

The work from \citet{Neustadt} considers a nonlinear $\varphi$, yet his discussion on bang-bang inputs would require us to show that $co(\varphi(W_c)) = co(\varphi(V_c))$. Since $\varphi$ is not linear, such a task is not trivial and in fact it amounts to proving that inputs in $V_c$ can do as much as inputs in $W_c$, i.e., we would need to prove the bang-bang principle.

Two more works \citep{Aronsson, Glashoff} consider bang-bang properties for systems with nonlinear dependency on the input. However, both of them require conditions that are not satisfied in our case. Work contained in \citep{Aronsson} needs the subsystem $\dot x = Cw$ to be controllable, while \citep{Glashoff} requires $T_M$ defined in Lemma~\ref{lemma: T continuous} in Appendix~\ref{apx:continuity} to be concave in $w_c$.
Thus, even if bang-bang theory seems like a natural approach to restrict the constraint space from $W_c$ to $V_c$ in \eqref{eq:T_M^* with W_c and U_c}, we had to establish our own optimization result, namely Theorem~\ref{thm:minimum on the vertices}. We can now prove that the maximum of \eqref{eq:T_M^* with W_c and U_c} is achieved on $V_c$.

\begin{prop}\label{prop:w on a vertex}
    For a resilient system and $d \in \mathbb{R}_*^n$, the maximum of \eqref{eq:T_M^* with W_c and U_c} is achieved with a constant input $w^* \in V_c$, i.e., its components are $w_i^* := \pm u_{max}$ for all $i \in [p]$.
\end{prop}
\begin{pf}
    We introduce sets $X := \big\{-Cw_c : w_c \in W_c \big\}$ and $Y := \big\{Bu_c : u_c \in U_c \big\}$. Then, using $\lambda = \frac{1}{T}$ in \eqref{eq:T_M^* with W_c and U_c} we have
    \begin{equation*}
        \frac{1}{T_M^*(d)} = \min_{x\, \in\, X} \big\{ \underset{y\, \in\, Y}{\max} \big\{ \lambda \geq 0 : y-x = \lambda d \big\} \big\}.
    \end{equation*}
    Since $\lambda \geq 0$, we can write $\lambda = |\lambda| = \frac{\| \lambda d\|}{\|d\|} = \frac{\|y - x\|}{\|d\|}$. Then, our problem of interest becomes
     \begin{equation}\label{eq:T_M^*(d) eq with x,y}
        \frac{1}{\|d\|} \min_{x\, \in\, X} \Big\{ \underset{y\, \in\, Y}{\max} \big\{ \|y-x\| : y-x \in \mathbb{R}^+ d \big\} \Big\}.
    \end{equation}
    
    Sets $U_c$ and $W_c$ as defined in \eqref{eq:constant sets U and W} are hypercubes in $\mathbb{R}^m$ and $\mathbb{R}^p$ respectively, and thus they are polytopes. Sets $X$ and $Y$ are defined as images of $W_c$ and $U_c$ under a linear transformation, so they are polytopes of $\mathbb{R}^n$ \citep{inf_dim_analysis}.
    
    To apply Theorem~\ref{thm:minimum on the vertices}, we need to show that $X \subset Y$.
    Since the system is resilient, for all $w_c \in W_c$ and all $d_0 \in \mathbb{R}^n$ there exists $u_c \in U_c$ and $T \geq 0$ such that $(Bu_c + Cw_c)T = d_0$. 
    Then, for $x = -Cw_c \in X$, $x \neq 0$ and $d_0 = x$ there exists $y \in Y$ and $T > 0$ such that $(y-x)T = x$. Then, $y = \lambda x$ with $\lambda := 1 + 1/T > 1$. Since $Y$ is convex, $0 \in Y$ and $\lambda x \in Y$ then $x \in Y$. Thus, $X \subset Y$.
    
    We can now apply Theorem~\ref{thm:minimum on the vertices} and conclude that the minimum $x^*$ of \eqref{eq:T_M^*(d) eq with x,y} must be realized on a vertex of $X$. Now, we want to show that $x^*$ belongs to the image of $V_c$ by $C$.
    
    Let $w_c \in W_c$ such that $x^* = -Cw_c$. If $w_c \in V_c$ we are done. 
    Otherwise, two possibilities remain.
    In the first case $w_c$ is on the boundary of the hypercube $W_c$ and then we take $F$ to be the surface of lowest dimension of $\partial W_c$ such that $w_c \in F$ and $\dim F \geq 1$.
    The other possibility is that $w_c \in W_c^\circ$; we then define $F := W_c$.
    Thus, in both cases $V_c \cap F \neq \emptyset$ and $F$ is convex.
    Then, we take $v \in V_c \cap F$ and $a := v - w_c \in F$. Since $\dim F \geq 1$ and $w_c \in F$, there exists some $\alpha > 0$ such that $w_c \pm \alpha a \in F$. Then
    \begin{equation*}
        x^* = -Cw_c = -C \Bigg(\frac{1}{2}(w_c + \alpha a) + \frac{1}{2}(w_c - \alpha a) \Bigg) = \frac{1}{2} x_+ + \frac{1}{2}x_-,
    \end{equation*}
    with $x_\pm := -C(w_c \pm \alpha a)$. Since $x^*$ is a vertex of $X$ and $x_\pm \in X$, according to our definition of vertices $x^* = x_+ = x_-$. Then, $x^* - x_+ = \alpha Ca = 0$, which yields $Ca = 0$ because $\alpha > 0$. Thus, $-Cv = -C(w_c + a) = x^*$ and $v \in V_c$.
    Therefore, the maximum of \eqref{eq:T_M^* with W_c and U_c} is achieved on $V_c$.  $\quad \blacksquare$
\end{pf}

We have reduced the constraint set of \eqref{eq:malfunctioning reach time} from an infinite-dimensional set $W$ to a finite set $V_c$ of cardinality $2^p$, with $p$ being the number of malfunctioning actuators.
Following Propositions \ref{prop:u cst}, \ref{prop:w cst} and \ref{prop:w on a vertex}, the malfunctioning reach time can now be calculated with
\begin{equation}\label{eq:optimization problem}
    T_M^*(d) = \max_{w_c\, \in\, V_c} \left\{ \underset{u_c\, \in\, U_c}{\min} \big\{ T \geq 0 : \big(Bu_c + Cw_c\big) T = d \big\} \right\}.
\end{equation}

It is logic to wonder if the minimum of \eqref{eq:optimization problem} could be restricted to the vertices of $U_c$, just like we did for the maximum over $W_c$. However, that is not possible. Indeed, $w_c$ is chosen freely in $W_c$ in order to make $T_M^*$ as large as possible. On the other hand, $u_c$ is chosen to counteract $w_c$ and make $Bu_c + Cw_c$ collinear with $d$. This constraint could not be fulfilled for all $d \in \mathbb{R}^n$ if $u_c$ was only chosen among the vertices of $U_c$.

Similarly to the nominal reach time, $T_M^*$ is also linear in the target distance.

\begin{prop}\label{prop:proportional malfunctioning reach time}
    The malfunctioning reach time $T_M^*$ is an absolutely homogeneous function of $d$, i.e., $T_M^*(\lambda d) = | \lambda |\ T_M^*(d)$ for $d \in \mathbb{R}^n$, $\lambda \in \mathbb{R}$.
\end{prop}

\begin{pf}
    Because of the minimax structure of \eqref{eq:optimization problem}, scaling like in the proof of Proposition~\ref{prop:proportional nominal reach time} is not sufficient to prove the homogeneity of $T_M^*(d)$.
    
    \vspace{2mm}
    
    According to Remark~\ref{rmk:d=0}, for $d = 0$ we have $T_M^*(d) = 0$, so $T_M^*$ is absolutely homogeneous at $d = 0$.
    
    Let $d \in \mathbb{R}_*^n$, $Y := \big\{Bu_c : u_c \in U_c \big\}$, $w_c \in W_c$ and $x = Cw_c$. Consider the function $y^*(x, d) := \arg \underset{y\, \in\, Y}{\min} \big\{ T \geq 0 : (y+x)T = d \big\}$. Note that $Bu_d^*(w_c) + Cw_c = y^*(x,d) + x$, with $u_d^*$ defined in Proposition~\ref{prop:u cst}. Then, with $T_M$ defined in Lemma~\ref{lemma: T continuous} of Appendix~\ref{apx:continuity}, we have $\big(Bu_d^*(w_c) + Cw_c\big) T_M(w_c, d) = d$, i.e., $y^*(x,d) = \frac{1}{T_M(w_c, d)}d - x$. For $\lambda > 0$,
    we define $\alpha(\lambda) := \frac{\lambda}{T_M(w_c, \lambda d)} - \frac{1}{T_M(w_c, d)}$,
    % there exists $\alpha(\lambda) \in \mathbb{R}$
    such that $y^*(x, \lambda d) - y^*(x,d) = \alpha(\lambda) d$.
    
    The polytope $Y$ of $\mathbb{R}^n$ has a finite number of faces, so we can choose $d \in \mathbb{R}_*^n$ not collinear with any face of $Y$. Since $Y$ is convex, the ray $\big\{ y^*(x,d) + \alpha d : \alpha \in \mathbb{R} \big\}$ intersects with $\partial Y$ at most twice. 
    Since $y^*(x,d) \in \partial Y$, one intersection happens at $\alpha = 0$. If there exists another intersection, it occurs for some $\alpha_0 \neq 0$.
    Since $y^*(x, \lambda d) \in \partial Y$, we have $y^*(x,d) + \alpha(\lambda) d \in \partial Y$. Then, $\alpha(\lambda) \in \{0, \alpha_0\}$ for all $\lambda > 0$.
    
    According to Lemma~\ref{lemma: T continuous}, $T_M$ is continuous in $d$, so $\alpha$ is continuous in $\lambda$ but its codomain is finite. Therefore, $\alpha$ is constant and we know that $\alpha(1) = 0$. So $\alpha$ is null for all $\lambda > 0$, leading to 
    % $Bu^*(w_c, d) = Bu^*(w_c, \lambda d)$ 
    $T_M(w_c, \lambda d) = \lambda T_M(w_c, d)$
    for $\lambda > 0$ and $d$ not collinear with any face of $\partial Y$. Since the dimension of the faces of $\partial Y$ is at most $n-1$ in $\mathbb{R}^n$ and $T_M$ is continuous in $d$, the homogeneity of $T_M$ holds on the whole of $\mathbb{R}^n$.
    Note that $T_M^*(d) = \underset{w_c\, \in\, W_c}{\max} T_M(w_c, d)$. Thus, $\lambda T_M^*(d) = T_M^*(\lambda d)$ for $\lambda > 0$ and $d \in \mathbb{R}^n$.  
    
    We now extend this result to negative $\lambda$. For $d \in \mathbb{R}_*^n$ and $w_c \in W_c$,
    \begin{align*}
        Bu_c^*(-w_c, -d) &= \frac{-d}{T_M(-w_c, -d)} + Cw_c = \arg \underset{y\, \in\, Y}{\min} \big\{ T \geq 0 : (y-x)T = -d \big\} \\
        &= \arg \underset{y\, \in\, Y}{\min} \big\{ T \geq 0 : (-y+x)T = d \big\} = -Bu_c^*(w_c, d) = -\left( \frac{d}{T_M(w_c, d)} - Cw_c\right),
    \end{align*}
    because $Y$ is symmetric.
    Therefore, $T_M(-w_c, -d) = T_M(w_c, d)$. Using the symmetry of $W_c$ we obtain $\underset{w_c\, \in\, W_c}{\max} T_M(-w_c, -d) = \underset{w_c\, \in\, W_c}{\max} T_M(w_c, -d) = T_M^*(-d)$. Thus, $T_M^*(-d) = T_M^*(d)$. Then for $\lambda < 0$, $T_M^*(\lambda d) = T_M^*( -|\lambda|d) = T_M^*( |\lambda|d ) = |\lambda| T_M^*(d)$.  $\quad \blacksquare$
\end{pf}

We can now combine the initial and malfunctioning dynamics in order to evaluate the quantitative resilience of the system.

\section{Quantitative Resilience}\label{section:r_q}

Quantitative resilience is defined in \eqref{eq:r_q} as the infimum of $T_N^*(d) / T_M^*(d)$ over $d \in \mathbb{R}^n$. Using Proposition~\ref{prop:proportional nominal reach time} and Proposition~\ref{prop:proportional malfunctioning reach time} we reduce this constraint to $d \in \mathbb{S}$.
Focusing on the loss of control over a single actuator we will simplify tremendously the computation of $r_q$. 
In this setting, we can determine the optimal $d \in \mathbb{S}$ by noting that the effects of the undesirable inputs are the strongest along the direction described by the malfunctioning actuator. This intuition is formalized below.

\begin{theorem}\label{thm:direction maximizing r}
    For a resilient system following \eqref{eq:splitted ODE} with $C$ a single column matrix, the direction $d$ maximizing the ratio of reach times $t(d)$ is collinear with the direction $C$, i.e., $\underset{d\, \in\, \mathbb{S}}{\max}\ t(d) = t(C)$.
\end{theorem}
\begin{pf}
    We fix $d \in \mathbb{S}$ and we will evaluate the ratio of reach times $t(d)$ in the direction $d$. 
    Since $C$ has a single column, $W_c = [-u_{max}, u_{max}]$.
    Then, according to Proposition~\ref{prop:w on a vertex}, the worst undesirable input is $w^*(d) = \pm u_{max}$ for the direction $d$. 
    Using the same transformation as in \eqref{eq:linear program}, we rewrite the malfunctioning reach time as
    \begin{equation*}
        T_M^*(d) = \underset{u_c\, \in\, U_c}{\min} \big\{ T : \big(Bu_c + Cw^*(d)\big) T = d\big\} = \frac{1}{\underset{u_c\, \in\, U_c}{\max} \big\{ \lambda : Bu_c + Cw^*(d) = \lambda d\big\}  }.
    \end{equation*}
    Let $Y := \big\{ Bu_c : u_c \in U_c \big\}$ and $x_M^*(d) := C w^*(d) = \pm C u_{max}$. 
    Since $\lambda \geq 0$ and $\|d\| = 1$ we have $\lambda = \| \lambda d\| = \|y + x_M^*(d)\|$. These simplifications lead to
    \begin{equation}\label{eq:T_M^* with x, y}
        T_M^*(d) = \frac{1}{\underset{y\, \in\, Y}{\max} \big\{ \|y + x_M^*(d)\| : y + x_M^*(d) \in \mathbb{R}^+ d\big\} }.
    \end{equation}
    We focus on the nominal reach time and proceed to the separation of $\bar{B} = [B\ C]$ in \eqref{eq:linear program}:
    \begin{equation}\label{eq:T_N^* with x, y}
        \frac{1}{T_N^*(d)} = \underset{\bar{u}\, \in\, \bar{U}_c}{\max} \big\{ \lambda : \bar{B}\bar{u} = \lambda d\big\} = \underset{\substack{u_c\, \in\, U_c \\ w_c\, \in\, W_c}}{\max} \big\{ \lambda : Bu_c + Cw_c = \lambda d \big\} = \underset{\substack{y\, \in\, Y \\ x\, \in\, X}}{\max} \big\{ \|y+x\| : y + x \in \mathbb{R}^+ d \big\},
    \end{equation}
    with $X := \big\{Cw_c : w_c \in W_c \big\}$.
    We can now gather \eqref{eq:T_M^* with x, y} and \eqref{eq:T_N^* with x, y} into
    \begin{equation*}
        t(d) = \frac{T_M^*(d)}{T_N^*(d)} = \frac{\underset{x\, \in\, X,\, y\, \in\, Y}{\max} \big\{ \|x + y\| : x + y \in \mathbb{R}^+ d\big\}}{\underset{y\, \in\, Y}{\max} \big\{ \|x_M^*(d) + y\| : x_M^*(d) + y \in \mathbb{R}^+ d\big\} } = r_{X, Y}(d),
    \end{equation*}
    with $r_{X,Y}$ defined in \eqref{eq:r_(X,Y)}.
     
    In the proof of Proposition~\ref{prop:w on a vertex} we showed that sets $X$ and $Y$ are polytopes verifying $X \subset Y$ and the resilience of the system states that for all $d \in \mathbb{R}^n$ and $x \in X$ there exists $y \in Y$ such that $x + y \in \mathbb{R}_*^+ d$. Then, $\dim Y = n$. To apply Theorem~\ref{thm:varying x_M and x_N} we need to prove that $X \subset Y^\circ$.
    
    Assume for contradiction purposes that there exists $x_1 \in X \cap \partial Y$. Take $d = -x_1$, then the best input is $y = -x_1 \in \partial Y$ because $Y$ is symmetric. Then, $x_1 + y = 0 \notin \mathbb{R}_*^+ d$, which contradicts the resilience of the system. Therefore, $X \cap \partial Y = \emptyset$, i.e., $X \subset Y^\circ$. Since $U_c$ and $W_c$ are symmetric, so are $X$ and $Y$. Because $C$ is a single column $\dim X = 1$ and $x_M^*(d) \in \partial X = \big\{ \pm Cu_{max} \big\}$.
    
    We can then apply Theorem~\ref{thm:varying x_M and x_N} and conclude that $\underset{d\, \in\, \mathbb{S}}{\max}\ t(d) = t(Cu_{max}) = t(C)$, since $t$ is invariant by scaling according to Propositions~\ref{prop:proportional nominal reach time} and \ref{prop:proportional malfunctioning reach time}.     $\quad \blacksquare$
\end{pf}

Then, to calculate the quantitative resilience $r_q$ we only need to evaluate $T_N^*(C)$ and $T_M^*(C)$. The computation load can be even further reduced with the following result.

\begin{theorem}\label{thm:computation of r_q}
    For a resilient system losing control over a single nonzero column $C$, $r_q = r_{max}$, where
    \begin{equation}\label{eq:r_max}
        r_{max} := \frac{\lambda^* - u_{max}}{\lambda^* + u_{max}} \qquad \text{and} \qquad \lambda^* := \underset{\upsilon \, \in\, U_c}{\max} \big\{ \lambda : B \upsilon = \lambda C \big\}.
    \end{equation}
\end{theorem}
\begin{pf}
    Let $\bar{u} \in \bar{U}_c$, $u \in U_c$ and $w \in W_c$ be the arguments of the optimization problems \eqref{eq:T_N^* simplified} and \eqref{eq:optimization problem} for $d = C \neq 0$. 
    We split $\bar{u} = [u_B\ u_C]^\top$ such that $u_B \in U_c$ and $u_C \in W_c$. Then,
    \begin{equation}\label{eq:T_N^*(C) and T_M^*(C)}
        \bar{B} \bar{u}\, T_N^*(C) = Bu_B\, T_N^*(C) + Cu_C\, T_N^*(C) = C \qquad \text{and} \qquad Bu\, T_M^*(C) + Cw\, T_M^*(C) = C.
    \end{equation}
    We consider the loss of a single actuator, thus $W_c = [-u_{max}, u_{max}] \subset \mathbb{R}$ which makes $Cw T_M^*(C)$ and $Cu_C T_N^*(C)$ collinear with $C$. From Proposition~\ref{prop:w on a vertex}, we know that $w = \pm u_{max}$. Since $w$ maximizes $T_M^*(C)$ in \eqref{eq:T_N^*(C) and T_M^*(C)}, we obviously have $w = -u_{max}$. On the contrary, $u_C$ is chosen to minimize $T_N^*(C)$ in \eqref{eq:T_N^*(C) and T_M^*(C)}, so $u_C = +u_{max}$.
    
    According to \eqref{eq:T_N^*(C) and T_M^*(C)}, $Bu_B$ and $Bu$ are then also collinear with $C$. The control inputs $u_B$ and $u$ are chosen to minimize respectively $T_N^*(C)$ and $T_M^*(C)$ in \eqref{eq:T_N^*(C) and T_M^*(C)}. Therefore, they are both solutions of the same optimization problem:
    \begin{equation*}\label{}
        \tau^* = \underset{\upsilon \, \in\, U_c}{\min} \big\{ \tau : B \upsilon \tau = C \big\} \qquad \text{with} \qquad u = u_B = \arg \underset{\upsilon \, \in\, U_c}{\min} \big\{ \tau : B \upsilon \tau = C \big\}.
    \end{equation*}
    We transform this problem into a linear one using the transformation $\lambda = \frac{1}{\tau}$:
    \begin{equation*}\label{}
        \lambda^* = \underset{\upsilon \, \in\, U_c}{\max} \big\{ \lambda : B \upsilon = \lambda C \big\} \qquad \text{with} \qquad u = u_B = \arg \underset{\upsilon \, \in\, U_c}{\max} \big\{ \lambda : B \upsilon = \lambda C \big\}.
    \end{equation*}
    By combining all the results, \eqref{eq:T_N^*(C) and T_M^*(C)} simplifies into:
    \begin{equation*}
        C(\lambda^* + u_{max}) T_N^*(C) = C \quad \text{and} \quad C(\lambda^* - u_{max}\big) T_M^*(C) = C.
    \end{equation*}
    Following Theorem~\ref{thm:direction maximizing r}, $r_q = \frac{T_N^*(C)}{T_M^*(C)} = \frac{\lambda^* - u_{max}}{\lambda^* + u_{max}} = r_{max}$.    $\quad \blacksquare$
\end{pf}

We introduced quantitative resilience as the solution of four nonlinear nested optimization problems and with Theorem~\ref{thm:computation of r_q} we reduced $r_q$ to the solution of a single linear optimization problem.
We can then quickly calculate the maximal delay caused by the loss of control of a given actuator.

So far, all our results need the system to be resilient. However, based on the work \citep{journal_paper} verifying the resilience of a system is not an easy task. Besides, as explained in Section~\ref{section:preliminaries}, the resilience criteria established in \citep{journal_paper} cannot be applied to this paper because of a difference of setting in the set of allowable control inputs. Proposition~\ref{prop:resilient full rank} establishes only a necessary condition for resilience. The following proposition produces a necessary and sufficient condition.

\begin{prop}\label{prop:resilience = full rank + finite T_M^*(C)}
    A system following \eqref{eq:original ODE} is resilient to the loss of control over a column $C$ if and only if it is controllable and $T_M^*(C)$ is finite.
\end{prop}
\begin{pf}
    First, assume that the system \eqref{eq:original ODE} is resilient. Then, according to Proposition~\ref{prop:resilient full rank}, the system $\dot x(t) = Bu(t)$ is controllable. Since $Im(B) \subset Im(\bar{B})$, the system \eqref{eq:original ODE} is controllable a fortiori. If $C \neq 0$, then following Proposition~\ref{prop:w cst}, $T_M^*(C)$ is finite. If $C = 0$, then $T_M^*(C)$ is also finite according to Remark~\ref{rmk:d=0}.
    
    \vspace{3mm}
    
    Now, assume that the system \eqref{eq:original ODE} is controllable and $T_M^*(C)$ is finite. 
    If $C = 0$, then $rank(B) = rank(\bar{B}) = n$. For any $w \in W_c$ and any $d \in \mathbb{R}_*^n$, there exists $u \in \mathbb{R}_*^m$ such that $Bu = d$. Define $u_c := \frac{u}{\|u\|_\infty} u_{max}$ and $T := \frac{\|u\|_\infty}{u_{max}}$, then $(B u_c + Cw)T = B u_c T = d$ and $u_c \in U_c$, so the system is resilient.
    
    For $C \neq 0$, because $T_M^*(C)$ is finite, $T_M(w,C)$ is positive and finite for $w \in W_c = [-u_{max}, u_{max}]$, with $T_M(\cdot,\cdot)$ of Lemma~\ref{lemma: T continuous} in Appendix~\ref{apx:continuity}. There exists $u_w \in U_c$ such that $(Bu_w + Cw)T_M(w,C) = C$.
    Then, $Bu_w = C \left( \frac{1}{T_M(w,C)} - w\right)$.
    Thus, $C \in Im(B)$ and we define $u_C = \left(\frac{1}{T_M(w,C)} -w\right)^{-1} u_w$, then $Bu_C = C$. For all $w \in W_c$, we have $u_w \in U_c$ and more specifically if $w = -u_{max}$, then
    \begin{equation}\label{eq:norm u_C less than u_max}
          \left(\frac{1}{T_M(-u_{max},C)} + u_{max}\right) \|u_C\|_\infty = \|u_w\|_\infty \leq u_{max}.
    \end{equation}
    
    We will now prove that $B$ is full rank. Indeed, $rank(\bar{B}) = \dim \big( Im([B\ C]) \big) = \dim \big( Im(B) \big)$ since $C \in Im(B)$ and $\bar{B} = [B\ C]$.
    Thus, $rank(B) = rank(\bar{B}) = n$. 
    
    Then, for any $d \in \mathbb{R}_*^n$, there exists $u_d \in \mathbb{R}_*^m$ such that $B u_d = d$. For any $w \in W_c$ and for $\lambda := \frac{\|u_C\|_\infty}{\|u_d\|_\infty T_M(-u_{max},C)} > 0$ we have $B(-w u_C + \lambda u_d) + Cw = \lambda d$. The norm of $u_\lambda := -w u_C + \lambda u_d$ is
    \begin{equation*}
        \|u_\lambda \|_\infty \leq |w| \|u_C\|_\infty + \lambda \|u_d\|_\infty = |w| \|u_C\|_\infty + \frac{\|u_C\|_\infty}{\|u_d\|_\infty T_M(-u_{max},C)} \|u_d\|_\infty \leq u_{max},
    \end{equation*}
    according to \eqref{eq:norm u_C less than u_max}. To sum up, we have found that for all $d \in \mathbb{R}_*^n$ and all $w \in W_c$, there exist $\lambda > 0$ and $u_\lambda \in U_c$ such that $Bu_\lambda + Cw = \lambda d$, i.e., the system is resilient to the loss of column $C$.  $\quad \blacksquare$
\end{pf}

The intuition behind Proposition~\ref{prop:resilience = full rank + finite T_M^*(C)} is that a resilient system must fulfill two conditions: being able to reach any state, this is controllability, and doing so in finite time despite the worst undesirable inputs, which corresponds to $T_M^*(C)$ being finite.

Our goal is to relate resilience and quantitative resilience through the value of $r_{max}$. To breach the gap between this desired result and Proposition~\ref{prop:resilience = full rank + finite T_M^*(C)}, we evaluate the requirements on the ratio $\frac{T_N^*(C)}{T_M^*(C)}$ for a system to be resilient.

\begin{cor}\label{cor:resilience T_N(C)/T_M(C)}
    A system following \eqref{eq:original ODE} is resilient to the loss of control over a column $C$ if and only if it is controllable and $\frac{T_N^*(C)}{T_M^*(C)} \in (0, 1]$.
\end{cor}
\begin{pf}
    First, assume that the system \eqref{eq:original ODE} is resilient. Then, according to Proposition~\ref{prop:resilient full rank}, it is controllable. For $d \in \mathbb{R}_*^n$, following Propositions~\ref{prop:unperturbed time} and \ref{prop:w cst}, $T_N^*(d)$ and $T_M^*(d)$ are both finite and positive. Using the separation $\bar{B} = [B\ C]$ and $\bar{u} = [u_B\ u_C]$, we have
    \begin{align*}
        T_N^*(d) = \underset{\bar{u}\, \in\, \bar{U}_c}{\min} \big\{ T \geq 0 : \bar{B}\bar{u}T = d\big\} &= \underset{u_C\, \in\, W_c}{\min} \big\{ \underset{u_B\, \in\, U_c}{\min} \big\{ T \geq 0 : (Bu_B + Cu_C)T = d \big\} \\
        &\leq \underset{w_c\, \in\, W_c}{\max} \big\{ \underset{u_c\, \in\, U_c}{\min} \big\{ T \geq 0 : (Bu_c + Cw_c)T = d \big\} = T_M^*(d).
    \end{align*}
    If $C \in \mathbb{R}_*^n$, we then have $0 < \frac{T_N^*(C)}{T_M^*(C)} \leq 1$. If $C = 0$, following Remark~\ref{rmk:d=0} we have $\frac{T_N^*(0)}{T_M^*(0)} = 1$.
    
    \vspace{3mm}
    
    Now, assume that the system is controllable and $\frac{T_N^*(C)}{T_M^*(C)} \in (0, 1]$. 
    If $C = 0$, then $T_M^*(C) = 0$ according to Remark~\ref{rmk:d=0}. We conclude with Proposition~\ref{prop:resilience = full rank + finite T_M^*(C)} that the system is resilient.
    
    Now for the case where $C \neq 0$, let $d \in \mathbb{R}_*^n$.
    Since the system following \eqref{eq:original ODE} is controllable, $T_N^*(C)$ is finite. Since $C \neq 0$, we have $T_N^*(C) > 0$. If $T_M^*(C) = +\infty$, then $\frac{T_N^*(C)}{T_M^*(C)} = 0$, which contradicts the assumption. By definition, $T_M^*(C) \geq 0$, thus $T_M^*(C)$ is finite. Then, according to Proposition~\ref{prop:resilience = full rank + finite T_M^*(C)}, the system is resilient.  $\quad \blacksquare$
\end{pf}

Theorem~\ref{thm:computation of r_q} allows us to compute $r_q$ for resilient systems with a single linear optimization.
We now want to extend that result to non-resilient systems, by showing that $r_{max}$ also indicates whether the system is resilient.

\begin{cor}\label{cor:resilience lambda}
    A system following \eqref{eq:original ODE} is resilient to the loss of control over a nonzero column $C$ if and only if it is controllable and $r_{max} \in (0, 1]$.
\end{cor}
\begin{pf}
    For a resilient system with $C \neq 0$, following Theorem~\ref{thm:computation of r_q} we have $r_q = \frac{T_N^*(C)}{T_M^*(C)} = r_{max}$. Then, according to Corollary~\ref{cor:resilience T_N(C)/T_M(C)} the resilient system is controllable and $r_q \in (0,1]$.
    
    \vspace{3mm}
    
    Now assume that the system is controllable and $r_{max} \in (0, 1]$. We will study $\lambda^*$ introduced in \eqref{eq:r_max}.
    If $\lambda^* + u_{max} < 0$, then by the definition of $r_{max}$ in \eqref{eq:r_max}, $\lambda^* - u_{max} \geq \lambda^* + u_{max}$, leading to the impossible conclusion that $-u_{max} \geq u_{max}$. If $\lambda^* + u_{max} = 0$, then $r_{max} = \frac{-2u_{max}}{0} = \infty$, contradicting $r_{max} \in (0,1]$. Therefore, $\lambda^* + u_{max} > 0$. Let $u^* \in U_c$ such that $Bu^* = \lambda^* C$. For $w \in W_c$, we define $T_w := \frac{1}{\lambda^* + w}$, so that $(Bu^* + Cw)T_w = C$. Note that $T_w$ is positive and finite because $\lambda^* + w \geq \lambda^* - u_{max} > 0$.
    Notice that $T_M^*(C) \leq \underset{w\, \in\, W_c}{\max} T_w = \frac{1}{\lambda^* - u_{max}}$, so $T_M^*(C)$ is finite. Then, Proposition~\ref{prop:resilience = full rank + finite T_M^*(C)} states that the system is resilient.    $\quad \blacksquare$
\end{pf}

We now have all the tools to assess the resilience and quantitative resilience of a driftless system. If $\bar{B}$ is not full rank, the system following \eqref{eq:original ODE} is not controllable and there is no need to go further. Otherwise, we compute the ratio $r_{max}$ and using Corollary~\ref{cor:resilience lambda} we assess whether the system is resilient. If it is, Theorem~\ref{thm:computation of r_q} states that $r_{max} = r_q$, so we have already computed the quantitative resilience of the system. If it is not resilient, then $r_q = 0$.
We will now apply this method to two numerical examples.

\section{Numerical Examples}\label{section:examples}

Our first example considers a linearized model of a low-thrust spacecraft performing orbital maneuvers. We study the resilience of the spacecraft with respect to the loss of control over some thrust frequencies.
Our second example features an opinion dynamics scenario where two agents are influenced by five different sources. 
We study how the loss of control over one of the sources affects the opinion shaping of the agents.

\subsection{Linear Quadratic Trajectory Dynamics}

We consider a low-thrust spacecraft in orbit around a celestial body. Because of the complexity of nonlinear low-thrust dynamics \citet{ko15} established a linear model for the spacecraft dynamics using Fourier thrust acceleration components. Given an initial state and a target state, the model simulates the trajectory of the spacecraft in different orbit maneuvers, such as an orbit raising or a plane change. The states of this linear model are the orbital elements:
\begin{equation*}
    x := \begin{bmatrix}a \\ e \\ i \\ \Omega \\ \omega \\ M\end{bmatrix} \quad \begin{array}{l}
         \text{semi-major axis,}\\
    \text{eccentricity,}\\
    \text{inclination,}\\
    \text{longitude of the ascending node,}\\
    \text{argument of perigee,}\\
    \text{mean anomaly.}
    \end{array}
\end{equation*}
Because of the periodic motion of the spacecraft, the thrust acceleration vector $F$ can be expressed in terms of its Fourier coefficients $\alpha$ and $\beta$:
\begin{align*}
    F &= F_R \hat{r} + F_W \hat{w} + F_S (\hat{w} \times \hat{r}) \\
    F_{R,W,S} &= \sum_{k=0}^{\infty} \big(\alpha^{R,W,S}_k \cos{kE}+\beta^{R,W,S}_k \sin{kE} \big),
\end{align*}
where $F_R$ is the radial thrust acceleration, $F_W$ is the circumferential thrust acceleration, $F_S$ is the normal thrust acceleration and $E$ is the eccentric anomaly.
The work \citep{trajectory_dynamics} determined that only 14 Fourier coefficients affect the average trajectory, and we use those coefficients as the input $\bar{u}$: 
\begin{equation*}
    \bar{u} = \left[\arraycolsep=4.4pt\begin{array}{cccccccccccccc}
    \alpha^R_0 &  \alpha^R_1 &  \alpha^R_2 &  \beta^R_1 &  \alpha^S_0 &  \alpha^S_1 &  \alpha^S_2 &  \beta^S_1 &  \beta^S_2 &  \alpha^W_0 &  \alpha^W_1 &  \alpha^W_2 &  \beta^W_1 &  \beta^W_2
    \end{array}\right]^\top .
\end{equation*}
The Fourier coefficients considered in \citep{trajectory_dynamics} have a magnitude of order $10^{-7}$, so we can safely assume that for our case the Fourier coefficients are bounded by $u_{max} = 1$.
Following \citep{ko15}, the state-space form of the system dynamics is $\dot{x} = \bar{B}(x)\bar{u}$.
We calculate $\bar{B}(x)$ in Appendix~\ref{apx:bar B} using the averaged variational equations for the orbital elements given in \citep{trajectory_dynamics}.
We implement the orbit raising scenario presented in \citep{ko15}, with the orbital elements of the initial and target orbits listed in Table~\ref{tbl:ITstates}.
\begin{table}[H] 
    \begin{center}
    \caption{Inital and Target States for Raising Maneuver}
    \label{tbl:ITstates}
    \begin{tabular}{cc|c|c} \hline \hline
        \multicolumn{2}{c|}{Parameters} & initial & target \\ \hline
        $a$\, &[\rm{km}] & 6678 & 7345 \\ \hline
        $e$\, &[\, -\, ] & 0.67 & 0.737 \\ \hline
        $i$\, &[\rm{degrees}] & 20 & 22 \\ \hline
        $\Omega$\, &[\rm{degrees}] & 20 & 22 \\ \hline
        $\omega$\, &[\rm{degrees}] & 20 & 22 \\ \hline
        $M$\, &[\rm{degrees}] & 20 & 20 \\ \hline
    \end{tabular}
    \end{center}
\end{table}

We approximate $\bar{B}(x)$ as a constant matrix $\bar{B}$ taken at the initial state. The resulting matrix is: $\bar{B} = 10^{-6}\times$
\begin{equation*}
\left[\arraycolsep=4.7pt
   \begin{array}{cccccccccccccc}
        0 & 0 & 0 & 18314 & 40583 & 0 & 0 & 0 & 0 & 0 & 0 & 0 & 0 & 0 \\
        0 & 0 & 0 & 1.1 & -3.4 & 2.3 & -0.4 & 0 & 0 & 0 & 0 & 0 & 0 & 0 \\
        0 & 0 & 0 & 0 & 0 & 0 & 0 & 0 & 0 & -5.2 & 3.8 & -0.9 & -0.7 & 0.2 \\
        0 & 0 &	0 &	0 &	0 &	0 &	0 &	0 &	0 &	-5.5 & 4 & -0.9 & 5.6 & -1.9 \\
        3 &	-2.7 & 0 & 0 & 0 & 0 & 0 & 4.7 & -1 & 5.2 &	-3.8 & 1.3 & -5.6 & 1.9 \\
        -12.3 & 7.2 & -0.9 & 0 & 0 & 0 & 0 & -3.5 & 0.8 & 0 & 0 & 0 & 0 & 0
    \end{array} \right].
\end{equation*}
We immediately notice that the two coefficients on the first row of $\bar{B}$ are significantly larger than all the other coefficients of $\bar{B}$. This difference of magnitude in $\bar{B}$ reflects the difference of magnitude in the state $x$, where the semi-major axis $a$ is significantly larger than any other element as can be seen in Table~\ref{tbl:ITstates}. 

Losing control over one of the 14 Fourier coefficients means that a certain frequency of the thrust vector cannot be controlled. 
Since the coefficients $\bar{B}_{1,5}$ and $\bar{B}_{6,1}$ have a magnitude significantly larger than coefficients of respectively the first and last row of $\bar{B}$, we have the intuition that the system is not resilient to the loss of the $1^{st}$ or the $5^{th}$ Fourier coefficient.

We will now assess the resilience of the system using the method described at the end of Section~\ref{section:r_q}. The matrix $\bar{B}$ is full rank, so $\dot x = \bar{B}\bar{u}$ is controllable. We denote with $r_{max}$ and $r_q$ the vectors whose components are respectively $r_{max}(j)$ and $r_q(j)$ for the loss of the frequency $j \in \{1, \hdots, 14\}$
\begin{align*}
    r_{max} = \left[\arraycolsep=4.2pt\begin{array}{cccccccccccccc}
    -0.2 & 0.34 & 0.9 & -0.004 & -0.38 & 0.15 & 0.83 & -0.32 & 0.71 & -0.06 & 0.24 & 0.2 & -0.5 & 0.5
    \end{array}\right].
\end{align*}
Since the $1^{st}$, $4^{th}$, $5^{th}$, $8^{th}$, $10^{th}$, and $13^{th}$ values of $r_{max}$ are negative, according to Corollary~\ref{cor:resilience lambda} the system is not resilient to the loss of control over any one of these six corresponding frequencies. Their associated $r_q$ is zero. This result validates our intuition about the $1^{st}$ and $5^{th}$ frequencies.
Corollary~\ref{cor:resilience lambda} also states the resilience of the spacecraft to the loss over any one of the $2^{nd}$, $3^{rd}$, $6^{th}$, $7^{th}$, $9^{th}$, $11^{th}$, $12^{th}$ and $14^{th}$ frequency because their $r_{max}$ belongs to $(0,1]$. Then, using Theorem~\ref{thm:computation of r_q} we deduce that
\begin{align*}
    r_q = \left[\arraycolsep=4.8pt\begin{array}{cccccccccccccc}
    0 & 0.34 & 0.9 & 0 & 0 & 0.15 & 0.83 & 0 & 0.71 & 0 & 0.24 & 0.2 & 0 & 0.5
    \end{array}\right].
\end{align*}

Since $r_q(3)$, $r_q(7)$ and $r_q(9)$ are close to $1$, the loss of one of these three frequency would not delay significantly the system.
The lowest positive value of $r_q$ occurs for the $6^{th}$ frequency, $r_q(6) = 0.15$. Its inverse, $\frac{1}{r_q(6)} = 6.8$ means that the malfunctioning system can take up to 6.8 times longer than the initial system to reach a target.

The specific maneuver described in Table~\ref{tbl:ITstates} leads to $d = x_{goal} - x_0 = \big(667,\, 0.067,\, 2,\, 2,\, 2,\, 2\big) $. We compute the associated time ratios $t(d)$ using \eqref{eq:linear program} and \eqref{eq:optimization problem} for the loss over each column of $\bar{B}$:
\begin{equation}\label{eq:t(d) specific target}
   t(d) =  \left[\begin{array}{cccccccccccccc}
    1.1 & 1.2 & 1.1 & 1 & \infty & 1 & 151.1 & \infty & 151.1 & \infty & 151.1 & 151.1 & \infty & 151.1
    \end{array}\right].
\end{equation}
Then, losing control over one of the first four frequencies will barely increase the time required for the malfunctioning system to reach the target compared with the initial system.
However, after the loss over the $7^{th}$, $9^{th}$, $11^{th}$, $12^{th}$, or the $14^{th}$ frequency of the thrust vector, the undesirable input can multiply the maneuver time by a factor of up to $151.1$.
If one of the $5^{th}$, $8^{th}$, $10^{th}$, or the $13^{th}$ frequency is lost, then some undesirable inputs can render the maneuver impossible to perform.

When computing $r_q$, we have seen that the system is not resilient to the loss of the $1^{st}$ or the $4^{th}$ frequency. Yet, the specific target described in Table \ref{tbl:ITstates} happens to be reachable for the same loss since the $1^{st}$ and $4^{th}$ components of $t(d)$ in \eqref{eq:t(d) specific target} are finite. Indeed, $r_q$ speaks only about a target for which the undesirable inputs cause maximal possible delay.

\subsection{Opinion Dynamics}

Opinion dynamics study how a group of agents shapes their opinions \citep{opinion_classic}. We are interested in scenarios where agents are affected by outside opinion drivers. 
The influence of an outside opinion source $u$ can be studied with $x(t+1) = x(t) + \mu \varepsilon \big( u(t) - x(t) \big)$, a modified Deffuant model \citep{opinion_ext} where $\mu$ is a convergence parameter and $\varepsilon$ encodes the strength of the opinion source $u$. For our purpose, we will consider $u$ as an input to the system and switch to a continuous time model: $\dot{x}(t) = \mu \varepsilon \big( u(t) - x(t) \big)$.
We assume that agents have no interactions with each other, and the influence of an opinion driver is independent of the agent's opinion. Then, $\dot{x}(t) = \mu \varepsilon\, u(t)$.

We refer to the outside sources as \emph{channels}. An example is a consumer of multiple media sources with different levels of trust towards different media. The agents opinions are solely determined by the controller of the channels. 
The dynamic model is then a driftless system: $\dot{x}(t) = \bar{B} \bar{u}(t)$. 

The controller is using its channels to steer the opinion of each agent towards a target set.

For instance, the controller could be a political party financing advertisments to sway the opinion of voters in swing states during election campaigns \citep{media_politics}.
Or the controller could be a worldwide media conglomerate such as the News Corporation \citep{Murdoch}.
The COVID-19 pandemic has minimized direct interactions, hence making our setting more realistic.
An extreme variant of this scenario is illustrated by the episode "Fifteen Million Merits" of the Black Mirror series \citep{Black_Mirror}. 

A perturbing event, e.g., loss of influence, foreign acquisition of a news channel, or a new board of directors, causes one of the channels to become uncontrollable and to produce undesirable inputs. The controller has still access to this channel and is informed in real time of its content, while being unable to modify it.

We consider $n = 2$ agents both having initially a neutral opinion: $x_0 = (0,\ 0)$. Then, the target is $d = x_{goal} \in \mathbb{R}^2$. For instance, $d = (1,\ 1)$ is a \emph{consensus target}, while $d = (-1,\ 1)$ is a \emph{polarization target}.
The components of $\bar{B}$, denoted by $\bar{B}_{i,j} \in [-1,1]$ reflect the influence of channel $j$ over agent $i$. Using as a guideline the resilience criterion from \citep{journal_paper}, we pick $m+p = 2n+1 = 5$ different channels.
For instance, consider
\begin{equation*}
    \bar{B} = \begin{bmatrix} 
    0.8 & -0.9 & 0.5 & -0.5 & 0 \\
    0.9 & -0.8 & -0.4 & 0.4 & 0.1
    \end{bmatrix}.
\end{equation*}
In this setting, both agents trust channel 1 but not channel 2, they have diverging opinion on channels 3 and 4, while they are not strongly influenced by channel 5. 

We compute $r_{max}$ for the loss of control over each single channel:
\begin{equation*}
    r_{max} = \begin{bmatrix} 0.02 & 0.04 & 0.14 & 0.14 & 0.91 \end{bmatrix}.
\end{equation*}
Since $rank(\bar{B}) = 2$ and all the values of $r_{max}$ belong to $(0,1]$, according to Corollary~\ref{cor:resilience lambda}, the system is resilient to the loss of control over any channel. 
Using Theorem~\ref{thm:computation of r_q} we have $r_q = r_{max}$.
Because neither agent is significantly influenced by channel 5, the resilience to its loss is greater than the resilience to the loss of channel 1, which is trusted by both agents.

The inverse of $r_q$ tells us how much extra time the malfunctioning system needs to reach some opinion state compared to the initial system:
\begin{equation}\label{eq:inv r_q}
    \frac{1}{r_q} = \begin{bmatrix} 39.5 & 26.3 & 7.2 & 7.2 & 1.1 \end{bmatrix}.
\end{equation}
From Theorem~\ref{thm:direction maximizing r}, we know that $\frac{1}{r_q(j)} = t(\bar{B}_j)$, which is the ratio of reach times in the direction $\bar{B}_j$, the $j^{th}$ column of $\bar{B}$. Thus, if $x_{goal} = \bar{B}_1 = (0.8,\ 0.9)$, then the loss of control over channel 1 can increase the time to reach this target by a factor up to $39.5$.
On the other hand, the loss of control over channel 5 has a negligible impact on the time to reach any target.

We now choose the target to be $x_{goal} = (1,\ 1)$ and compare how the loss of each channel affects the delay to reach this target.
Intuitively, since this is a consensus target, losing control over the channel 1 or 2 will have a considerable impact, while the loss of the other channels should not be significant.
Indeed, when calculating $t(d)$ for the loss of each channel we obtain
\begin{equation*}
    t(d) = \begin{bmatrix} 39.5 & 26.3 & 1.0 & 1.0 & 1.1 \end{bmatrix},
\end{equation*}
which confirms our intuition. 

If the controller has polarization objectives, for instance $d = (-1,\ 1)$, then losing control of channel 3 or 4 should be problematic, while the others should have a smaller impact.
Indeed,
\begin{equation*}
    t(d) = \begin{bmatrix} 2.6 & 1.7 & 7.1 & 7.1 & 1.1 \end{bmatrix},
\end{equation*}
so the loss of channel 3 or 4 causes the most delay for a polarization target.

\begin{figure}[htbp!]\label{fig:failure_2}
\centering
\begin{subfigure}{.49\textwidth}
    \centering
    \includegraphics[scale = 0.35]{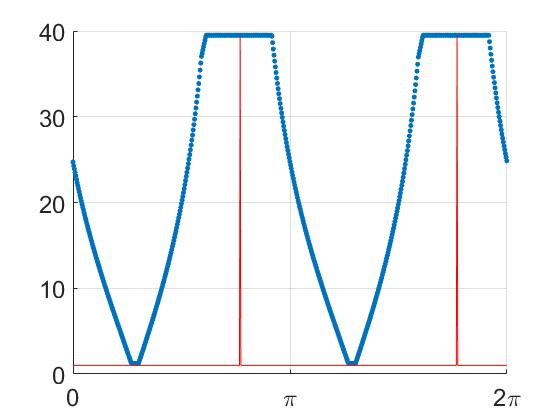}
    \caption{Loss of control over channel 1.}
    \label{fig:channel 1}
\end{subfigure}
\begin{subfigure}{.49\textwidth}
    \centering
    \includegraphics[scale = 0.35]{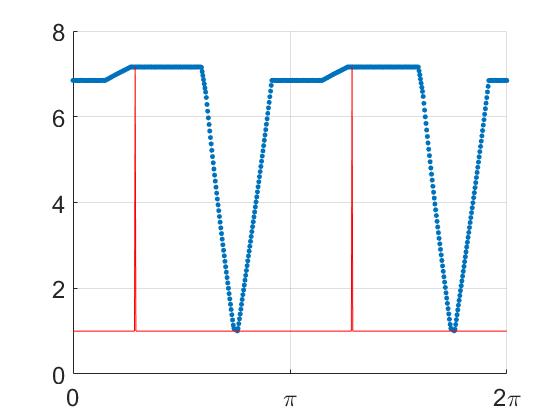}
    \caption{Loss of control over channel 3.}
    \label{fig:channel 3}
\end{subfigure}
\caption{Ratio of reach times $t(d)$ for different loss of control.}
\end{figure}

To illustrate Theorem~\ref{thm:minimum of r(d) collinear with x} we compute the ratio $t(d)$ for all directions $d \in \mathbb{S}$ parametrized by an angle $\beta \in [0, 2\pi]$. The red spike shows when $d$ is collinear with the direction $C$. As can be seen on Figure~\ref{fig:channel 1} for the loss of the first channel, the spike coincides with the maximum of $t(d)$, located at $39.5$ as announced in \eqref{eq:inv r_q}.

The two red spikes correspond to the directions $C$ and $-C$ and note that $t$ takes the same value for these two directions $t(C) = t(-C)$. Indeed, we showed in the proof of Theorem~\ref{thm:varying x_M and x_N} that $r_{X,Y}$ is an even function and in Theorem~\ref{thm:direction maximizing r} we proved that $r_{X,Y}(d) = t(d)$.

Similarly, on Figure~\ref{fig:channel 3} for the loss of channel 3 the maximal ratio $t(d)$ is $7.1$ as calculated in \eqref{eq:inv r_q} and is reached when $d$ is collinear with $C = \pm \bar{B}_3$.

\section{Conclusion and Future Work}

To quantify the drop in performance caused by the loss of control authority over actuators, this paper introduced the notion of quantitative resilience for control systems.
Relying on bang-bang control theory and on three novel optimization results, we transformed a nonlinear problem consisting of four nested optimizations into a single linear problem.
This simplification leads to a computationally efficient algorithm to verify resilience and calculate the quantitative resilience of driftless systems.

There are three promising avenues of future work. 
Previous work on resilient systems only considered $\mathcal{L}_2$ input bounds, while we worked here with $\mathcal{L}_{\infty}$ bounds. Then, the first direction of work is to build a proper resilience theory concerning these inputs. 
Secondly, we have only considered driftless systems because of the complexity of the subject. However, future work should be able to extend the concept of quantitative resilience to non-driftless linear systems.
Finally, noting that Theorems~\ref{thm:direction maximizing r} and \ref{thm:computation of r_q} only concern the loss of a single actuator, our third direction of work is to extend these results to the simultaneous loss of multiple actuators.

\bibliographystyle{IEEEtranSN}
\bibliography{refs.bib}

\appendix

\section{Proof of Theorem~\ref{thm:minimum of r(d) collinear with x}}\label{apx:proofs}

\begin{lemma}\label{lemma:r(d) cst on faces}
    When $d$, $y^+$ and $y^-$ all intersect the same face of $\partial Y$, the ratio $r(d)$ is constant.
\end{lemma}

\begin{pf}
    We define the lengths
    \begin{equation}\label{eq:delta}
        \delta^+ := \|y^+ + x\| - D \qquad \text{and} \qquad \delta^- := D - \|y^- - x\|.
    \end{equation}
    Then, $r(d) = \frac{D + \delta^+ }{D - \delta^-}$. The sign of $\delta^\pm$ depends on whether $y^\pm$ is inside or outside, as illustrated on Figure~\ref{fig:illustration of r(d)}.

    \begin{figure}[htbp!]
        \centering
        \begin{tikzpicture}[scale = 0.8]
            
            \draw[|->] (12,0) -- (13,0);
            \node at (12.5, 0.5) {$d$};
            \draw (-2, 0) -- (7.6, 0);
            \draw[dotted] (-2, 0) -- (-2, -2.5);
            \draw[dotted] (9.25, 0) -- (9.25, -2.5);
            \draw[<->] (-2, -2.5) -- (9.25, -2.5);
            \node at (4, -2.25) {$D$};
            \draw[red] (9.25, 0) -- (10.8, 0);
            \node at (10, 0.25) {\textcolor{red}{$\delta^+$}};
            \draw[blue] (7.6, 0) -- (9.25, 0);
            \node at (8.5, -0.25) {\textcolor{blue}{$\delta^-$}};
            
             \draw (8.3, 3) -- (10, -2.5);
            \node at (8.8, 2.8) {$\partial Y$};
            
            \draw[->, red] (-2, 0) -- (-1, 2);
            \node at (-0.8, 1.5) {\textcolor{red}{$x$}};
            \draw[<-, blue] (-3, -2) -- (-2, 0);
            \node at (-3, -1) {\textcolor{blue}{$-x$}};
            
            \draw[blue] (-2,0) -- (8.6, 2);
            \node at (4, 1.5) {\textcolor{blue}{$y^-$}};
            \draw[->, blue] (8.6, 2) -- (7.6, 0);
            \node at (7.5,1) {\textcolor{blue}{$-x$}};
            
            \draw[red] (-2,0) -- (9.8,-2);
            \node at (4, -0.6) {\textcolor{red}{$y^+$}};
            \draw[->, red] (9.8, -2) -- (10.8, 0);
            \node at (10.8, -1) {\textcolor{red}{$x$}};
            
            \draw[<-] (-1.2,0) arc (0:65:0.8);
            \node at (-1.05, 0.5) {$\beta$};
            \draw[->] (8.75, 0) arc (180:110:0.5);
            \node at (8.65, 0.4) {$\alpha$};
            \draw[<-] (10.05, -1.5) arc (80:120:0.5);
            \node at (9.9, -1.2) {$\gamma$};
            \draw[<-] (7.9,0) arc (0:65:0.3);
            \node at (8, 0.4) {$\beta$};
            \draw[<-] (10.5, 0) arc (180:245:0.3);
            \node at (10.3, -0.3) {$\beta$};
            \draw[->] (9.65, 0) arc (0:-70:0.4);
            \node at (9.65, -0.5) {$\alpha$};
            
        \end{tikzpicture}
        \caption{Evaluating $r(d)$ on a face of $\partial Y$}
        \label{fig:illustration of r(d)}
    \end{figure}
    
    Because $y^+$, $y^-$ and $D$ all intersect the same face of $\partial Y$ as illustrated on Figure~\ref{fig:illustration of r(d)}, the two triangles bounded by $\partial Y$, $\delta^\pm$ and $\pm x$ are congruent. Using the law of sines in these triangles, we have $\frac{\|x\|}{\sin \alpha} = \frac{\delta^+}{\sin \gamma} = \frac{\delta^-}{\sin \gamma}$. Then, $\delta^+ = \delta^- =: \delta > 0$ and $\gamma = \pi - \alpha - \beta$. Thus, $\frac{\delta}{D} = \frac{\|x\| \sin (\alpha + \beta)}{D \sin \alpha}$. As we have seen before, the representation of $\gamma$ on Figure~\ref{fig:illustration of r(d)} is only accurate when $\alpha + \beta \in [0, \pi)$.
    
    When $\alpha + \beta \in [\pi, 2\pi)$, we instead refer to Figure~\ref{fig:y^- leading and outside}. In this setting $\delta^+ = \delta^- =: \delta < 0$ and $\chi = 2\pi - \alpha - \beta$. We similarly use the sine law in the triangles of sides $\partial Y$, $\pm x$ and $\delta^\pm$:
    \begin{equation*}
        \frac{\|x\|}{\sin \alpha} = \frac{-\delta}{\sin \chi} = \frac{-\delta}{\sin (2\pi - \alpha - \beta)} = \frac{\delta}{\sin(\alpha + \beta)}.
    \end{equation*}
    The sine law uses lengths that must be positive, which explains the minus sign in front of $\delta$. Therefore, the expression $\frac{\delta}{D} = \frac{\|x\| \sin(\alpha + \beta)}{D\sin \alpha}$ holds for all values of $\alpha + \beta$. Noticing that $r(d) = \frac{1 + \frac{\delta}{D} }{1 - \frac{\delta}{D}}$ we can now evaluate $r(d)$.
    
    \vspace{2mm}
    
    We will prove that the ratio $\delta / D$ is the same for two directions $d_1 \in \mathbb{S}$ and $d_2 \in \mathbb{S}$ when their respective $D$, $y^+$ and $y^-$ all intersect the same face of $\partial Y$, as illustrated on Figure~\ref{fig:ratio constant on faces}. We also define $\delta_1 := \|y_1^+ + x\| - D_1$ and $\delta_2 := \|y_2^+ + x\| - D_2$.
    
    \begin{figure}[htbp!]
        \centering
        \begin{tikzpicture}[scale = 0.8]
            
            \draw (8.375, 3) -- (10.8, -3.5);
            \node at (10, 0) {$\partial Y$};
            
            \draw[<-, blue] (-2.5, -1) -- (-2, 0);
            \draw[->, red] (-2, 0) -- (-1.5, 1);
            
            \draw[|->] (10, 2.2) -- (11, 2.42);
            \node at (10.5, 2.7) {$d_1$};
            \draw (-2, 0) -- (8.75, 2);
            \node at (4, 1.5) {$D_1$};
            
            \draw[<-] (-1.2, 0.16) arc (0:65:0.7);
            \node at (-1.1, 0.9) {$\beta_1$};
            \draw[->] (8.3, 1.95) arc (190:120:0.5);
            \node at (8.05, 2.4) {$\alpha_1$};
           
            \draw[|->] (12,-3.3) -- (13,-3.52);
            \node at (12.5, -3) {$d_2$};
            \draw (-2, 0) -- (10.6, -3);
            \node at (4, -1) {$D_2$};
            
            \draw[<-] (-1.5, -0.15) arc (-10:65:0.5);
            \node at (-1.5, -0.4) {$\beta_2$};
            \draw[->] (10.1, -2.9) arc (170:106:0.5);
            \node at (9.8, -2.5) {$\alpha_2$};
            
            \draw[<-] (0, -0.5) arc (-25:26:1);
            \node at (1,0) {$\beta_2 - \beta_1$};
            
        \end{tikzpicture}
        \caption{Ratio $r(d)$ is constant on a face of $\partial Y$}
        \label{fig:ratio constant on faces}
    \end{figure}

    The sum of the angles of the triangle in Figure~\ref{fig:ratio constant on faces} is
    \begin{equation}\label{eq:alpha + beta = cst on a face}
        (\beta_2 - \beta_1) + \alpha_2 + (\pi - \alpha_1) = \pi \qquad \text{so} \qquad \beta_2 + \alpha_2 = \beta_1 + \alpha_1.
    \end{equation}
    Therefore, $\alpha + \beta$ is constant on faces of $\partial Y$. 
    We also use the sine law in the triangle in Figure~\ref{fig:ratio constant on faces} and obtain
    \begin{equation*}
        \frac{D_1}{\sin \alpha_2} = \frac{D_2}{\sin (\pi - \alpha_1)} = \frac{D_2}{\sin \alpha_1}.
    \end{equation*}
    Then,
    \begin{equation*}
        \frac{\delta_1}{D_1} = \frac{\|x\| \sin(\alpha_1 + \beta_1)}{D_1 \sin \alpha_1} = \frac{\|x\| \sin(\alpha_2 + \beta_2)}{D_2 \sin \alpha_2} = \frac{\delta_2}{D_2}.
    \end{equation*}
    Hence, $r(d_1) = r(d_2)$, the ratio $r(d)$ is constant when $d$, $y^+$ and $y^-$ are on the same face of $\partial Y$. Thus, the variations of $r(d)$ only occur when $d$ is crossing a vertex. $\quad \blacksquare$
\end{pf}

\vspace{2mm}

\begin{lemma}\label{lemma:leading and outside}
    The following statements are true:
    \begin{itemize}
        \item If $\beta \in (0, \pi)$, then $y^+$ is leading. If $\beta \in (\pi, 2\pi)$, then $y^-$ is leading.
        \item If $\alpha + \beta \in (0, \pi)$, then $y^+$ is outside. If $\alpha + \beta \in (\pi, 2\pi)$, then $y^-$ is outside.
    \end{itemize}
\end{lemma}

\begin{pf}
    We assume that $D$, $y^+$ and $y^-$ all intersect the same face of $\partial Y$. The objective of this part is to learn the values of the angle $\beta$ for which $y^\pm$ is leading or trailing and outside or inside.
    
    Figure~\ref{fig:y^+ leading and outside} represents the situation where the vector $y^+$ is leading and outside, while $y^-$ is trailing and inside. We want to determine for which values of $\beta$ this situation arises.
    
    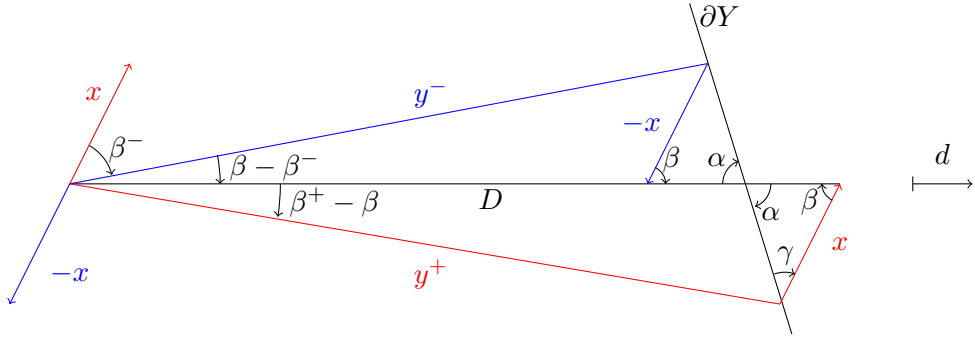
\begin{figure}[htbp!]
        \centering
        \begin{tikzpicture}[scale = 0.8]
            
            \draw[|->] (12,0) -- (13,0);
            \node at (12.5, 0.5) {$d$};
            \draw (-2, 0) -- (10.8, 0);
            \node at (5, -0.25) {$D$};
            
            \draw (8.3, 3) -- (10, -2.5);
            \node at (8.8, 2.8) {$\partial Y$};
            
            \draw[->, red] (-2, 0) -- (-1, 2);
            \node at (-1.6, 1.5) {\textcolor{red}{$x$}};
            \draw[<-, blue] (-3, -2) -- (-2, 0);
            \node at (-2, -1.5) {\textcolor{blue}{$-x$}};
            
            \draw[blue] (-2,0) -- (8.6, 2);
            \node at (4, 1.5) {\textcolor{blue}{$y^-$}};
            \draw[->, blue] (8.6, 2) -- (7.6, 0);
            \node at (7.5,1) {\textcolor{blue}{$-x$}};
            
            \draw[red] (-2,0) -- (9.8,-2);
            \node at (4,-1.5) {\textcolor{red}{$y^+$}};
            \draw[->, red] (9.8, -2) -- (10.8, 0);
            \node at (10.8, -1) {\textcolor{red}{$x$}};
            
            \draw[<-] (-1.3, 0.15) arc (15:60:0.8);
            \node at (-1.05, 0.6) {$\beta^-$};
            \draw[->] (8.85, 0) arc (180:110:0.4);
            \node at (8.75, 0.4) {$\alpha$};
            \draw[<-] (10.05, -1.5) arc (70:110:0.5);
            \node at (9.9, -1.2) {$\gamma$};
            \draw[<-] (7.9,0) arc (0:65:0.3);
            \node at (8, 0.4) {$\beta$};
            \draw[<-] (10.5, 0) arc (180:245:0.3);
            \node at (10.3, -0.3) {$\beta$};
            \draw[->] (9.65, 0) arc (0:-70:0.4);
            \node at (9.65, -0.5) {$\alpha$};
            
            \draw[<-] (0.5, 0) arc (0:11:2.5);
            \node at (1.4, 0.3) {$\beta - \beta^-$};
            
            \draw[->] (1.5, 0) arc (0:-9:3.5);
            \node at (2.4, -0.3) {$\beta^+ - \beta$};
            
        \end{tikzpicture}
        \caption{Illustration of $y^+$ leading and outside with $y^-$ trailing and inside.}
        \label{fig:y^+ leading and outside}
    \end{figure}

    We apply the sine law in the two triangles of Figure~\ref{fig:y^+ leading and outside} bounded by $y^\pm$, $\pm x$ and $D$:
    \begin{equation*}
        \frac{\|-x\|}{\sin(\beta - \beta^-)} = \frac{\|y^-\|}{\sin(\pi-\beta)} = \frac{\|y^-\|}{\sin \beta} \qquad \text{and} \qquad \frac{\|x\|}{\sin(\beta^+ - \beta)} = \frac{\|y^+\|}{\sin \beta}.
    \end{equation*}
    
    Then, we have $\|x\| \sin \beta = \|y^-\| \sin(\beta - \beta^-) = \|y^+\| \sin(\beta^+ - \beta)$. Since the three norms are positive, the three sine functions have the same sign. Since we assumed that $y^+$  is leading, we have $0 \leq \beta^- < \beta < \beta^+ \leq 2\pi$. Then, $\beta - \beta^- > 0$ and $\beta^+ - \beta > 0$. 
    
    For contradiction purposes, assume that $\beta - \beta^- > \pi$, then $\sin(\beta - \beta^-) <  0$ and so $\sin(\beta^+ - \beta) < 0$, which leads to $\beta^+ - \beta >  \pi$. Then, $\beta^+ - \beta^- > 2\pi$, but that is impossible since $\beta^\pm \in [0, 2\pi)$. Therefore, $\beta - \beta^- \in (0, \pi)$. Thus, $\sin(\beta - \beta^-) >  0$, which leads to $\sin \beta > 0$ and then $\beta \in (0,\pi)$. 
    
    To sum up, when $y^+$ is leading we have $\beta \in (0, \pi)$. Now we study the other case, when $y^-$ is leading as represented on Figure~\ref{fig:y^- leading and outside} and we want to find the range of $\beta$ where this situation occurs.
    
     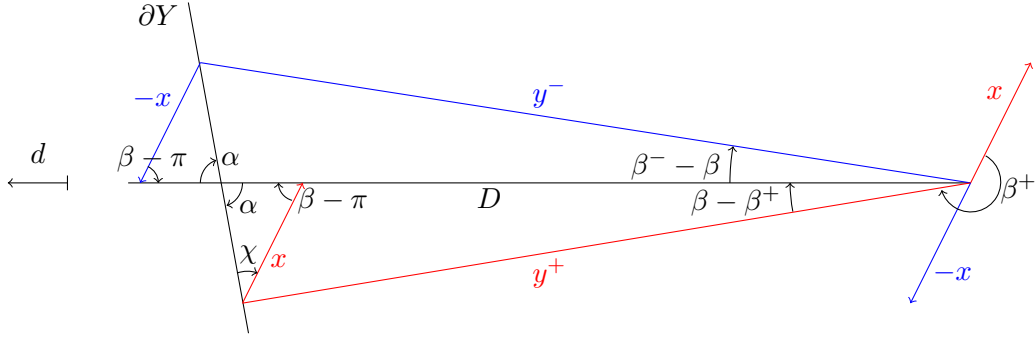
\begin{figure}[htbp!]
        \centering
        
        \begin{tikzpicture}[scale = 0.8]
            
            \draw[|->] (-3,0) -- (-4,0);
            \node at (-3.5, 0.5) {$d$};
            \draw (-2, 0) -- (12, 0);
            \node at (4, -0.25) {$D$};
            
            \draw (-1, 3) -- (0, -2.5);
            \node at (-1.5, 2.8) {$\partial Y$};
            
            \draw[->, red] (12, 0) -- (13, 2);
            \node at (12.4, 1.5) {\textcolor{red}{$x$}};
            \draw[<-, blue] (11, -2) -- (12, 0);
            \node at (11.7, -1.5) {\textcolor{blue}{$-x$}};
            
            \draw[blue] (12,0) -- (-0.8, 2);
            \node at (5, 1.5) {\textcolor{blue}{$y^-$}};
            \draw[->, blue] (-0.8, 2) -- (-1.8, 0);
            \node at (-1.6, 1.4) {\textcolor{blue}{$-x$}};
            
            \draw[red] (12,0) -- (-0.1, -2);
            \node at (5,-1.5) {\textcolor{red}{$y^+$}};
            \draw[->, red] (-0.1, -2) -- (0.9, 0);
            \node at (0.5, -1.3) {\textcolor{red}{$x$}};
            
            \draw[->] (12.25, 0.45) arc (60:-164:0.5);
            \node at (12.8, -0.1) {$\beta^+$};
            \draw[->] (-0.1, 0) arc (0:-70:0.4);
            \node at (0, -0.4) {$\alpha$};
            \draw[<-] (0.15, -1.5) arc (70:110:0.5);
            \node at (0, -1.2) {$\chi$};
            \draw[<-] (-1.5, 0) arc (0:65:0.3);
            \node at (-1.6, 0.4) {$\beta - \pi$};
            \draw[<-] (0.5, 0) arc (180:255:0.3);
            \node at (1.4, -0.3) {$\beta - \pi$};
            \draw[->] (-0.8, 0) arc (180:110:0.4);
            \node at (-0.3, 0.4) {$\alpha$};
            
            \draw[<-] (9, 0) arc (180:191:2.5);
            \node at (8.1, -0.3) {$\beta - \beta^+$};
            
            \draw[->] (8, 0) arc (180:170:3.5);
            \node at (7.1, 0.3) {$\beta^- - \beta$};
            
        \end{tikzpicture}
        \caption{Illustration of $y^-$ leading and outside, while $y^+$ is trailing and inside.}
        \label{fig:y^- leading and outside}
    \end{figure}
    
    We apply the sine law in the triangles delimited by $y^\pm$, $D$ and $\pm x$:
    \begin{equation*}
        \frac{\|-x\|}{\sin(\beta^- - \beta)} = \frac{\|y^-\|}{\sin(\beta - \pi)} = \frac{-\|y^-\|}{\sin \beta} \qquad \text{and} \qquad \frac{\|x\|}{\sin(\beta - \beta^+)} = \frac{\|y^+\|}{\sin \big(\pi - (\beta - \pi)\big)} = \frac{-\|y^+\|}{\sin \beta}.
    \end{equation*}
    Then, we have $\|x\| \sin \beta = - \|y^-\| \sin (\beta^- - \beta) = -\|y^+\| \sin(\beta - \beta^+)$. 
    Since $y^-$ is leading we have $0 \leq \beta^+ < \beta < \beta^- < 2\pi$. Therefore $\beta^- - \beta > 0$ and $\beta - \beta^+ > 0$.
    
    Assume for contradiction purposes that $\beta^- - \beta > \pi$. Then, $\frac{\|y^-\|}{\|y^+\|}\sin(\beta^- -  \beta) < 0$, and so, $\sin(\beta - \beta^+) < 0$. Thus, $\beta - \beta^+ > \pi$, which leads to the impossible conclusion that $\beta^- - \beta^+ > 2\pi$. Therefore, $\beta^- - \beta \in (0, \pi)$, so $-\frac{\|y^-\|}{\|x\|}\sin(\beta^- - \beta) < 0$, i.e., $\beta  \in (\pi, 2\pi)$.
    
    To sum up, when $y^-$ is leading we have $\beta \in (\pi, 2\pi)$. We also know that $y^+$ leading implies $\beta \in (0, \pi)$, and for all $\beta \in (0, \pi) \cup (\pi, 2\pi)$ either $y^+$ or $y^-$ must be leading. We deduce that the converse of the two implications proved above are true: if $\beta \in (0, \pi)$, then $y^+$ is leading, and if $\beta \in (\pi, 2\pi)$, then $y^-$ is leading.
    
    \vspace{3mm}
    
    Now, we want to determine the range of values of the angle $\alpha + \beta$ for which $y^\pm$ is outside. Based on Figure~\ref{fig:y^+ leading and outside} where $y^+$ is outside, we have $\alpha + \beta \in (0, \pi)$.
    Then, based on Figure~\ref{fig:y^- leading and outside} where $y^+$ is inside and $y^-$ is outside, we have $\alpha + \beta \in (\pi, 2\pi)$.
    
    The only situation where neither $y^+$ nor $y^-$ is outside occurs when $\|y^+ + x\| = D = \|y^- - x\|$, i.e., at the vertices $v_\pi$ and $v_{2\pi}$, i.e., when $\alpha + \beta \in \{ \pi, 2\pi \}$. For all other values of $\alpha + \beta$, either $y^+$ or $y^-$ is outside. We deduce that if $\alpha + \beta \in (0, \pi)$, then $y^+$ is outside, and if $\alpha + \beta \in (\pi, 2\pi)$, then $y^-$ is outside.   $\quad \blacksquare$    
\end{pf}

\vspace{5mm}

\begin{lemma}\label{lemma:alpha + beta reparametrization}
    The following statements are true:
    \begin{itemize}
        \item if $\alpha + \beta \in (\alpha_0, \pi)$, then $y^+$ is leading and outside,
        \item if $\alpha + \beta \in (\pi, \alpha_0 + \pi)$, then $y^+$ is leading and inside,
        \item if $\alpha + \beta \in (\alpha_0 + \pi, 2\pi)$, then $y^-$ is leading and outside,
        \item if $\alpha + \beta \in (2\pi, \alpha_0 + 2\pi)$, then $y^-$ is leading and inside.
    \end{itemize}
\end{lemma}
\begin{pf}
    We have taken the convention that the angles are positively oriented in the clockwise orientation.
    According to \eqref{eq:alpha + beta = cst on a face}, the angle $\alpha + \beta$ is constant on a face of $\partial Y$. When $d$ crosses a vertex of external angle $\varepsilon$ as represented on Figure~\ref{fig:y^+ crossing a vertex}, the value of $\alpha$ has a discontinuity of $+\varepsilon$. Let $q$ be the number of vertices of $\partial Y$ and $\varepsilon_i$ the external angle of the $i^{th}$ vertex $v_i$. Since $Y \cap \mathcal{P}$ is a polygon, $\sum_{i = 1}^q \varepsilon_i = 2\pi$.
    We can then represent the evolution of $\alpha + \beta$ as a function of $\beta$ with Figure~\ref{fig:graph of alpha + beta}.
    
    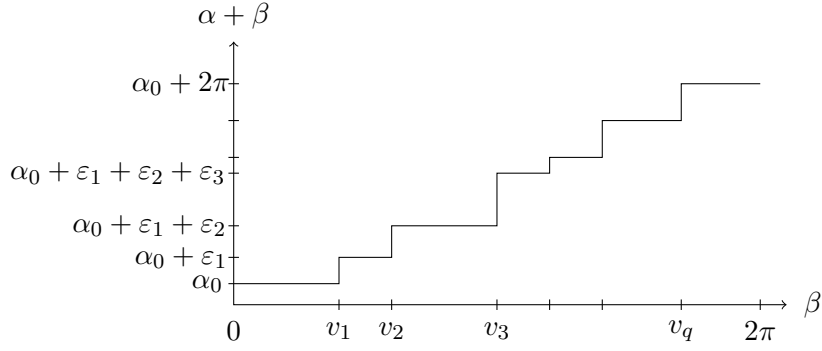
\begin{figure}[htbp!]
        \centering
        \begin{tikzpicture}[scale = 0.7]
            \draw[<->] (0, 5) -- (0, 0) -- (10.5, 0);
            \node at (11, 0) {$\beta$};
            \node at (0, 5.5) {$\alpha + \beta$};
            \node at (0, -0.5) {$0$};
            \draw (10, -0.1) -- (10, 0.1);
            \node at (10, -0.5) {$2\pi$};
            
            \draw (-0.1, 0.4) -- (2, 0.4) -- (2, 0.9) -- (3, 0.9) -- (3, 1.5) -- (5, 1.5) -- (5, 2.5) -- (6, 2.5) -- (6, 2.8) -- (7, 2.8) -- (7, 3.5) -- (8.5, 3.5) -- (8.5, 4.2) -- (10, 4.2);
            
            \draw (2, -0.1) -- (2, 0.1);
            \node at (2, -0.5) {$v_1$};
            \draw (3, -0.1) -- (3, 0.1);
            \node at (3, -0.5) {$v_2$};
            \draw (5, -0.1) -- (5, 0.1);
            \node at (5, -0.5) {$v_3$};
            \draw (6, -0.1) -- (6, 0.1);
            \draw (7, -0.1) -- (7, 0.1);
            \draw (8.5, -0.1) -- (8.5, 0.1);
            \node at (8.5, -0.5) {$v_q$};
            
            \node at (-0.5, 0.4) {$\alpha_0$};
            \draw (-0.1, 0.9) -- (0.1, 0.9);
            \node at (-1, 0.9) {$\alpha_0 + \varepsilon_1$};
            \draw (-0.1, 1.5) -- (0.1, 1.5);
            \node at (-1.6, 1.5) {$\alpha_0 + \varepsilon_1 + \varepsilon_2$};
            \draw (-0.1, 2.5) -- (0.1, 2.5);
            \node at (-2.2, 2.5) {$\alpha_0 + \varepsilon_1 + \varepsilon_2+\varepsilon_3$};
            \draw (-0.1, 2.8) -- (0.1, 2.8);
            \draw (-0.1, 3.5) -- (0.1, 3.5);
            \draw (-0.1, 4.2) -- (0.1, 4.2);
            \node at (-1, 4.2) {$\alpha_0 + 2\pi$};
        
        \end{tikzpicture}
        \caption{Evolution of $\alpha + \beta$ with $\beta$ increasing clockwise in $[0, 2\pi)$.}
        \label{fig:graph of alpha + beta}
    \end{figure}
    
    Recall that $\alpha_0$ is the value of $\alpha$ when $\beta = 0$. After a whole revolution $\alpha + \beta = \alpha_0 + 2\pi$. So there are two vertices $v_{\pi}$ and $v_{2\pi}$ where $\alpha + \beta$ first crosses $\pi$ and then $2\pi$. 
    In the eventuality that $\alpha + \beta = \pi$ or $2\pi$ on a face, we define $v_\pi$ or $v_{2\pi}$ as the vertex preceding the face. This face is parallel with the span of $x$. Thus $\|y^+ + x\| = \|y^- - x\| = D$, so $y^+$ and $y^-$ are neither outside nor inside. The ratio is $r(d) = 1$ on this face.
   
    Because of the monotonic evolution of $\alpha + \beta$ as a function of $\beta$, we can use $\alpha + \beta$ instead of $\beta$ to parametrize the directions $d$.  The interval $\beta \in (0, \pi)$ is the same as $\alpha + \beta \in (\alpha_0, \alpha_0 + \pi)$ and the interval $\beta \in (\pi, 2\pi)$ is the same as $\alpha + \beta \in (\alpha_0 + \pi, \alpha_0 + 2\pi)$.
    Then, the bullet list established in Lemma~\ref{lemma:leading and outside} can be rewritten as claimed in this lemma.     $\quad \blacksquare$
\end{pf}

\vspace{2mm}

\begin{lemma}\label{lemma: y^+ crossing}
    The ratio $r(d)$ decreases when the leading vector $y^+$ is outside for a vertex crossing.
\end{lemma}
\begin{pf}
    The leading vector $y^+$ is outside and crosses a vertex while $\beta$ increases. We separate the vertex crossing into two parts: when only $y^+$ has crossed, and when both $d$ and $y^+$ have crossed the vertex. Since we do not yet consider the vertices $v_\pi$ and $v_{2\pi}$, the leading vector is outside before and after the vertex.
    Let $\varepsilon$ be the external angle of the vertex between the faces $F_1$ and $F_2$ of $\partial Y$ as shown on Figure~\ref{fig:y^+ crossing a vertex}. 
    
    \begin{figure}[htbp!]
        \centering
        \begin{tikzpicture}[scale = 0.7]
            
            \draw (-5, 0) -- (0.2,0) -- (5, -2.1);
            \draw[dashed] (0,0) -- (5,0);
            \node at (-4.5, 0.3) {$F_1$};
            \node at (5, -1.8) {$F_2$};
            
            \draw[->] (4,0) arc (0:-22:4);
            \node at (4.3, -1) {$\varepsilon$};
            
            \draw (-2.8, -3) -- (-2, -2);
            \draw (1.3, 2) -- (1.45, 2.2);
            \node at (-2.2, -2.6) {$d$};
            \draw[red] (-1, -3) -- (2.5, -1);
            \node at (-0.1, -2.2) {\textcolor{red}{$y^+$}};
            \draw[blue] (-4.75, -3) -- (-4, 0);
            \node at (-4, -2) {\textcolor{blue}{$y^-$}};
            
            \draw[->, red] (2.5, -1) -- (0.5, 1);
            \node at (1.4, 0.5) {\textcolor{red}{$x$}};
            \draw[->, blue] (-4, 0) -- (-2, -2);
            \node at (-2.8, -0.5) {\textcolor{blue}{$-x$}};
            
            \draw[->, red] (3.3, 0) -- (1.3, 2);
            \node at (2.5, 1.2) {\textcolor{red}{$x$}};
            \draw[green] (2.5, -1) -- (3.3, 0);
            \node at (3, -0.8) {\textcolor{green}{$l$}};
            \draw[green] (0.5, 1) -- (1.3, 2);
            \node at (0.8, 1.8) {\textcolor{green}{$l$}};
           
           \draw[<-] (-1.8, -1.8) arc (60:114:0.5);
           \node at (-2, -1.4) {$\beta$};
           \draw[<-] (2.67, -0.75) arc (60:114:0.5);
           \node at (2.2, -0.5) {$\beta$};
            
            \draw[<-] (-0.8, 0) arc (180:230:0.5);
            \node at (-1, -0.2) {$\alpha$};
            \draw[blue] (-2, -2) -- (-0.3, 0);
            \node at (-0.7, -1) {\textcolor{blue}{$\delta^-$}};
            \draw[red] (-0.3, 0) -- (0.5, 1);
            \node at (-0.1, 0.6) {\textcolor{red}{$\delta^+$}};
            
            \draw[<-] (2.8, 0) arc (180:230:0.5);
            \node at (2.6, -0.2) {$\alpha$};
            
            \draw[dotted] (1.5, 0) -- (1.5, -2.5);
            \draw[dotted] (3.3, 0) -- (3.3, -2.5);
            \draw[<->] (1.5, -2.5) -- (3.3, -2.5);
            \node at (2.4, -2.2) {$m$};
            
        \end{tikzpicture}
        \caption{Part I of the crossing of a vertex by $y^+$ leading and outside as $\beta$ increases.}
        \label{fig:y^+ crossing a vertex}
    \end{figure}
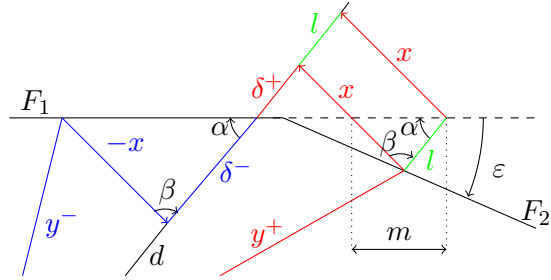
    
    Because $y^+$ does not intersect $F_1$ anymore, $\delta^+$ is shorter than $\delta^-$. We define $l := \delta^- - \delta^+$. Notice that the two green segments of length $l$ in Figure~\ref{fig:y^+ crossing a vertex} are parallel. We parametrize the position of $y^+$ on $F_2$ with the length $m$ as defined on Figure~\ref{fig:y^+ crossing a vertex}. When $y^+$ is at the vertex $m = 0$, and $m$ increases with $\beta$. Using the sine law we can link the loss $l$ with the distance $m$
    \begin{equation}\label{eq:loss crossing part 1}
        \frac{m}{\sin \beta} = \frac{l}{\sin (\pi - \alpha - \beta)} = \frac{l}{\sin(\alpha + \beta)}.
    \end{equation}
    
     We calculate the ratio $r(d)$ as a function of $r_{F_1}$, which is the value of $r(d)$ on the face $F_1$:
    \begin{equation}\label{eq:r(d) r_F_1}
        r(d) = \frac{D + \delta^+}{D - \delta^-} = \frac{D + \delta^- - l}{D - \delta^-} = r_{F_1} - \frac{l}{D - \delta^-} = r_{F_1} - \frac{m \sin (\alpha + \beta)}{(D - \delta^-) \sin(\beta)}.
    \end{equation}
    By definition the length $m$ is positive. Since $x \notin \partial Y$ but $y^- \in \partial Y$, we have $D - \delta^- = \|y^- - x \| > 0$. 
    
    We have seen previously that for $y^+$ to be leading and outside we need $\alpha + \beta \in (\alpha_0, \pi)$. In that case $\sin(\alpha + \beta) > 0$ and $\sin(\beta) > 0$.
    Therefore, the term subtracted from $r_{F_1}$ is positive, i.e., $r(d) < r_{F_1}$. 
    
    \vspace{2mm}
    
    We can now tackle the second part of the crossing, when $y^+$ and $d$ both have crossed the vertex as illustrated on Figure~\ref{fig:y^+ and d crossing a vertex}.
    
    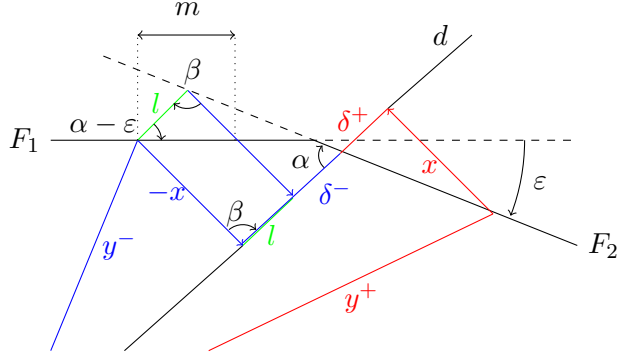
\begin{figure}[htbp!]
        \centering
        \begin{tikzpicture}[scale = 0.7]
            
            \draw (-5, 0) -- (0,0) -- (5, -2);
            \draw[dashed] (-4, 1.6) -- (0,0) -- (5,0);
            \node at (-5.5, 0) {$F_1$};
            \node at (5.5, -2) {$F_2$};
            
            \draw[->] (4,0) arc (0:-21:4);
            \node at (4.3, -0.8) {$\varepsilon$};
            
            \draw (-3.6, -4) -- (-1.35, -2);
            \draw (1.4, 0.6) -- (3, 2);
            \node at (2.4, 2) {$d$};
            \draw[red] (-2, -4) -- (3.4, -1.4);
            \node at (0.9, -3) {\textcolor{red}{$y^+$}};
            \draw[blue] (-5, -4) -- (-3.35, 0);
            \node at (-3.7, -2) {\textcolor{blue}{$y^-$}};
            
            \draw[->, red] (3.4, -1.4) -- (1.4, 0.6);
            \node at (2.2, -0.5) {\textcolor{red}{$x$}};
            \draw[->, blue] (-3.35, 0) -- (-1.35, -2);
            \node at (-2.8, -1) {\textcolor{blue}{$-x$}};
            
            \draw[->, blue] (-2.4, 0.95) -- (-0.4, -1.05);
            % \node at (-1.4, -0.5) {\textcolor{green}{$x_m$}};
            \draw[green] (-3.35, 0) -- (-2.4, 0.95);
            \node at (-3, 0.7) {\textcolor{green}{$l$}};
            \draw[green] (-1.35, -2) -- (-0.4, -1.1);
            \node at (-0.8, -1.8) {\textcolor{green}{$l$}};
           
            \draw[<-] (-2.9, 0) arc (0:50:0.4);
            \node at (-4, 0.3) {$\alpha - \varepsilon$};
            \draw[->] (-2.15, 0.7) arc (-50:-130:0.4);
            \node at (-2.3, 1.3) {$\beta$};
            \draw[<-] (-1.1, -1.7) arc (50:130:0.4);
            \node at (-1.5, -1.4) {$\beta$};
            
            \draw[<-] (0.16, -0.1) arc (140:230:0.3);
            \node at (-0.25, -0.4) {$\alpha$};
            \draw[blue] (0.5, -0.25) -- (-1.35, -1.95);
            \node at (0.4, -1) {\textcolor{blue}{$\delta^-$}};
            \draw[red] (0.5, -0.25) -- (1.4, 0.6);
            \node at (0.75, 0.4) {\textcolor{red}{$\delta^+$}};
            
            \draw[dotted] (-3.35, 0) -- (-3.35, 2);
            \draw[dotted] (-1.5, 0) -- (-1.5, 2);
            \draw[<->] (-3.35, 2) -- (-1.5, 2);
            \node at (-2.4, 2.5) {$m$};
        \end{tikzpicture}
        \caption{Part II of the crossing of a vertex by $y^+$ leading and outside as $\beta$ increases.}
        \label{fig:y^+ and d crossing a vertex}
    \end{figure}
    
    Because $y^-$ does not intersect $F_2$, $\delta^-$ is longer than $\delta^+$. As before, let $l := \delta^- - \delta^+$. Using the sine law, we can relate $l$ to $m$ and express the ratio $r(d)$:
    \begin{equation}\label{eq:gain crossing part 2}
        r(d) = \frac{D + \delta^+}{D - \delta^-} = \frac{D + \delta^+}{D - \delta^+ - l} \qquad \text{with} \qquad \frac{m}{\sin \beta} = \frac{l}{\sin(\pi - \beta - \alpha + \varepsilon)} = \frac{l}{\sin(\alpha + \beta - \varepsilon)}.
    \end{equation}
    Since $y^+$ is leading and outside on $F_2$ we have $\alpha + \beta \in (\alpha_0, \pi)$, so $\sin(\beta) > 0$.
    If $\alpha$ was still measured between $d$ and $F_1$, then its value would be $\alpha_{F_1} = \alpha - \varepsilon$. Since we are not considering the crossing of $v_\pi$ or $v_{2\pi}$, $y^+$ is also leading and outside on $F_1$. Then, $\alpha_{F_1} + \beta \in (\alpha_0 + \pi)$, i.e., $\alpha + \beta - \varepsilon \in (\alpha_0, \pi)$.
    This yields $\sin(\alpha + \beta - \varepsilon) > 0$, which makes $l > 0$, because the length $m$ is positive by definition.
    Note that $r_{F_2} = \frac{D + \delta^+}{D - \delta^+}$, which leads to $r(d) > r_{F_2}$. 
    Thus, the ratio $r(d)$ decreases during the crossing of a vertex when $y^+$ is leading and outside. $\quad \blacksquare$
\end{pf}

\vspace{2mm}

\begin{lemma}\label{lemma: y^- crossing}
    The ratio $r(d)$ decreases when the leading vector $y^-$ is outside for a vertex crossing. 
\end{lemma}

\begin{pf}
    The leading vector $y^-$ is outside and crosses a vertex while $\beta$ increases. We separate the vertex crossing into two parts: when only $y^-$ has crossed, and when both $d$ and $y^-$ have crossed the vertex. Since we do not yet consider the vertices $v_\pi$ and $v_{2\pi}$, the leading vector is outside before and after the vertex.
    Let $\varepsilon$ be the external angle of the vertex between the faces $F_1$ and $F_2$ of $\partial Y$ as shown on Figure~\ref{fig:y^- crossing a vertex}.
    
    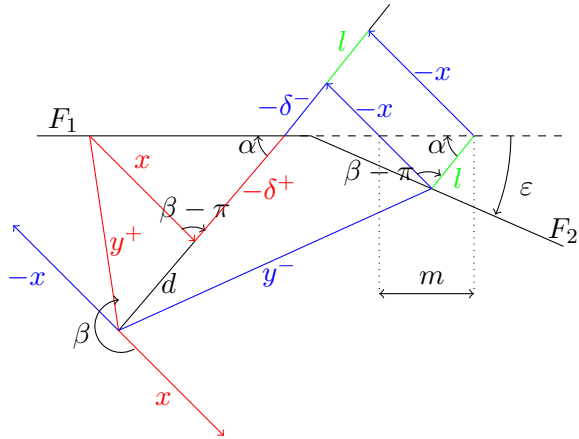
\begin{figure}[htbp!]
        \centering
        \begin{tikzpicture}[scale = 0.7]
            
            \draw (-5, 0) -- (0.2,0) -- (5, -2.1);
            \draw[dashed] (0,0) -- (5,0);
            \node at (-4.5, 0.3) {$F_1$};
            \node at (5, -1.8) {$F_2$};
            
            \draw[->] (4,0) arc (0:-22:4);
            \node at (4.3, -1) {$\varepsilon$};
            
            \draw (1.3, 2) -- (1.7, 2.5);
            \node at (-2.5, -2.8) {$d$};
            \draw (-3.45, -3.7) -- (-2, -2);
            
            \node at (-0.4, -2.6) {\textcolor{blue}{$y^-$}};
            \draw[blue] (-3.45, -3.7) -- (2.5, -1);
            \node at (-3.3, -2) {\textcolor{red}{$y^+$}};
            \draw[red] (-3.45, -3.7) -- (-4, 0);
            
            \draw[<-, blue] (-5.45, -1.7)  -- (-3.45, -3.7);
            \node at (-5.2, -2.7) {\textcolor{blue}{$-x$}};
            \draw[->, red] (-3.45, -3.7)  -- (-1.45, -5.7);
            \node at (-2.6, -5) {\textcolor{red}{$x$}};
            \draw[->] (-3.15, -4.05) arc (-60:-265:0.5);
            \node at (-4.15, -3.8) {$\beta$};
            
            \draw[->, blue] (2.5, -1) -- (0.5, 1);
            \node at (1.4, 0.5) {\textcolor{blue}{$-x$}};
            \draw[->, red] (-4, 0) -- (-2, -2);
            \node at (-3, -0.5) {\textcolor{red}{$x$}};
            
            \draw[->, blue] (3.3, 0) -- (1.3, 2);
            \node at (2.5, 1.2) {\textcolor{blue}{$-x$}};
            \draw[green] (2.5, -1) -- (3.3, 0);
            \node at (3, -0.8) {\textcolor{green}{$l$}};
            \draw[green] (0.5, 1) -- (1.3, 2);
            \node at (0.8, 1.8) {\textcolor{green}{$l$}};
           
           \draw[<-] (-1.8, -1.8) arc (60:114:0.5);
           \node at (-2, -1.4) {$\beta-\pi$};
           \draw[<-] (2.67, -0.75) arc (60:114:0.5);
           \node at (1.5, -0.7) {$\beta-\pi$};
            
            \draw[<-] (-0.8, 0) arc (180:230:0.5);
            \node at (-1, -0.2) {$\alpha$};
            \draw[red] (-2, -2) -- (-0.3, 0);
            \node at (-0.6, -1) {\textcolor{red}{$-\delta^+$}};
            \draw[blue] (-0.3, 0) -- (0.5, 1);
            \node at (-0.3, 0.6) {\textcolor{blue}{$-\delta^-$}};
            
            \draw[<-] (2.8, 0) arc (180:230:0.5);
            \node at (2.6, -0.2) {$\alpha$};
            
            \draw[dotted] (1.5, 0) -- (1.5, -3);
            \draw[dotted] (3.3, 0) -- (3.3, -3);
            \draw[<->] (1.5, -3) -- (3.3, -3);
            \node at (2.5, -2.7) {$m$};
            
        \end{tikzpicture}
        \caption{Part I of the crossing of a vertex by $y^-$ leading and outside as $\beta$ increases.}
        \label{fig:y^- crossing a vertex}
    \end{figure}
    
    Since $y^-$ is outside and $y^+$ is inside, by definition \eqref{eq:delta}, $\delta^+ < 0$ and $\delta^- < 0$. We keep $l := \delta^- - \delta^+$ like in the previous case. The distance $m$ also increases monotonically with $\beta$ as $y^-$ goes further away from the vertex. We apply the sine law in the same triangle as before:
    \begin{equation*}
        \frac{m}{\sin (\beta-\pi)} = \frac{l}{\sin \big(\pi - \alpha - (\beta-\pi)\big)} = \frac{l}{\sin(2\pi - \alpha -\beta)} = \frac{-m}{\sin \beta} = \frac{-l}{\sin(\alpha + \beta)}.
    \end{equation*}
    Thus, the relation linking $m$ and $l$ is the same whether $y^+$ or $y^-$ is leading: $l = \frac{m \sin(\alpha + \beta)}{\sin \beta}$. Besides, $\delta^+$ and $\delta^-$ are also related through the same equation with $l$. Therefore, the ratio $r(d)$ as a function of $r_{F_1}$ is the same as previously:
    \begin{equation*}
        r(d) = \frac{D + \delta^+}{D - \delta^-} = \frac{D + \delta^- - l}{D - \delta^-} = r_{F_1} - \frac{l}{D - \delta^-} = r_{F_1} - \frac{m \sin (\alpha + \beta)}{(D - \delta^-) \sin(\beta)}.
    \end{equation*}
    For the same reasons as above $m > 0$ and $D - \delta^- > 0$. 
    We have established previously that to have $y^-$ leading and outside we need $\alpha + \beta \in (\alpha_0 + \pi, 2\pi)$. In this situation $\sin(\alpha + \beta) < 0$ and $\sin \beta < 0$. Therefore, the term subtracted from $r_{F_1}$ is positive, so $r(d) < r_{F_1}$.

    \vspace{2mm}
    
    Now we consider the second part of the crossing, when both $y^-$ and $d$ are on $F_2$, as illustrated on Figure~\ref{fig:y^- and d crossing a vertex}.
    
    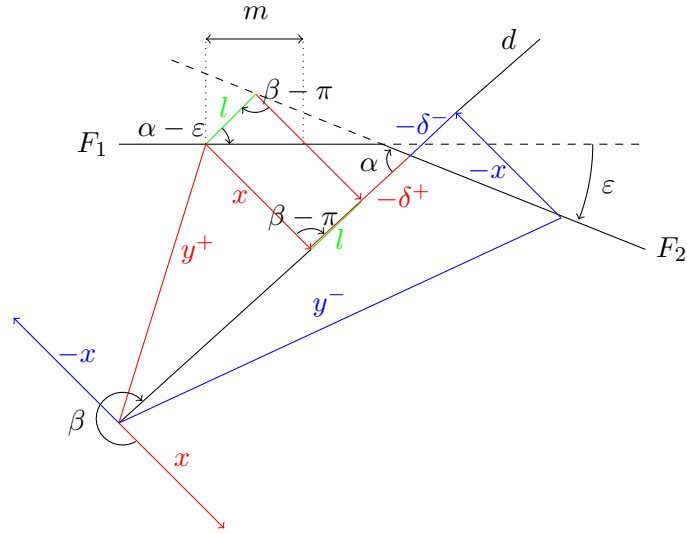
\begin{figure}[htbp!]
        \centering
        \begin{tikzpicture}[scale = 0.7]
            
            \draw (-5, 0) -- (0,0) -- (5, -2);
            \draw[dashed] (-4, 1.6) -- (0,0) -- (5,0);
            \node at (-5.5, 0) {$F_1$};
            \node at (5.5, -2) {$F_2$};
            
            \draw[->] (4,0) arc (0:-21:4);
            \node at (4.3, -0.8) {$\varepsilon$};
            
            \draw (-5, -5.3) -- (-1.35, -2);
            \draw (1.4, 0.6) -- (3, 2);
            \node at (2.4, 2) {$d$};
            \draw[blue] (-5, -5.3) -- (3.4, -1.4);
            \node at (-1, -3) {\textcolor{blue}{$y^-$}};
            \draw[red] (-5, -5.3) -- (-3.35, 0);
            \node at (-3.5, -2) {\textcolor{red}{$y^+$}};
            \draw[<-, blue] (-7, -3.3) -- (-5, -5.3);
            \node at (-5.8, -4) {\textcolor{blue}{$-x$}};
            \draw[->, red] (-5, -5.3) -- (-3, -7.3);
            \node at (-3.8, -6) {\textcolor{red}{$x$}};
            \draw[->] (-4.68, -5.65) arc (-60:-318:0.5);
            \node at (-5.8, -5.3) {$\beta$};
            
            \draw[->, blue] (3.4, -1.4) -- (1.4, 0.6);
            \node at (2, -0.5) {\textcolor{blue}{$-x$}};
            \draw[->, red] (-3.35, 0) -- (-1.35, -2);
            \node at (-2.7, -1) {\textcolor{red}{$x$}};
            
            \draw[->, red] (-2.4, 0.95) -- (-0.4, -1.05);
            % \node at (-1.4, -0.5) {\textcolor{green}{$x$}};
            \draw[green] (-3.35, 0) -- (-2.4, 0.95);
            \node at (-3, 0.7) {\textcolor{green}{$l$}};
            \draw[green] (-1.35, -2) -- (-0.4, -1.1);
            \node at (-0.8, -1.8) {\textcolor{green}{$l$}};
           
            \draw[<-] (-2.9, 0) arc (0:50:0.4);
            \node at (-4, 0.3) {$\alpha - \varepsilon$};
            \draw[->] (-2.15, 0.7) arc (-50:-130:0.4);
            \node at (-1.6, 1) {$\beta - \pi$};
            \draw[<-] (-1.1, -1.7) arc (50:130:0.4);
            \node at (-1.5, -1.4) {$\beta - \pi$};
            
            \draw[<-] (0.16, -0.1) arc (140:230:0.3);
            \node at (-0.25, -0.4) {$\alpha$};
            \draw[red] (0.5, -0.25) -- (-1.35, -1.95);
            \node at (0.4, -1) {\textcolor{red}{$-\delta^+$}};
            \draw[blue] (0.5, -0.25) -- (1.4, 0.6);
            \node at (0.75, 0.4) {\textcolor{blue}{$-\delta^-$}};
            
            \draw[dotted] (-3.35, 0) -- (-3.35, 2);
            \draw[dotted] (-1.5, 0) -- (-1.5, 2);
            \draw[<->] (-3.35, 2) -- (-1.5, 2);
            \node at (-2.4, 2.5) {$m$};
        \end{tikzpicture}
        \caption{Part II of the crossing of a vertex by $y^-$ leading and outside as $\beta$ increases.}
        \label{fig:y^- and d crossing a vertex}
    \end{figure}
    
    Since $y^+$ is inside and $y^-$ outside, we have $\delta^+ < 0$ and $\delta^- < 0$ according to \eqref{eq:delta}. Their length on Figure~\ref{fig:y^- and d crossing a vertex} is then given by $-\delta^+$ and $-\delta^-$ respectively. 
    Because $y^+$ does not yet intersects $F_2$, $-\delta^+$ is longer than $-\delta^-$. We also reuse $l := \delta^- - \delta^+$. Using the sine law, we can relate $l$ to $m$ 
    \begin{equation*}
        \frac{m}{\sin (\beta - \pi)} = \frac{l}{\sin\big(\pi - (\alpha - \varepsilon) - (\beta - \pi) \big)} = \frac{-m}{\sin \beta} = \frac{-l}{\sin(\alpha + \beta - \varepsilon)} \quad so \quad l = \frac{m \sin(\alpha + \beta - \varepsilon)}{\sin \beta}.
    \end{equation*}
    As previously $m > 0$. Since $y^-$ is leading and outside on $F_1$ and on $F_2$, we have $\alpha + \beta -\varepsilon \in (\alpha_0 + \pi, 2\pi)$ and $\alpha + \beta \in (\alpha_0 + \pi, 2\pi)$. Therefore, $\sin(\alpha + \beta - \varepsilon) < 0$ and $\sin \beta < 0$. Thus $l > 0$.
    Then, we can express the ratio $r(d)$:
    \begin{equation*}
        r(d) = \frac{D + \delta^+}{D - \delta^-} = \frac{D + \delta^+}{D - \delta^+ - l} > \frac{D + \delta^+}{D - \delta^+} = r_{F_2},
    \end{equation*}
    with $r_{F_2}$ the value of $r(d)$ on the face $F_2$. Therefore, $r(d)$ decreases during the crossing of a vertex when $y^-$ is leading and outside.
    $\quad \blacksquare$
\end{pf}

\vspace{3mm}

\begin{lemma}\label{lemma:crossing with leading inside}
    The ratio $r(d)$ increases when the leading vector is inside for a vertex crossing.
\end{lemma}
\begin{pf}
    We base this reasoning on the proof of Lemma~\ref{lemma: y^+ crossing} where $y^+$ was leading and outside, but it could be done similarly based on Lemma~\ref{lemma: y^- crossing}.
    We now assume that the angles are positive when oriented counterclockwise. With this change of orientation, Figure~\ref{fig:y^+ and d crossing a vertex} represents $y^-$ leading and inside after crossing a vertex from face $F_2$ to $F_1$, while $d$ and the trailing vector $y^+$ are still on $F_2$. The figure is the same, so \eqref{eq:gain crossing part 2} still holds. Since $y^-$ is leading and inside on $F_1$ and $F_2$, we have $\alpha + \beta - \varepsilon \in (2\pi, \alpha_0 + 2\pi)$ and $\alpha + \beta \in (2\pi, \alpha_0 + 2\pi)$. Then, $\sin(\alpha + \beta - \varepsilon) < 0$ and $\sin(\beta) < 0$ which leads to $r(d) > r_{F_2}$, i.e., $r(d)$ is increasing during the first part of that crossing.
    
    The second part of the crossing is illustrated by Figure~\ref{fig:y^+ crossing a vertex}. It represents $d$ and $y^-$ leading and inside having both crossed the vertex from $F_2$ to $F_1$ and $y^+$ is trailing and still on $F_2$. Similarly, \eqref{eq:loss crossing part 1} and \eqref{eq:r(d) r_F_1} still holds. The difference is again in the range of angles, $\alpha + \beta \in (2\pi, \alpha_0 + 2\pi)$. Then, $\sin(\alpha + \beta) < 0$ and $\sin(\beta) < 0$, which leads to $r(d) < r_{F_1}$. Therefore, $r(d)$ is also increasing during the second part of this crossing.
    
    The same method can be applied to the proof of Lemma~\ref{lemma: y^- crossing} to show that when $y^+$ is leading and inside, $r(d)$ increases at the vertices crossings.     $\quad \blacksquare$
\end{pf}

\vspace{3mm}

\begin{lemma}\label{lemma:v pi crossing}
    The ratio $r(d)$ decreases when crossing the vertices $v_\pi$ and $v_{2\pi}$.
\end{lemma}

\begin{pf}
    As can be seen on Figure~\ref{fig:global view}, before the vertices $v_\pi$ and $v_{2\pi}$ the leading vector is outside, but comes inside after crossing the vertex. Because of this feature Lemma~\ref{lemma: y^- crossing} does not apply to the crossing of the vertices $v_\pi$ and $v_{2\pi}$.
    
    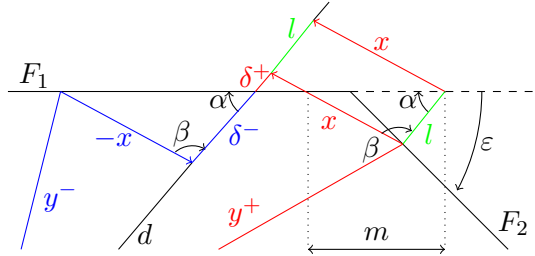
\begin{figure}[htbp!]
        \centering
        \begin{tikzpicture}[scale = 0.7]
            
            \draw (-5, 0) -- (1.5, 0) -- (4.5, -3);
            \draw[dashed] (1.5, 0) -- (5,0);
            \node at (-4.5, 0.3) {$F_1$};
            \node at (4.6, -2.5) {$F_2$};
            
            \draw[->] (4,0) arc (0:-28:4);
            \node at (4.1, -1) {$\varepsilon$};
            
            \draw (-2.9, -3) -- (-1.5, -1.35);
            \draw (0.8, 1.35) -- (1.1, 1.7);
            \node at (-2.4, -2.7) {$d$};
            \draw[red] (-1, -3) -- (2.5, -1);
            \node at (-0.5, -2.3) {\textcolor{red}{$y^+$}};
            \draw[blue] (-4.75, -3) -- (-4, 0);
            \node at (-4, -2) {\textcolor{blue}{$y^-$}};
            
            \draw[->, red] (2.5, -1) -- (0, 0.35);
            \node at (1.1, -0.6) {\textcolor{red}{$x$}};
            \draw[->, blue] (-4, 0) -- (-1.5, -1.35);
            \node at (-3, -0.9) {\textcolor{blue}{$-x$}};
            
            \draw[->, red] (3.3, 0) -- (0.8, 1.35);
            \node at (2.1, 0.9) {\textcolor{red}{$x$}};
            \draw[green] (2.5, -1) -- (3.3, 0);
            \node at (3, -0.8) {\textcolor{green}{$l$}};
            \draw[green] (0, 0.35) -- (0.8, 1.35);
            \node at (0.4, 1.2) {\textcolor{green}{$l$}};
           
           \draw[<-] (-1.25, -1.1) arc (60:130:0.5);
           \node at (-1.7, -0.8) {$\beta$};
           \draw[<-] (2.67, -0.75) arc (60:130:0.5);
           \node at (1.9, -1.1) {$\beta$};
            
            \draw[<-] (-0.8, 0) arc (180:230:0.5);
            \node at (-1, -0.2) {$\alpha$};
            \draw[blue] (-1.5, -1.35) -- (-0.3, 0);
            \node at (-0.5, -0.8) {\textcolor{blue}{$\delta^-$}};
            \draw[red] (-0.3, 0) -- (0, 0.35);
            \node at (-0.3, 0.3) {\textcolor{red}{$\delta^+$}};
            
            \draw[<-] (2.8, 0) arc (180:230:0.5);
            \node at (2.6, -0.2) {$\alpha$};
            
            \draw[dotted] (0.7, 0) -- (0.7, -3);
            \draw[dotted] (3.3, 0) -- (3.3, -3);
            \draw[<->] (0.7, -3) -- (3.3, -3);
            \node at (2, -2.7) {$m$};
            
        \end{tikzpicture}
        \caption{Part I of the crossing of $v_\pi$ as $\beta$ increases.}
        \label{fig:y^+ crossing v pi}
    \end{figure}
    
    The first part of the crossing of $v_\pi$ is illustrated on Figure~\ref{fig:y^+ crossing v pi}. Notice that the situation is very similar to the one described by Figure~\ref{fig:y^+ crossing a vertex}. Indeed, equations \eqref{eq:loss crossing part 1} and \eqref{eq:r(d) r_F_1} also hold for this case. Thus, $r(d)$ decreases when $y^+$ crosses $v_\pi$. 
    
    \vspace{2mm}
    
    The second part of the crossing of $v_\pi$ as illustrated on Figure~\ref{fig:y^+ and d crossing v pi} is also similar to Figure~\ref{fig:y^+ and d crossing a vertex}. The difference is that $y^+$ is inside and thus $\delta^+ < 0$. Since $y^-$ is inside, $\delta^- > 0$. If $y^-$ was already intersecting $F_2$ it would be of same length as $\delta^+$. We have as usual $l := \delta^- - \delta^+$.
    
    \begin{figure}[htbp!]
        \centering
        \begin{tikzpicture}[scale = 0.7]
            
            \draw (-4, 0.2) -- (0.9, 0.4) -- (4.8, -3);
            \draw[dashed] (-2, 2.7) -- (0.9, 0.4) -- (5, 0.5);
            \node at (-4.5, 0.3) {$F_1$};
            \node at (5, -2.7) {$F_2$};
            
            \draw[->] (3.5, 0.5) arc (10:-35:2.6);
            \node at (3.8, -0.4) {$\varepsilon$};
            
            \draw (-2.6, -4) -- (0.6, -1);
            \draw (1.55, -0.2) -- (4, 2);
            \node at (3.4, 2) {$d$};
            \draw[red] (-1, -4) -- (3.5, -1.85);
            \node at (1.8, -3) {\textcolor{red}{$y^+$}};
            \draw[blue] (-4, -4) -- (-1.9, 0.3);
            \node at (-3.6, -2) {\textcolor{blue}{$y^-$}};
            
            \draw[->, red] (3.5, -1.85) -- (1.2, -0.55);
            \node at (2.2, -1.5) {\textcolor{red}{$x$}};
            \draw[->, blue] (-1.9, 0.3) -- (0.6, -1);
            \node at (-1.1, -0.7) {\textcolor{blue}{$-x$}};
            \draw[->, blue] (-0.55, 1.5) -- (1.95, 0.2);
            
            \draw[green] (-1.9, 0.3) -- (-0.55, 1.5);
            \node at (-1.5, 1) {\textcolor{green}{$l$}};
            % \draw[red] (-1.35, -2) -- (-0.4, -1.1);
            % \node at (-0.8, -1.8) {\textcolor{red}{$g$}};
           
            \draw[<-] (-1.5, 0.3) arc (0:45:0.4);
            \node at (-2.6, 0.6) {$\alpha - \varepsilon$};
            \draw[->] (-0.2, 1.35) arc (-50:-130:0.4);
            \node at (-0.4, 0.8) {$\beta$};
            \draw[<-] (0.8, -0.85) arc (50:130:0.4);
            \node at (0.4, -0.5) {$\beta$};
            
            \draw[<-] (1.4, -0.03) arc (150:210:0.3);
            \node at (1.1, -0.2) {$\alpha$};
            \draw[blue] (1.55, -0.18) -- (0.6, -1);
            \node at (1.2, -1) {\textcolor{blue}{$\delta^-$}};
            \draw[red] (1.57, -0.24) -- (1.2, -0.55);
            \node at (2, -0.4) {\textcolor{red}{$-\delta^+$}};
            
            \draw[dotted] (-1.9, 0.3) -- (-1.9, 2);
            \draw[dotted] (1.5, 0.45) -- (1.5, 2.2);
            \draw[<->] (-1.9, 2) -- (1.5, 2.2);
            \node at (-0.1, 2.4) {$m$};
        \end{tikzpicture}
        \caption{Part II of the crossing of $v_\pi$ as $\beta$ increases.}
        \label{fig:y^+ and d crossing v pi}
    \end{figure}
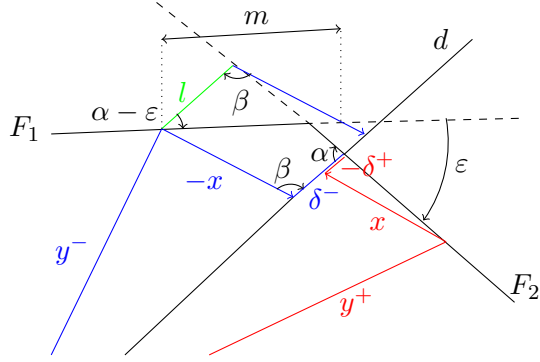
    
    We parametrize how far $y^-$ is from $v_\pi$ using the length $m$ defined on Figure~\ref{fig:y^+ and d crossing v pi}. The sine law gives
    \begin{equation*}
        \frac{m}{\sin \beta} = \frac{l}{\sin \big(\pi - (\alpha - \varepsilon) - \beta\big)} = \frac{l}{\sin (\alpha + \beta - \varepsilon)}.
    \end{equation*}
    Since $m > 0$, and $\alpha + \beta - \varepsilon < \pi$ and $\beta \in (0, \pi)$ we have that $l > 0$. We calculate the ratio $r(d)$:
    \begin{equation*}
        r(d) = \frac{D + \delta^+}{D - \delta^-} = \frac{D + \delta^+}{D - \delta^+ - l} > \frac{D + \delta^+}{D - \delta^+} = r_{F_2},
    \end{equation*}
    because $l > 0$. Therefore, the ratio $r(d)$ decreases when crossing $v_\pi$. The crossing of $v_{2\pi}$ is identical except that $y^-$ is leading instead of $y^+$. $\quad \blacksquare$
\end{pf}

\section{Proof of Theorem~\ref{thm:varying x_M and x_N}}\label{apx:proof with x_N and x_M}

In the following four lemmas we reuse the notation introduced in Theorem~\ref{thm:minimum of r(d) collinear with x} and Appendix~\ref{apx:proofs}.

\begin{lemma}\label{lemma: x_N^*(d) = - x_M^*(d) on faces}
    If $d$, $y_N^*$ and $y_M^*$ all intersect the same face of $\partial Y$, then $x_N^*(d)$ and $x_M^*(d)$ are constant and opposite: $x_N^*(d) = -x_M^*(d) \in \partial X$ and $r_{X,Y}(d)$ is constant.
\end{lemma}
\begin{pf}
    We reuse the length $D$ and the angles $\alpha$, $\beta$ as illustrated on Figure~\ref{fig: x_N^*(d) = -x_M^*(d)}.
    Similarly to \eqref{eq:delta}, we also introduce $\delta_M(d) := D(d) - \|x_M^*(d) + y_M^*(d)\|$ and $\delta_N(d) := \|x_N^*(d) + y_N^*(d)\| - D(d)$.
    
    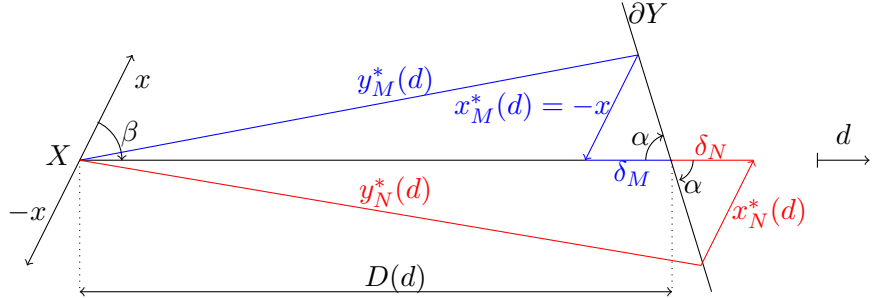
\begin{figure}[htbp!]
        \centering
        \begin{tikzpicture}[scale = 0.7]
            
            \draw[|->] (12,0) -- (13,0);
            \node at (12.5, 0.5) {$d$};
            \draw (-2, 0) -- (7.6, 0);
            \draw[dotted] (-2, 0) -- (-2, -2.5);
            \draw[dotted] (9.25, 0) -- (9.25, -2.5);
            \draw[<->] (-2, -2.5) -- (9.25, -2.5);
            \node at (4, -2.25) {$D(d)$};
            \draw[red] (9.25, 0) -- (10.8, 0);
            \node at (10, 0.25) {\textcolor{red}{$\delta_N$}};
            \draw[blue] (7.6, 0) -- (9.25, 0);
            \node at (8.5, -0.25) {\textcolor{blue}{$\delta_M$}};
            
             \draw (8.3, 3) -- (10, -2.5);
            \node at (8.8, 2.8) {$\partial Y$};
            
            \node at (-2.4, 0.1) {$X$};
            \draw[->] (-2, 0) -- (-1, 2);
            \node at (-0.8, 1.5) {\textcolor{black}{$x$}};
            \draw[<-] (-3, -2) -- (-2, 0);
            \node at (-3, -1) {\textcolor{black}{$-x$}};
            
            \draw[blue] (-2,0) -- (8.6, 2);
            \node at (4, 1.5) {\textcolor{blue}{$y_M^*(d)$}};
            \draw[->, blue] (8.6, 2) -- (7.6, 0);
            \node at (6.6, 1) {\textcolor{blue}{$x_M^*(d) = -x$}};
            
            \draw[red] (-2,0) -- (9.8,-2);
            \node at (4, -0.6) {\textcolor{red}{$y_N^*(d)$}};
            \draw[->, red] (9.8, -2) -- (10.8, 0);
            \node at (11.1, -1) {\textcolor{red}{$x_N^*(d)$}};
            
            \draw[<-] (-1.2,0) arc (0:65:0.8);
            \node at (-1.05, 0.5) {$\beta$};
            \draw[->] (8.75, 0) arc (180:110:0.5);
            \node at (8.65, 0.4) {$\alpha$};
            
            \draw[->] (9.65, 0) arc (0:-70:0.4);
            \node at (9.65, -0.5) {$\alpha$};
            
        \end{tikzpicture}
        \caption{Illustration of $x_N^*(d)$ and $x_M^*(d)$, when $d$ intersects the interior of a face of $\partial Y$.}
        \label{fig: x_N^*(d) = -x_M^*(d)}
    \end{figure}
    
    We know from Theorem~\ref{thm:minimum on the vertices} that $x_M^*(d) \in \big\{x, -x\big\}$ for all $d \in \mathbb{S}$. In the case illustrated on Figure~\ref{fig: x_N^*(d) = -x_M^*(d)}, $x_M^*(d) = -x$ because it maximizes $\delta_M$, while in general we only know that $\|x_M^*(d)\| = \|x\|$. 
    
    If $\alpha + \beta \in \{\pi, 2\pi\}$, then $X$ is parallel with a face of $\partial Y$ making $x_N^*$ and $x_M^*$ not uniquely defined. Regardless, we can still take $x_N^*(d) = - x_M^*(d) \in \partial X$.
    Otherwise, $x_N^*$ and $x_M^*$ are uniquely defined. Since $x_N^*(d) \in X$, $x_M^*(d) \in X$ for all $d \in \mathbb{S}$ and $\dim X = 1$, vectors $x_N^*(d)$ and $x_M^*(d)$ are always collinear. We then use Thales's theorem and obtain $\delta_N(d) = \delta_M(d) \frac{\|x_N^*(d)\|}{\|x_M^*(d)\|} = \delta_M(d) \frac{\|x_N^*(d)\|}{\|x\|}$. Since $x_N^*(d)$ is chosen to maximize $\delta_N$ and is independent from $\delta_M$ it must have the greatest norm, so $x_N^*(d) \in \partial X$. 
    Due to $\alpha + \beta \notin \{\pi, 2\pi\}$, $\|x+y\|$ is not constant.
    Because $x_N^*(d)$ is chosen to maximize $\|x+y\|$ while $x_M^*(d)$ is minimizing it, we have $x_N^*(d) \neq x_M^*(d)$. Since they both belong in $\partial X = \big\{-x, x\big\}$, we have $x_N^*(d) = - x_M^*(d)$.

    According to Lemma~\ref{lemma: continuity of x_N}, the coupled functions $\big( x_N^*, y_N^* \big) (d) = \arg \underset{x\, \in\, X,\, y\, \in\, Y}{\max} \big\{ \|x+y\| : x+y \in \mathbb{R}^+ d \big\}$ are continuous. Since $x_N^*(d) \in \big\{x, -x\big\}$, $x_N^*(d)$ is constant on the faces of $\partial Y$ and so is $x_M^*(d)$. Then, $r_{X,Y}(d) = r(d)$ is constant on the faces of $\partial Y$ according to Lemma~\ref{lemma:r(d) cst on faces}.    $\quad \blacksquare$
\end{pf}

 \vspace{3mm}

\begin{lemma}\label{lemma: x_N^*(d) = - x_M^*(d) before v_pi}
    During the crossing vertices before $v_\pi$ as $\beta$ increases, $x_N^*(d)$ and $x_M^*(d)$ are constant and opposite: $x_N^*(d) = -x_M^*(d) \in \partial X$.
\end{lemma}
\begin{pf}
    We study the crossing of a vertex $v$ of angle $\varepsilon$ between the faces $F_1$ and $F_2$ of $\partial Y$.
    In Theorem~\ref{thm:minimum of r(d) collinear with x} $x$ was fixed, while here the vectors $x_M^*(d)$ and $x_N^*(d)$ depend on $d$, so we need a new definition for vertex crossings. For each vertex $v$ we introduce $x_v$ the vector collinear with $x$, going from $v$ to the ray directed by $d$, as illustrated on Figure~\ref{fig: x_v} and we say that the crossing of $v$ is ongoing as long as $\|x_v\| < \|x\|$. We also define $\delta_v := \|v + x_v\| - D$.
    
    \begin{figure}[htbp!]
        \centering
        \begin{tikzpicture}[scale = 0.7]
            
            \draw (-6, 0) -- (0, 0) -- (5, -2.1);
            \draw[dashed] (0.1, 0) -- (5,0);
            \node at (-5.5, 0.3) {$F_1$};
            \node at (5, -1.8) {$F_2$};
            \node at (0, -0.3) {$v$};
            
            \draw[->] (4,0) arc (0:-22:4);
            \node at (4.3, -1) {$\varepsilon$};
            
            \draw (-4.5, -3) -- (-3.4, -2);
            \draw (-0.6, 0.6) -- (1.3, 2.4);
            \node at (1.5, 2.2) {$d$};
            
            \draw[red] (-2, -3) -- (2, -0.8);
            \node at (0.5, -2.2) {\textcolor{red}{$y_N^*$}};
            \draw[->, red] (2, -0.8) -- (0, 1.2);
            \node at (1.4, 0.5) {\textcolor{red}{$x_N^*$}};
            \draw[red, dotted] (-1.3, 0) -- (-2.1, 0.8);
            \draw[red, dotted] (0, 1.2) -- (-0.8, 2);
            \draw[red, <->] (-2.1, 0.8) -- (-0.8, 2);
            \node at (-1.7, 1.7) {\textcolor{red}{$\delta_N$}};
            
            \draw[blue] (-6.5, -2) -- (-5.4, 0);
            \node at (-5.6, -1.3) {\textcolor{blue}{$y_M^*$}};
            \draw[->, blue] (-5.4, 0) -- (-3.4, -2);
            \node at (-4.1, -0.7) {\textcolor{blue}{$x_M^*$}};
            \draw[blue] (-3.4, -2) -- (-1.3, 0);
            \node at (-1.9, -1.2) {\textcolor{blue}{$\delta_M$}};
           
           \draw[->, green] (0, 0) -- (-0.6, 0.6);
           \node at (0, 0.4) {\textcolor{green}{$x_v$}};
           \draw[green] (-1.3, 0) -- (-0.6, 0.6);
           \node at (-1.1, 0.6) {\textcolor{green}{$\delta_v$}};
           
           \draw[<-] (-3.2, -1.8) arc (60:114:0.5);
           \node at (-3.4, -1.4) {$\beta$};
            
            \draw[<-] (-1.9, 0) arc (180:230:0.5);
            \node at (-2.1, -0.3) {$\alpha$};
            
        \end{tikzpicture}
        \caption{Illustration of $x_v$ during the crossing of a vertex $v$, with $y_N^*$ leading.}
        \label{fig: x_v}
    \end{figure}
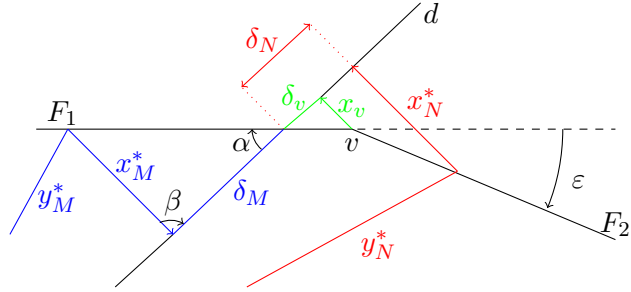

    Before starting the crossing of $v_\pi$ we have $\alpha + \beta \in (\alpha_0, \pi)$. Then, as can be seen on Figure~\ref{fig: x_N^*(d) = -x_M^*(d)}, $y_N^*$ is leading and outside, so it reaches the vertex before $y_M^*$ and $d$. The length of $x_N^*(d)$ can vary to maximize $\delta_N$, so $y_N^*$ could still intersect $F_1$, even if the crossing is ongoing. We have seen in Lemma~\ref{lemma: x_N^*(d) = - x_M^*(d) on faces} that if $y_N^*$ is still on $F_1$, then it must be the furthest possible to maximize $\delta_N$, in that case $y_N^* = v$. Otherwise, $y_N^*$ intersects $F_2$. We want to establish a criterion to distinguish these two possible scenarios.
    
    We first consider the scenario where $y_N^* = v$ and $x_N^*(d) = x_v$. We take $y \in F_2 \backslash \{v\}$ such that $y + x \in \mathbb{R}^+d$ as represented on Figure~\ref{fig: y_N in v} and we define $\delta := \|x+y\| - D$.
    
    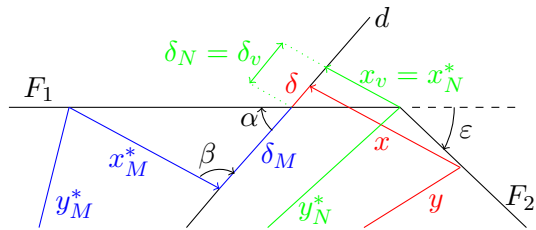
\begin{figure}[htbp!]
        \centering
        \begin{tikzpicture}[scale = 0.8]
            
            \draw (-5, 0) -- (1.5, 0) -- (3.6, -2);
            \draw[dashed] (1.7, 0) -- (3.5,0);
            \node at (-4.5, 0.3) {$F_1$};
            \node at (3.5, -1.5) {$F_2$};
            
            \draw[->] (2.4,0) arc (0:-28:1.5);
            \node at (2.6, -0.4) {$\varepsilon$};
            
            \draw (-2.05, -2) -- (-1.5, -1.35);
            \draw (0, 0.35) -- (1, 1.5);  
            \node at (1.2, 1.5) {$d$};
            
            \draw[red] (0.9, -2) -- (2.5, -1);
            \node at (2.1, -1.6) {\textcolor{red}{$y$}};
            \draw[->, red] (2.5, -1) -- (0, 0.35);
            \node at (1.2, -0.6) {\textcolor{red}{$x$}};
            \draw[red] (-0.3, 0) -- (0, 0.35);
            \node at (-0.3, 0.4) {\textcolor{red}{$\delta$}};
            
            \draw[green] (-0.7, -2) -- (1.5, 0);
            \node at (0.1, -1.7) {\textcolor{green}{$y_N^*$}};
            \draw[green, ->] (1.5, 0) -- (0.25, 0.675);
            \node at (1.7, 0.5) {\textcolor{green}{$x_v = x_N^*$}};
            \draw[green, dotted] (-0.3, 0) -- (-1, 0.4);
            \draw[green, dotted] (0.25, 0.675) -- (-0.45, 1.075);
            \draw[green, <->] (-1, 0.4) -- (-0.45, 1.075);
            \node at (-1.6, 0.9) {\textcolor{green}{$\delta_N = \delta_v$}};
            
            \draw[blue] (-4.5, -2) -- (-4, 0);
            \node at (-3.9, -1.6) {\textcolor{blue}{$y_M^*$}};
            \draw[->, blue] (-4, 0) -- (-1.5, -1.35);
            \node at (-3, -0.9) {\textcolor{blue}{$x_M^*$}};
            \draw[blue] (-1.5, -1.35) -- (-0.3, 0);
            \node at (-0.5, -0.8) {\textcolor{blue}{$\delta_M$}};
           
           \draw[<-] (-1.25, -1.1) arc (60:130:0.5);
           \node at (-1.7, -0.8) {$\beta$};
            
            \draw[<-] (-0.8, 0) arc (180:230:0.5);
            \node at (-1, -0.2) {$\alpha$};
            
        \end{tikzpicture}
        \caption{Illustration of the crossing scenario where $y_N^* = v$.}
        \label{fig: y_N in v}
    \end{figure}
    
    Since $\delta_N$ must be maximized by the choice of $y_N^*$ and $y \neq y_N^*$, we have $\delta < \delta_N = \delta_v$. But $\|x\| > \|x_v\|$, so the line segment corresponding to $x$ crosses the interior of $Y$. Focusing on this part of Figure~\ref{fig: y_N in v} we obtain Figure~\ref{fig: zoom x inside}.
    
    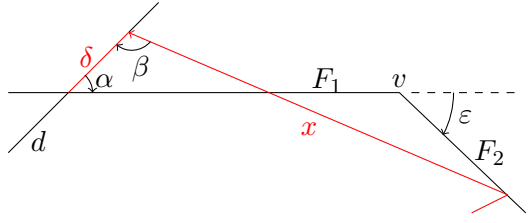
\begin{figure}[htbp!]
        \centering
        \begin{tikzpicture}[scale = 0.8]
            
            \draw (-5, 0) -- (1.5, 0) -- (3.6, -2);
            \draw[dashed] (1.7, 0) -- (3.5,0);
            \node at (0.3, 0.2) {$F_1$};
            \node at (3, -1) {$F_2$};
            \node at (1.5, 0.2) {$v$};
            
            \draw[->] (2.4,0) arc (0:-28:1.5);
            \node at (2.6, -0.4) {$\varepsilon$};
            
            \draw (-5, -1) -- (-4, 0);
            \draw (-3, 1) -- (-2.5, 1.5);
            \node at (-4.5, -0.8) {$d$};
            \draw[red] (-4, 0) -- (-3, 1);
            \node at (-3.7, 0.6) {\textcolor{red}{$\delta$}};
            
            \draw[red, ->] (3.3, -1.7) -- (-3, 1);
            \node at (0, -0.6) {\textcolor{red}{$x$}};
            \draw[red] (2.7, -2) -- (3.3, -1.7);
            
            \draw[<-] (-3.2, 0.8) arc (230:320:0.4);
            \node at (-2.8, 0.4) {$\beta$};
            
            \draw[<-] (-3.6, 0) arc (0:47:0.4);
            \node at (-3.4, 0.2) {$\alpha$};

        \end{tikzpicture}
        \caption{Illustration of the line segment corresponding to $x$ crossing the interior of $Y$ in Figure~\ref{fig: y_N in v}.}
        \label{fig: zoom x inside}
    \end{figure}
    
    Two of the angles of the triangle delimited by $F_1$, $F_2$ and $x$ are $\pi - \alpha - \beta$ and $\pi - \varepsilon$. Therefore, their sum is in $(0, \pi)$ and thus $\alpha + \beta + \varepsilon > \pi$. Since we assumed that $\alpha + \beta \in (\alpha_0, \pi)$, the vertex $v$ must in fact be $v_\pi$ for this scenario to happen.

    Thus, the crossing of a vertex preceding $v_\pi$ follows the second scenario as depicted on Figure~\ref{fig: x_v} with $y_N^* \in F_2$. We study Figure~\ref{fig: zoom x_v x_N^*} which is a more detailed view of Figure~\ref{fig: x_v}, with $\delta_0$ depending solely on $d$ and $\varepsilon$. 
    
    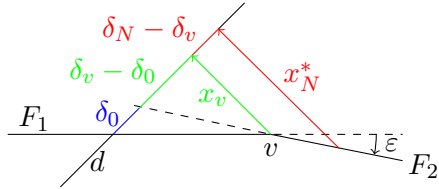
\begin{figure}[htbp!]
        \centering
        \begin{tikzpicture}[scale = 0.7]
            
            \draw (-5, 0) -- (0, 0) -- (2.5, -0.5);
            \draw[dashed] (-2.6, 0.55) -- (0, 0) -- (2.5,0);
            \node at (-4.5, 0.3) {$F_1$};
            \node at (2.9, -0.6) {$F_2$};
            \node at (0, -0.3) {$v$};
            
            \draw[->] (2,0) arc (0:-11:2);
            \node at (2.3, -0.2) {$\varepsilon$};
            
            \draw (-4, -1) -- (-3, 0);
            \node at (-3.3, -0.5) {$d$};
            \draw[blue] (-3, 0) -- (-2.5, 0.5);
            \node at (-3.1, 0.4) {\textcolor{blue}{$\delta_0$}};
            \draw[green] (-2.5, 0.5) -- (-1.5, 1.5);
            \node at (-3, 1.2) {\textcolor{green}{$\delta_v - \delta_0$}};
            \draw[red] (-1.5, 1.5) -- (-1, 2);
            \node at (-2.3, 2) {\textcolor{red}{$\delta_N - \delta_v$}};
            \draw (-1, 2) -- (-0.5, 2.5);
            
            \draw[green, ->] (0,0) -- (-1.5, 1.5);
            \node at (-1.1, 0.7) {\textcolor{green}{$x_v$}};
            \draw[red, ->] (1.3, -0.27) -- (-1, 2);
            \node at (0.6, 1.1) {\textcolor{red}{$x_N^*$}};
            
        \end{tikzpicture}
        \caption{Illustration of $x_v$ and $x_N^*$ in Figure~\ref{fig: x_v}.}
        \label{fig: zoom x_v x_N^*}
    \end{figure}
    
    Since $x_v$ and $x_N^*(d)$ are collinear, we can apply Thales's theorem in Figure~\ref{fig: zoom x_v x_N^*} and obtain that $\delta_N - \delta_0 = (\delta_v - \delta_0) \frac{\|x_N^*(d)\|}{\|x_v(d)\|}$. Then, $\delta_N$ is maximized when $\|x_N^*(d)\|$ is maximal, so $x_N^*(d) \in \partial X$.
    Since $x_N^*$ and $x_M^*$ play opposite roles and are both in $\partial X = \big\{x, -x\big\}$, then $x_N^*(d) = -x_M^*(d)$. As in Lemma~\ref{lemma: x_N^*(d) = - x_M^*(d) on faces}, these vectors are constant since $x_N^*$ is continuous in $d$.    $\quad \blacksquare$
\end{pf}

 \vspace{5mm}

\begin{lemma}\label{lemma: x_N^*(d) crossing v_pi}
    During the crossing of $v_\pi$ and $v_{2\pi}$, the ratio $r_{X,Y}(d)$ reaches a local minimum.
\end{lemma}
\begin{pf}
    During the crossing of $v_\pi$, i.e., when $\| x_{v_\pi} \| < \|x\|$, we have $\alpha + \beta \leq \pi$ but $\alpha + \beta + \varepsilon > \pi$. The situation is illustrated on Figure~\ref{fig: y_N crossing v_pi}. We showed in Lemma~\ref{lemma: x_N^*(d) = - x_M^*(d) before v_pi} that $y_N^* = v_\pi$ and $x_N^*(d) = x_{v_\pi}$.
    
    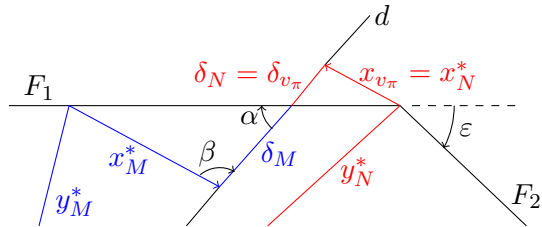
\begin{figure}[htbp!]
        \centering
        \begin{tikzpicture}[scale = 0.8]
            
            \draw (-5, 0) -- (1.5, 0) -- (3.6, -2);
            \draw[dashed] (1.7, 0) -- (3.5,0);
            \node at (-4.5, 0.3) {$F_1$};
            \node at (3.6, -1.5) {$F_2$};
            
            \draw[->] (2.4,0) arc (0:-28:1.5);
            \node at (2.6, -0.4) {$\varepsilon$};
            
            \draw (-2.05, -2) -- (-1.5, -1.35);
            \draw (0.25, 0.675) -- (1, 1.5);
            \node at (1.2, 1.5) {$d$};
            
            \draw[red] (-0.7, -2) -- (1.5, 0);
            \node at (0.8, -1.1) {\textcolor{red}{$y_N^*$}};
            \draw[red, ->] (1.5, 0) -- (0.25, 0.675);
            \node at (1.8, 0.5) {\textcolor{red}{$x_{v_\pi} = x_N^*$}};
            \draw[red] (-0.3, 0) -- (0.25, 0.675);
            \node at (-1, 0.5) {\textcolor{red}{$\delta_N = \delta_{v_\pi}$}};
            
            \draw[blue] (-4.5, -2) -- (-4, 0);
            \node at (-3.9, -1.6) {\textcolor{blue}{$y_M^*$}};
            \draw[->, blue] (-4, 0) -- (-1.5, -1.35);
            \node at (-3, -0.9) {\textcolor{blue}{$x_M^*$}};
            \draw[blue] (-1.5, -1.35) -- (-0.3, 0);
            \node at (-0.5, -0.8) {\textcolor{blue}{$\delta_M$}};
           
           \draw[<-] (-1.25, -1.1) arc (60:130:0.5);
           \node at (-1.7, -0.8) {$\beta$};
            
            \draw[<-] (-0.8, 0) arc (180:230:0.5);
            \node at (-1, -0.2) {$\alpha$};
            
        \end{tikzpicture}
        \caption{Crossing of $v_\pi$, with $y_N^* = v_\pi$.}
        \label{fig: y_N crossing v_pi}
    \end{figure}
    
    Once $d$ has crossed $v_\pi$, we still have $y_N^* = v_\pi$ to maximize $\delta_N$. Then, the equality $x_N^*(d) = x_{v_\pi}$ holds during the entire crossing of $v_\pi$, i.e., until $x_N^*(d) = -x$. This second part of the crossing is illustrated on Figure~\ref{fig: d passed v_pi}.
    
    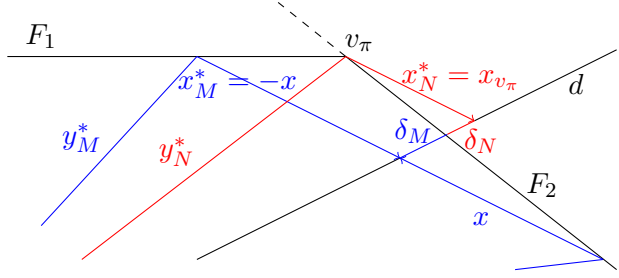
\begin{figure}[htbp!]
        \centering
        \begin{tikzpicture}[scale = 0.9]
            
            \draw (-5, 0) -- (0, 0) -- (4, -3.15);
            \node at (0.2, 0.2) {$v_\pi$};
            \draw[dashed] (0, 0) -- (-1, 0.8);
            \node at (-4.5, 0.3) {$F_1$};
            \node at (2.9, -1.9) {$F_2$};
            
            \draw (-2.2, -3) -- (0.8, -1.5);
            \draw (1.9, -0.95) -- (4, 0.1);
            \node at (3.4, -0.4) {$d$};
            
            \draw[blue, ->] (-2.2, 0) -- (0.8, -1.5);
            \node at (-1.6, -0.4) {\textcolor{blue}{$x_M^* = -x$}};
            \draw[blue] (-4.5, -2.5) -- (-2.2, 0);
            \node at (-3.9, -1.2) {\textcolor{blue}{$y_M^*$}};
            \draw[blue] (0.8, -1.5) -- (1.5, -1.15);
            \node at (1, -1.1) {\textcolor{blue}{$\delta_M$}};
            
            \draw[blue, ->] (3.8, -3) -- (0.8, -1.5);
            \node at (2, -2.4) {\textcolor{blue}{$x$}};
            \draw[blue] (2.5, -3.15) -- (3.8, -3);
            
            \draw[red, ->] (0, 0) -- (1.9, -0.95);
            \node at (1.7, -0.3) {\textcolor{red}{$x_N^* = x_{v_\pi}$}};
            \draw[red] (-3.9, -3) -- (0, 0);
            \node at (-2.5, -1.4) {\textcolor{red}{$y_N^*$}};
            \draw[red] (1.5, -1.15) -- (1.9, -0.95);
            \node at (2, -1.25) {\textcolor{red}{$\delta_N$}};
            
        \end{tikzpicture}
        \caption{Illustration of the endpoint of $y_M^*$ switching from $F_1$ to $F_2$.}
        \label{fig: d passed v_pi}
    \end{figure}
    
    Assume that during the entire crossing of $v_\pi$, $x_M^*(d) = -x$. Then, at the end of the crossing we will have $y_M^* = v_\pi$ and $x_M^*(d) = x_{v_\pi} = x_N^*(d)$, which contradicts the definitions of $x_M^*(d)$ and $x_N^*(d)$. Thus, $x_M^*(d)$ does not remain equal to $-x$ during the entire crossing. Since $x_M^* \in \big\{x, -x\big\}$, at some point $x_M^*$ switches to $x$ as $y_M^*$ switches from $F_1$ to $F_2$. This switching point is illustrated on Figure~\ref{fig: d passed v_pi}, and $y_M^*$ becomes the leading vector.
    
    By definition, $\|x_{v_\pi}\| < \|x\|$ during the crossing. We have showed that $x_N^*(d) = x_{v_\pi}$, so $\delta_N = \delta_{v_\pi}$. Now using Thales's theorem in Figure~\ref{fig: y_N crossing v_pi}, we have $\delta_{v_\pi} = \delta_M \frac{\|x_{v_\pi}\|}{\|x_M^*\|}$. Since $\|x_M^*\| = \|x\|$ we conclude that $\delta_{v_\pi} < \delta_M$ during the crossing.
    
    Also, $y_N^* \in v_\pi$ during the entire crossing, so $r_{X,Y}(d) = \frac{D + \delta_{v_\pi}}{D - \delta_M}$. During the first part of the crossing, $y_M^*$ intersects $F_1$, as represented on Figure~\ref{fig: y_N crossing v_pi}. If $y_N^*$ was further on the dashed line of Figure~\ref{fig: y_N crossing v_pi}, the ratio would be $r_{F_1} = \frac{D + \delta_M}{D - \delta_M}$, which is the value of $r_{X,Y}$ on $F_1$. However, since $\delta_{v_\pi} < \delta_M$, we have $r_{X,Y}(d) < r_{F_1}$.
    
    During the second part of the crossing, $y_M^* \in F_2$ and $x_M^*(d) = x$. If $y_N^*$ was on the dashed prolongation of $F_2$ in Figure~\ref{fig: d passed v_pi}, we would have $r_{F_2} = \frac{D + \delta_M}{D - \delta_M}$, value of $r_{X,Y}$ on $F_2$. However, since $\delta_{v_\pi} < \delta_M$, we have $r_{X,Y}(d) < r_{F_2}$. Thus, $r_{X,Y}$ reaches a local minimum during the crossing of $v_\pi$. 
    
    During the crossing of $v_{2\pi}$ we also have $y_N^*(d) = v_{2\pi}$ and $r_{X,Y}$ reaching a local minimum because functions $y_N^*$, $y_M^*$, $x_N^*$ and $x_M^*$ are odd.    $\quad \blacksquare$
\end{pf}

\vspace{3mm}

\begin{lemma}\label{lemma: x_N^*(d) = - x_M^*(d) on regular vertices}
   During the crossing of a vertex other than $v_\pi$ and $v_{2\pi}$, $x_N^*(d)$ and $x_M^*(d)$ are constant and opposite: $x_N^*(d) = -x_M^*(d) \in \partial X$.
\end{lemma}
\begin{pf}
    After the crossing of $v_\pi$, $\alpha + \beta \in (\pi, \alpha_0 + \pi)$ and $y_M^*$ is leading and inside as established in Lemma~\ref{lemma: x_N^*(d) crossing v_pi}. Thus, $y_M^*$ is the first to reach vertex $v$, but since $\|x_M^*\| = \|x\|$ we cannot have $y_M^* = v$ during the entire crossing. 
    In Lemma~\ref{lemma: x_N^*(d) = - x_M^*(d) on faces} we showed that $x_N^*$ is continuous in $d$. Thus, $x_N^*(d)$ cannot switch like $x_M^*(d)$ to take the lead. Instead, $x_N^*(d)$ is trailing during the crossing as illustrated on Figure~\ref{fig: crossing after v_pi}.
    
    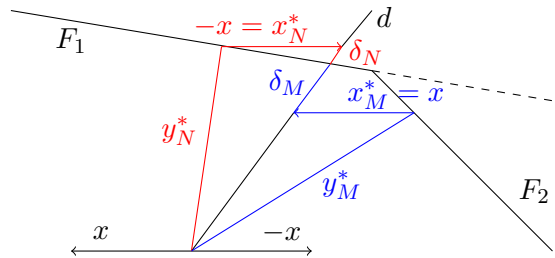
\begin{figure}[htbp!]
        \centering
        \begin{tikzpicture}[scale = 0.8]
            
            \draw[->] (0, 0) -- (-2, 0);
            \node at (-1.5, 0.3) {$x$};
            \draw[->] (0, 0) -- (2, 0);
            \node at (1.5, 0.3) {$-x$};
            
            \draw (-3, 4) -- (3, 3) -- (6, 0);
            \node at (-2, 3.5) {$F_1$};
            \node at (5.7, 1) {$F_2$};
            % \node at (3, 3.2) {$v$};
            \draw[dashed] (3, 3) -- (6, 2.5);
            
            \draw (0, 0) -- (1.7, 2.3);
            \draw (2.5, 3.4) -- (3, 4);
            \node at (3.2, 3.9) {$d$};
            
            \draw[red] (0, 0) -- (0.5, 3.4);
            \node at (-0.2, 2) {\textcolor{red}{$y_N^*$}};
            \draw[red, ->] (0.5, 3.4) -- (2.5, 3.4);
            \node at (1, 3.7) {\textcolor{red}{$-x = x_N^*$}};
            \draw[red] (2.5, 3.4) -- (2.3, 3.1);
            \node at (2.9, 3.3) {\textcolor{red}{$\delta_N$}};
            
            \draw[blue] (0, 0) -- (3.7, 2.3);
            \node at (2.5, 1.1) {\textcolor{blue}{$y_M^*$}};
            \draw[blue, ->] (3.7, 2.3) -- (1.7, 2.3);
            \node at (3.4, 2.6) {\textcolor{blue}{$x_M^* = x$}};
            \draw[blue] (1.7, 2.3) -- (2.3, 3.1);
            \node at (1.6, 2.8) {\textcolor{blue}{$\delta_M$}};
            
        \end{tikzpicture}
        \caption{Crossing of a vertex $v$ after $v_\pi$.}
        \label{fig: crossing after v_pi}
    \end{figure}
    
    Since $y_N^* \in F_1$ during the crossing, we can apply Thales's theorem on Figure~\ref{fig: crossing after v_pi} and obtain that for a fixed $d$, $\delta_N$ is proportional to $\|x_N^*(d)\|$. Thus, $x_N^*(d) \in \partial X$ and, since $y_N^*$ is trailing, we have $x_N^*(d) = -x$ during the entire crossing. By the definitions of $x_N^*(d)$ and $x_M^*(d)$, we have $x_N^*(d) \neq x_M^*(d)$. Also, both $x_N^*(d)$ and $x_M^*(d)$ belong to $\partial X = \big\{x, -x\big\}$, then $x_M^*(d) = x$ during the entire crossing. Following Lemma~\ref{lemma: x_N^*(d) = - x_M^*(d) before v_pi} we have showed that $x_N^*(d)$ and $x_M^*(d)$ are constant and opposite for the crossing of all vertices encountered when $\beta \in (0, \pi)$, except for $v_\pi$. 
    
    Since functions $x_N^*$ and $x_M^*$ are odd, $x_N^*(d)$ and $x_M^*(d)$ are also constant and opposite for all the vertices encountered when $\beta \in (\pi, 2\pi)$ except for $v_{2\pi}$.    $\quad \blacksquare$
\end{pf}

 \vspace{5mm}

\section{Continuity of Extrema}\label{apx:continuity}

\begin{lemma}\label{lemma: T continuous}
    For a resilient system following \eqref{eq:splitted ODE}, $T_M(w_c, d) := \underset{u_c\, \in\, U_c}{\min} \big\{ T \geq 0 : (Bu_c + Cw_c)T = d\big\}$ is continuous in $w_c \in W_c$ and $d \in \mathbb{R}_*^n$.
\end{lemma}
\begin{pf}
    We define set $Y := \big\{ Bu_c : u_c \in U_c \big\}$. Then, $T_M(w_c, d) = \underset{y\, \in\, Y}{\min} \big\{ T \geq 0 : (y + Cw_c)T = d\big\}$. 
    We define a set-valued function $\varphi$ for $w_c \in W_c$ and $d \in \mathbb{R}_*^n$
    \begin{align}\label{eq:constraint set}
        \varphi(w_c, d) &:= \big\{y \in Y : \exists\ T > 0 : (y + Cw_c)T = d\big\} = \big\{y \in Y : y + Cw_c \in \mathbb{R}^+ d\big\} \nonumber \\
        &= Y \cap \big(  \mathbb{R}^+ d - \{Cw_c\} \big),
    \end{align}
    where $\mathbb{R}^+ d - \{Cw_c\} = \big\{ \lambda d - Cw_c : \lambda \geq 0 \big\}$. We call graph of $\varphi$ the set
    \begin{equation*}\label{eq:graph phi}
        \text{Gr}\, \varphi := \big\{ (w_c, d, y) \in W_c \times \mathbb{R}_*^n \times Y : y \in \varphi(w_c, d) \big\}. 
    \end{equation*}
    We can now define the function $f : \text{Gr}\, \varphi \rightarrow \mathbb{R}_*^+$ as
    \begin{equation*}
        f(w_c, d, y) := T \qquad \text{such that}\ \big(y + Cw_c)T = d. 
    \end{equation*}
    Since $d \neq 0$, for $(w_c, d ,y) \in \text{Gr}\, \varphi$ we have $y + Cw_c \neq 0$. For all the non-zero components of $y + Cw_c$ indexed by $i \in [n]$ we have $f(w_c, d, y) = \frac{d_i}{y_i + C_iw_c}$, with $C_i$ the row $i$ of $C$. Therefore, $f$ is continuous in the components of $w_c$, $d$ and $y$, and $T_M(w_c, d) = \underset{y\, \in\, \varphi(w_c, d)}{\min}\ f(w_c, d, y) $. 
    
    The resilience of the system implies that for all $w_c \in W_c$ and all $d \in \mathbb{R}^n$, the set $\varphi(w_c, d)$ is nonempty. 
    Since $Y$ is compact and $\mathbb{R}^+d - \{Cw_c\}$ is closed, their intersection $\varphi(w_c, d)$ is compact for all $w_c \in W_c$ and all $d \in \mathbb{R}^n$. 
    Additionally, based on Lemma~\ref{lemma:phi continuous}, $\varphi$ satisfies the continuity definition 17.2 of \citep{inf_dim_analysis}.
    Thus, the conditions of the Berge Maximum Theorem~\citep{inf_dim_analysis} are satisfied, leading to the continuity of $T_M$ in $w_c$ and $d$.   $\quad \blacksquare$
\end{pf}

 \vspace{5mm}

\begin{lemma}\label{lemma:phi continuous}
    The set-valued function $\varphi$ defined in \eqref{eq:constraint set}  satisfies the continuity definition 17.2 of \citep{inf_dim_analysis}.
\end{lemma}
\begin{pf}
    We define $X :=  W_c \times \mathbb{R}^n$, and $Y := \big\{ Bu_c : u_c \in U_c \big\}$ so that the set-valued function is $\varphi : X \twoheadrightarrow Y$. On the space $X$ we introduce the norm $\|\cdot\|_X$ as $\|(w,d)\|_X = \|w\| + \|d\|$. Since $\|\cdot\|$ is the Euclidean norm, $\|\cdot\|_X$ is a norm on $X$.
    By the definition 17.2 of \citep{inf_dim_analysis}, we need to prove that $\varphi$ is both upper and lower hemicontinuous at all points of $X$.
    
    \vspace{2mm}
    
    First, using Lemma~17.5 of \citep{inf_dim_analysis} we will prove that $\varphi$ is lower hemicontinuous by showing that for an open subset $A$ of $Y$, $\varphi^l(A)$ is open. The lower inverse image of $A$ is defined in \citep{inf_dim_analysis} as
    \begin{align*}
        \varphi^l(A) &:= \big\{ x \in X : \varphi(x) \cap A \neq \emptyset \big\} = 
        \big\{(w,d) \in W_c \times \mathbb{R}^n : Y \cap (\mathbb{R}^+d - \{Cw\}) \cap A \neq \emptyset\big\} \\
        &= \big\{(w,d) \in W_c \times \mathbb{R}^n : \{\lambda d - Cw : \lambda \geq 0\} \cap A \neq \emptyset\big\},
    \end{align*}
    because $A \subset Y$.
    Let $x = (w,d) \in \varphi^l(A)$. Then, there exists $\lambda \geq 0$ such that $\lambda d - Cw \in A$. Since $A$ is open, there exists $\varepsilon > 0$ such that the ball $B_\varepsilon(\lambda d - Cw) \subset A$. Now let $x_1 = (w_1, d) \in X$ and denote $\varepsilon_w := \|w_1 - w\|$ and $\varepsilon_d := \|d_1 - d\|$. Then,
    \begin{align*}
        \|\lambda d_1 - Cw_1 - (\lambda d - Cw) \| &= \| \lambda (d_1 - d) - C(w_1 - w)\| \leq \lambda \varepsilon_d + \|C\| \varepsilon_w.
    \end{align*}
    Since $\lambda \geq 0$ and $\|C\| \geq 0$ are both fixed, we can choose $\varepsilon_d$ and $\varepsilon_w$ positive and small enough so that $\lambda \varepsilon_d + \|C\| \varepsilon_w \leq \varepsilon$.
    
    Then, we have showed that for all $x_1 = (w_1, d_1) \in X$ such that $\|x - x_1\|_X \leq \min(\varepsilon_d, \varepsilon_w)$, i.e., such that $\|w_1 - w\| \leq \varepsilon_w$ and $\|d_1 - d\| \leq \varepsilon_d$, we have $\lambda d_1 - Cw_1 \in B_\varepsilon(\lambda d - Cw) \subset A$, i.e., $x_1 \in \varphi^l(A)$. Therefore, $\varphi^l(A)$ is open, and so $\varphi$ is lower hemicontinuous.
    
    \vspace{3mm}
    
    To prove the upper hemicontinuity of $\varphi$, we will use Lemma~17.4 of \citep{inf_dim_analysis} and prove that for a closed subset $A$ of $Y$, the lower inverse image of $A$ is closed. Let $\{x_k\}$ be a sequence in $\varphi^l(A)$ converging to $x = (w,d) \in X$. We want to prove that $x \in \varphi^l(A)$.
    
    For each $k \geq 0$, we have $(w_k, d_k) = x_k$ and we define $\Lambda_k := \big\{ \lambda_k \geq 0 : \lambda_k d_k - Cw_k \in A \big\} \neq \emptyset$. Since $A$ is a closed subset of the compact set $Y$, then $A$ is compact. Thus $\Lambda_k$ has a minimum and a maximum; we denote them by $\lambda_k^{min}$ and $\lambda_k^{max}$ respectively. 
    
    Since sequences $\{d_k\}$ and $\{w_k\}$ converge, they are bounded. The set $A$ is also bounded, thus sequence $\{\lambda_k^{max}\}$ is bounded. Let $\lambda^{max} := \underset{k\, \geq\, 0}{\sup}\ \lambda_k^{max} > 0$.
    
    We define segments $S_k := \big\{\lambda d_k - Cw_k : \lambda \in [0, \lambda^{max}] \big\}$, and $S := \big\{\lambda d - Cw : \lambda \in [0, \lambda^{max}] \big\}$. These segments are all compact sets.
    We also introduce the sequences $a_k := \lambda_k^{min}d_k - Cw_k \in A \cap S_k$ and $b_k := \lambda_k^{min}d - Cw \in S$.
    
    Take $\varepsilon > 0$. Since $\{d_k\}$ and $\{w_k\}$ converge toward $d$ and $w$ respectively, there exists $N \geq 0$ such that for $k \geq N$, we have $\|d_k - d\| \leq \frac{\varepsilon}{2 \lambda^{max}}$ and $\|w_k - w\| \leq \frac{\varepsilon}{2 \|C\|}$. Then, for any $\lambda_k \in [0, \lambda^{max}]$
    \begin{equation*}
        \| \lambda_k d_k - Cw_k - (\lambda_k d - Cw)\| = \| \lambda_k(d_k - d) - C(w_k - w)\| \leq \lambda_k \frac{\varepsilon}{2 \lambda^{max}} + \|C\| \frac{\varepsilon}{2 \|C\|} \leq \varepsilon.
    \end{equation*}
    Since $\lambda_k^{min} \in [0, \lambda^{max}]$, we have $\|a_k - b_k\| \xrightarrow[k \rightarrow \infty]{} 0$. We define the distance between the sets $A$ and $S$
    \begin{equation*}
        dist(A,S) := \min \big\{ \|a - s_\lambda\| : a \in A,\ s_\lambda \in S\big\}.
    \end{equation*}
    The minimum exists because $A$ and $S$ are both compact and the norm is continuous. Since $a_k \in A$ and $b_k \in S$, we have $dist(A, S) \leq \|a_k - b_k\|$ for all $k \geq 0$. Therefore, $dist(A, S) = 0$. So, $A \cap S \neq \emptyset$, leading to $x \in \varphi^l(A)$. Then, $\varphi^l(A)$ is closed and so $\varphi$ is upper hemicontinuous.    $\quad \blacksquare$
\end{pf}

\vspace{5mm}

\begin{lemma}\label{lemma: lambda continuous}
    Let $X$ and $Y$ be two nonempty symmetric polytopes in $\mathbb{R}^n$ with $X \subset Y^\circ$. Then, $\lambda^*(x ,d) := \underset{y\, \in\, Y}{\max} \big\{ \|x + y\| : x + y \in \mathbb{R}^+ d \big\}$ is continuous in $x \in X$ and $d \in \mathbb{S}$.
\end{lemma}
\begin{pf}
    According to Proposition~\ref{prop: r_X,Y well-defined} (ii), whose proof does not rely on the current lemma, $\lambda^*$ is well-defined. We introduce the set-valued function $\varphi : X \times \mathbb{S} \twoheadrightarrow Y$ defined by $\varphi(x,d) := \big\{ y \in Y : x + y \in \mathbb{R}^+ d \big\} = Y \cap \big( \mathbb{R}^+ d - \{x\} \big)$, where $\mathbb{R}^+ d - \{x\} = \big\{ \lambda d - x : \lambda \geq 0\big\}$. If we take $x = Cw_c$, then $\varphi$ is the same as in \eqref{eq:constraint set}. 
    
    We define the graph of $\varphi$ as $\text{Gr}\, \varphi := \big\{ (x,d,y) \in X \times \mathbb{S} \times Y : y \in \varphi(x,d) \big\}$, and the continuous  function $f: \text{Gr}\, \varphi \rightarrow \mathbb{R}^+$ as $f(x,d,y) = \|x + y\|$. Set $X \times \mathbb{S}$ is compact and nonempty. Since $Y$ is compact and $\mathbb{R}^+ d - \{x\}$ is closed, their intersection $\varphi(x,d)$ is compact. The symmetry of $X$ and $Y$ leads to $-x \in \varphi(x,d)$, so $\varphi(x,d) \neq \emptyset$. According to Lemma~\ref{lemma:phi continuous}, $\varphi$ satisfies the continuity definition $17.2$ of \citep{inf_dim_analysis}. Then, we can apply the Berge Maximum Theorem \citep{inf_dim_analysis} and conclude that $\lambda^*$ is continuous in $x$ and $d$.    $\quad \blacksquare$
\end{pf}

\vspace{5mm}

\begin{lemma}\label{lemma: continuity of x_N}
    Let $X$ and $Y$ be two nonempty polytopes in $\mathbb{R}^n$. Then, the coupled functions $\big( x_N^*, y_N^* \big) (d) = \arg \underset{x\, \in\, X,\, y\, \in\, Y}{\max} \big\{ \|x+y\| : x+y \in \mathbb{R}^+ d \big\}$ are continuous in $d \in \mathbb{S}$.
\end{lemma}
\begin{pf}
    Let $Z := X + Y = \big\{x+y : x \in X, y \in Y\big\}$. Then $Z$ is a nonempty compact set. According to Proposition~\ref{prop: r_X,Y well-defined} (i), whose proof does not rely on the current lemma, $\underset{x\, \in\, X,\, y\, \in\, Y}{\max} \big\{ \|x+y\| : x+y \in \mathbb{R}^+ d \big\}$ exists and thus $\underset{z\, \in\, Z}{\max} \big\{ \|z\| : z \in \mathbb{R}^+ d \big\}$ is also well-defined.
    
    We introduce the set-valued function $\varphi : \mathbb{S} \rightarrow Z$ as $\varphi(d) = Z \cap \mathbb{R}^+d$ for $d \in \mathbb{S}$. The proof of Lemma~\ref{lemma:phi continuous} can be easily adapted to show that $\varphi$ is continuous as it it is defined very similarly to \eqref{eq:constraint set}. The graph of $\varphi$ is $\text{Gr}\, \varphi := \big\{ (z,d) \in Z \times \mathbb{S} : z \in \varphi(d) \big\}$. The function $f : \text{Gr}\, \varphi \rightarrow \mathbb{R}^+$ defined as $f(z,d) = \|z\|$ is obviously continuous. Then, $m(d) := \underset{z\, \in\, \varphi(d)}{\max}\big\{ f(z,d) \big\}$ is continuous by the Berge Maximum Theorem \citep{inf_dim_analysis}.

    We define $z(d) := m(d) d \in Z$ for $d \in \mathbb{S}$. This function is continuous since $m$ is continous, and $z(d) = \arg\underset{z\, \in\, Z}{\max} \big\{\|z\| : z \in \mathbb{R}^+d \big\}$. Since $z(d) = \big(x_N^*, y_N^*\big)(d)$, these functions are continuous.     $\quad \blacksquare$
\end{pf}

\section{Equation of Motion for the Low-Thrust Spacecraft}\label{apx:bar B}

The control matrix $\bar{B}$ can be written as
\begin{equation*}
    \bar{B}(x) := \begin{bmatrix}
    0_{2,3} & B_1(x) & 0_{2,2} & 0_{2,5} \\
    0_{2,3} & 0_{2,4} & 0_{2,2} & B_2(x) \\
    B_3(x) & 0_{2,4} & B_4(x) & B_5(x) \end{bmatrix} \in \mathbb{R}^{6 \times 14},
\end{equation*}
with $0_{i,j}$ denoting the null matrix with $i$ rows and $j$ columns.
We calculate the submatrices using the averaged variational equations for the orbital elements given in \citep{trajectory_dynamics}:

\begin{equation*}
    B_1(x) = \begin{bmatrix}  \sqrt{\frac{a^3}{\mu}}e & 2\sqrt{\frac{a^3}{\mu}}\sqrt{1-e^2} & 0 & 0\\
    \frac{1}{2}\sqrt{\frac{a}{\mu}}(1-e^2) & -\frac{3}{2}e\sqrt{\frac{a}{\mu}}\sqrt{1-e^2} & \sqrt{\frac{a}{\mu}}\sqrt{1-e^2} & -\frac{1}{4}e\sqrt{\frac{a}{\mu}}\sqrt{1-e^2}\end{bmatrix}
\end{equation*}

\begin{equation*}
     B_2(x) = \sqrt{\frac{a}{\mu}}\begin{bmatrix} -\frac{3}{2}e\cos{\omega}\frac{1}{\sqrt{1-e^2}} & \frac{1}{2}(1+e^2)\cos{\omega}\frac{1}{\sqrt{1-e^2}} & -\frac{1}{4}e\cos{\omega}\frac{1}{\sqrt{1-e^2}} & -\frac{1}{2}\sin{\omega} & \frac{1}{4}e\sin{\omega} \\
    -\frac{3}{2}e\sin{\omega}\frac{\csc{i}}{\sqrt{1-e^2}} & \frac{1}{2}(1+e^2)\sin{\omega}\frac{\csc{i}}{\sqrt{1-e^2}} & -\frac{1}{4}e\sin{\omega}\frac{\csc{i}}{\sqrt{1-e^2}} & \frac{1}{2}\cos{\omega}\csc{i} & -\frac{1}{4}e\cos{\omega}\csc{i} \end{bmatrix}
\end{equation*}

\begin{equation*}
    B_3(x) = \begin{bmatrix} \sqrt{\frac{a}{\mu}}\sqrt{1-e^2} & -\frac{1}{2e}\sqrt{\frac{a}{\mu}}\sqrt{1-e^2} & 0 \\
    -3\sqrt{\frac{a}{\mu}} & \sqrt{\frac{a}{\mu}}(\frac{3e}{2}+\frac{1}{2e}) & -\frac{1}{2}e^2\sqrt{\frac{a}{\mu}} \end{bmatrix}
\end{equation*}

\begin{equation*}
    B_4(x) = \sqrt{\frac{a}{\mu}}\begin{bmatrix}  \frac{1}{2}(2-e^2)\frac{1}{e} & -\frac{1}{4}  \\
    -\frac{1}{2e}(2-e^2)\sqrt{1-e^2} & \frac{1}{4}\sqrt{1-e^2}\end{bmatrix}
\end{equation*}

\begin{align*}
    B_5(x) = \cos{i}\sqrt{\frac{a}{\mu}}\begin{bmatrix}
   \frac{3}{2}e\sin{\omega}\frac{\csc{i}}{\sqrt{1-e^2}} & -\frac{1}{2}(1+e^2)\sin{\omega}\frac{\csc{i}}{\sqrt{1-e^2}} &    \frac{1}{4}e\sin{\omega}\frac{\csc{i}}{\sqrt{1-e^2}} & -\frac{1}{2}\csc{i} & \frac{1}{4}e\csc{i} \\
    0 & 0 & 0 & 0 & 0 \end{bmatrix}
\end{align*}
with $\mu = 3.986 \times 10^{14}\, \rm{m^3s^{-2}}$ being the standard gravitational parameter of the Earth.

\end{document}